\documentclass[12pt]{article}
\usepackage[margin=1in]{geometry}
\usepackage{amsmath,amssymb,amsfonts,amsthm}
\usepackage{setspace,booktabs,graphicx,xcolor}
\usepackage{placeins}

\usepackage[round]{natbib}
\usepackage[hidelinks]{hyperref}
\newtheorem{identificationassumption}{Assumption}

\newtheorem{assumption}{Assumption}

\newtheorem{proposition}{Proposition}
\newtheorem{lemma}{Lemma}
\newtheorem{theorem}{Theorem}

\theoremstyle{remark}
\newtheorem{remark}{Remark}
\newcommand{\E}{\mathbb E}

\newcommand{\R}{\mathbb R}
\newcommand{\ind}{\mathbf 1}
\newcommand{\Pn}{\mathbb P_n}
\newcommand{\Gn}{\mathbb G_n}

\newcommand{\norm}[1]{\lVert #1\rVert}

\newcommand{\Appendix}{\appendix}
\DeclareMathOperator{\Var}{Var}
\DeclareMathOperator{\Cov}{Cov}

\title{Testing Exogeneity in Nonparametric Instrumental Variable Models}
\author{Xiaojun Song\thanks{Department of Business Statistics and Econometrics, Guanghua School of Management, Peking University, Beijing, 100871, China. Email: \texttt{sxj@gsm.pku.edu.cn}.}\and Jichao Yuan\thanks{Department of Business Statistics and Econometrics, Guanghua School of Management, Peking University, Beijing, 100871, China. Email: \texttt{2201111030@stu.pku.edu.cn}.}}
\date{\today}

\begin{document}
\maketitle
\begin{abstract}
We propose a test of mean exogeneity in nonparametric instrumental
variable models. Under completeness, the null is equivalent to a
conditional moment restriction involving the ordinary regression
function, so the test avoids estimation of the structural function
and its ill-posed inverse problem. We construct a residual-marked
empirical process and analyze it through an auxiliary U-process of
order two. We derive its limiting distributions under the null and
Pitman-type local alternatives of order $n^{-1/2}$, and establish
consistency against fixed global alternatives. The analysis reveals
a nonparametric estimation effect that invalidates a direct
multiplier bootstrap. We propose a computationally simple multiplier 
bootstrap with a tailor-made indicator correction to account for 
this effect, and establish its asymptotic validity. 
Simulations show good size control and power gains
over existing methods in the nonparametric designs considered.
An application to a wage equation provides evidence against the
mean exogeneity of schooling.

\end{abstract}

\section{Introduction}\label{sec:intro}

Nonparametric instrumental variable (NPIV) models provide a flexible
framework for studying structural economic relationships with endogenous
regressors without imposing a finite-dimensional functional form.
Research on nonparametric identification and IV estimation includes
\citet{Roehrig1988}, \citet{Newey&Powell2003}, \citet{Ai&Chen2003},
\citet{Hall&Horowitz2005}, \citet{Blundell2007a},
\citet{Darolles_etal2011}, \citet{Chen&Pouzo2015},
\citet{Chen&Christensen2018}, and \citet{Chen_etal2025}; related
developments in hypothesis testing include \citet{Horowitz2006},
\citet{Santos2012}, \citet{Breunig2015}, and
\citet{Breunig&Chen2024}. Testing exogeneity is important because 
incorrectly treating a regressor as exogenous can invalidate 
structural conclusions, whereas unnecessarily using NPIV estimation 
can entail a substantial loss of precision.
The exogeneity-testing literature extends from \citet{Durbin1954},
\citet{Wu1973}, and \citet{Hausman1978} to the nonparametric
contributions of \citet{Blundell&Horowitz2007}, \citet{Breunig2015},
\citet{Feve_etal2018}, and \citet{Fu_etal2020}.

Testing exogeneity in a nonparametric instrumental variable (NPIV)
model requires distinguishing the structural regression from the 
ordinary conditional mean. Direct comparison of estimates of these 
functions is unattractive because recovering the structural regression 
entails an ill-posed inverse problem and typically yields slow 
convergence \citep{Hall&Horowitz2005,Blundell&Horowitz2007}. 
A test based on that comparison may therefore have low power
\citep{Blundell&Horowitz2007,Fu_etal2020}. Existing tests can be
broadly grouped into integrated-moment and series-based
implementations, with the latter using sieve approximations. Both
test conditional moments implied by exogeneity without estimating
the structural function. Within the first class, kernel-based tests
estimate the relevant densities and conditional means or quantiles
and use a nonsingular integral operator to form an integrated squared
statistic, as in \citet{Blundell&Horowitz2007} and
\citet{Fu_etal2020}. Series-based tests approximate the conditional
mean or quantile under exogeneity by sieves and aggregate the resulting
residual moments in a quadratic statistic, as in \citet{Breunig2015}
for mean exogeneity and \citet{Breunig2020} for quantile exogeneity.
We adopt the integrated-moment strategy and construct a test based on
the ordinary regression rather than an NPIV estimate of the structural
function.

Within the integrated-moment approach described above, 
our test differs from procedures based on nonsingular integral 
transformations by constructing the statistic directly from 
an indicator-weighted ICM process.
Originally developed for conventional model-specification 
testing by \citet{Bierens1982} and extended by \citet{bierens1990consistent} 
and \citet{Bierens&Ploberger1997}, the ICM principle represents a 
conditional moment restriction by a determining family of unconditional moments; 
indicator-weighted residual empirical processes, 
as in \citet{stute1997nonparametric} and \citet{Delgado&Manteiga2001}, 
provide a prominent implementation. 
\citet{Antoine&Lavergne2023} apply ICM to IV inference with a
nonparametric first stage, but its structural equation is linear, 
whereas our setting allows an unknown NPIV structural function. 
Relative to series/sieve tests, ICM avoids choosing a growing basis 
dimension and a spectral weighting sequence. 
Neither class uniformly dominates in power: the integrated-moment 
test of \citet{Fu_etal2020} has greater power against the class 
of local alternatives emphasized in its comparison, whereas the 
series test of \citet{Breunig2020} can dominate in other directions 
when its series length is appropriately chosen. 
Although ICM tests still involve tuning parameters, 
their global construction avoids building the test statistic 
directly from local estimators of the conditional moment and 
therefore does not require the restrictive bandwidth choices 
associated with traditional kernel-based tests.

We develop an indicator-based ICM test of mean exogeneity for 
an unknown NPIV structural function. We characterize the limiting 
behavior of the test process under the null, fixed alternatives, 
and $n^{-1/2}$ Pitman alternatives. The analysis reveals a 
nonparametric estimation effect, analogous to that studied by 
\citet{Song&Yuan2026} in nonparametric significance testing, 
that arises from replacing the regression errors with nonparametrically 
estimated residuals. This effect invalidates a direct multiplier 
bootstrap that perturbs only the fitted residual marks. We therefore 
propose a tailored multiplier bootstrap that corrects the indicator 
by its estimated conditional expectation given $W$. Unlike conventional 
wild-bootstrap procedures used in nonparametric specification and 
significance testing \citep{Hardle&Mammen1993,Gu_etal2007}, our 
procedure holds the nuisance estimates fixed across draws and requires 
only new multipliers and weighted sums, making it simpler and computationally 
less demanding.

The paper is organized as follows. Section~\ref{sec:model} constructs the observed process and test
statistics. Section~\ref{sec:test.stat} develops their asymptotic
theory, and Section~\ref{sec:altboot} constructs and justifies the
indicator-corrected multiplier bootstrap. Sections~\ref{sec:simulations}
and \ref{sec:application} report the numerical experiments and schooling
illustration. Section~\ref{sec:discussion} discusses the
relationship to significance testing and the role of strong
conditional exogeneity in alternative weighting schemes.

\section{Model, hypotheses, and test statistics}\label{sec:model}

\subsection{Model and hypotheses}

Consider
\begin{equation}\label{eq:stru.model}
 Y=g(X,Z_1)+U,\qquad \E[U\mid Z]=0\quad\text{a.s.},
\end{equation}
where $X\in\R^p$ denotes the potentially endogenous regressors,
$Z_1\in\R^{q_1}$ the included exogenous regressors, and
$Z_2\in\R^{q_2}$ the excluded instruments. Write
$W=(X',Z_1')'\in\R^r$, $Z=(Z_1',Z_2')'\in\R^q$,
$r=p+q_1$, and $q=q_1+q_2$. Instrument validity is maintained
throughout. The structural function solves
\begin{equation}\label{eq:ident}
 \E[Y-g(W)\mid Z]=0\quad\text{a.s.},
\end{equation}
or, equivalently,
\begin{equation}\label{eq:integral.eq}
 \int g(w)\,dF_{W\mid Z}(w\mid z)=\E[Y\mid Z=z].
\end{equation}
Let $G(w)=\E[Y\mid W=w]$. The hypothesis of interest is
\begin{equation}\label{eq:null}
 H_0:\ X\text{ is mean exogenous},
\end{equation}
meaning
\begin{equation}\label{eq:null1}
 H_0:\ \E[U\mid W]=0\quad\text{a.s.},
\end{equation}
or
\begin{equation}\label{eq:null2}
 H_0:\ g(W)=G(W)\quad\text{a.s.}
\end{equation}
We distinguish the structural error $U=Y-g(W)$ from the regression residual $\varepsilon=Y-G(W)$. The identity $\E[\varepsilon\mid W]=0$ holds under every data-generating process; $\varepsilon=U$ requires $H_0$.

\begin{identificationassumption}[Completeness]\label{ass:complete}
The operator $a\mapsto\E[a(W)\mid Z]$ is injective on the relevant
$L^2(P_W)$ space: $\E[a(W)\mid Z]=0$ a.s. implies $a(W)=0$ a.s.
The functions $g$ and $G$ belong to this space.
\end{identificationassumption}

The completeness condition as used by \citet{Newey&Powell2003} for identification of NPIV
models is needed here for the converse implication and for consistency
against all structural alternatives. It is, however, not required for 
deriving the stochastic expansions below. We also note that neither 
instrument validity nor \eqref{eq:null1} implies $\E[U\mid W,Z_2]=0$. 

Under the completeness condition, the following lemma provides an 
equivalent reformulation of the null hypothesis.

\begin{lemma}[Equivalence of null formulations]\label{lem:equiv}
Under model \eqref{eq:stru.model}, instrument validity, and
Assumption~\ref{ass:complete}, \eqref{eq:null1} and \eqref{eq:null2}
are equivalent to
\begin{equation}\label{eq:cmr}
 H_0:\ \E[Y-G(W)\mid Z]=0\quad\text{a.s.}
\end{equation}
\end{lemma}

\subsection{Construction of the test statistics}\label{sec:stat.construction}

Lower-orthant indicators, as used in residual empirical-process tests
by \citet{stute1997nonparametric} and \citet{koul1999nonparametric},
yield an equivalent family of moments:
\begin{equation}\label{eq:ucmr}
 H_0:\ \E[\varepsilon\ind(Z\le z)]=0
 \quad\text{for all }z\in\Pi.
\end{equation}
All inequalities are componentwise. We take $\Pi$ to be the compact
rectangular support of $Z$, including its coordinatewise upper
endpoints, and assume that the distribution of $Z$ has full support on
$\Pi$. Equivalently, one can index by all lower orthants in $\R^q$.
The rectangle must not omit directions needed to characterize the
conditional moment restriction. Other determining weight families
include the exponential weights of \citet{Bierens1982} and
\citet{bierens1990consistent} and the projected-covariate indicators
studied by \citet{escanciano2006consistent}. Their treatment requires
the corresponding entropy conditions. Projection of covariates in
this last weight family is distinct from the conditional-mean
correction of the bootstrap indicator below.

Let $K_h(w,v)$ be the normalized smoothing kernel defined below.
In the interior it equals $h^{-r}K((w-v)/h)$. The leave-one-out
regression and density estimates are
\begin{equation}\label{NW_G}
 \hat G_i=\frac{(n-1)^{-1}\sum_{j\ne i}K_h(W_i,W_j)Y_j}
 {\hat f_i},
\end{equation}
and
\begin{equation}\label{KDE_W}
 \hat f_i=\frac1{n-1}\sum_{j\ne i}K_h(W_i,W_j).
\end{equation}
For the definition of exceptional samples, replace $\hat f_i$ in
denominators by $\max\{\hat f_i,\tau_n\}$, where
$\tau_n\downarrow0$. Since the denominator replacement is required 
only with probability tending to zero, we use the same notation 
for the original and modified quantities.
With $\hat\varepsilon_i=Y_i-\hat G_i$, define
\begin{align}\label{eq:ts.exog}
 \hat T_n(z)
 &=\frac1n\sum_i\hat\varepsilon_i\ind(Z_i\le z)\notag\\
 &=\frac1{n(n-1)}\sum_{i\ne j}
 \frac{K_h(W_i,W_j)(Y_i-Y_j)\ind(Z_i\le z)}{\hat f_i}.
\end{align}
The second equality holds with probability tending to one.

Let $\hat P_{Z,n}=n^{-1}\sum_{j=1}^n\delta_{Z_j}$ denote the empirical
distribution of the instrument vector. The sample versions of the
Kolmogorov--Smirnov and Cram\'er--von Mises statistics are
\begin{equation}\label{eq:ts.s}
 T_n^{\rm KS}=\sqrt n\max_{1\le j\le n}|\hat T_n(Z_j)|,
 \qquad
 T_n^{\rm CvM}=\sum_{j=1}^n\hat T_n(Z_j)^2
 =n\int_\Pi\hat T_n(z)^2\,d\hat P_{Z,n}(z).
\end{equation}
Thus both statistics are evaluated at the observed instrument vectors;
no external marker grid or numerical integration measure is used. The
full-support condition ensures that these sample markers become dense
in $\Pi$ and that the empirical CvM average retains every nonzero
continuous population moment.

\section{Asymptotic theory}\label{sec:test.stat}

Throughout this section, let $G(w)=\E[Y\mid W=w]$ and define the
regression residual and the structural error by
$\varepsilon=Y-G(W)$ and $U=Y-g(W)$, respectively. Thus
$\varepsilon=U+g(W)-G(W)$ and $\E[\varepsilon\mid W]=0$; under
$H_0$, $g(W)=G(W)$ a.s. and hence $\varepsilon=U$. For observation
$i$, write $\varepsilon_i=Y_i-G(W_i)$. Also write $f=f_W$ and
$\chi=(Y,W,Z_2)$, and let $\Pn a=n^{-1}\sum_i a(\chi_i)$ and
$\Gn=\sqrt n(\Pn-P)$. Along triangular sequences, $\E$ denotes
expectation under the law relevant at each $n$; whenever a limiting
null law is involved, it is identified explicitly in words.

\subsection{Regularity conditions}\label{sec:ass}

The following conditions are
explicit sufficient conditions.

\begin{assumption}[Sampling, support, and moments]\label{A1}
The observations are i.i.d. The support $S_W$ is a compact rectangle,
$0<\underline f\le f(w)\le\overline f<\infty$ on $S_W$, and $W$
has a density. The joint distribution of $Z$ has full support on the
compact rectangle $\Pi$, and each coordinate has a continuous
distribution with bounded density. There is a constant
$M<\infty$ such that $\sup_{(w,z_2)\in S_{W,Z_2}}
\E[\exp(Y^2/M^2)\mid W=w,Z_2=z_2]\le2$ up to null sets, where
$S_{W,Z_2}$ is the support of $(W,Z_2)$.
All supports and constants are fixed. In particular, conditional
moments of every fixed order are uniformly bounded.\footnote{The
conditional exponential-moment restriction is a convenient sufficient
condition for the uniform kernel rates used below. It can be replaced
by sufficiently high conditional moments together with suitable
truncation conditions, but that extension is not pursued here.}
\end{assumption}

\begin{assumption}[Smoothness]\label{A2}
We adopt the notation $\mathcal L_\beta^\alpha$ from
\citet{Robinson1988} and \citet{Delgado&Manteiga2001}. It denotes the
class of functions $a$ that are
$(b-1)$-times continuously differentiable, where $b-1\le\beta\le b$,
whose partial derivatives through order $b-1$ have finite
$\alpha$th moments, and for which there exist $\rho>0$ and a function
$d_a$ with a finite $\alpha$th moment such that
\[
 \sup_{\norm{v-w}<\rho}
 \frac{|a(v)-Q_a(v,w)|}{\norm{v-w}^\beta}\le d_a(w),
\]
where $Q_a(v,w)$ is the Taylor polynomial in $v-w$ of degree $b-1$,
with coefficients given by the corresponding partial derivatives of
$a$ at $w$. When $\alpha=\infty$, the derivatives and $d_a$ are
uniformly bounded. For some even integer $s\ge2$, $f$ and $G$ belong
to $\mathcal L_s^\infty$, and, uniformly in $z_2$,
$F(z_2\mid\cdot)=F_{Z_2\mid W}(z_2\mid\cdot)=P(Z_2\le z_2\mid W=\cdot)$ and
$\psi(z_2,\cdot)=\E[Y\ind(Z_2\le z_2)\mid W=\cdot]$ belong to
$\mathcal L_s^\infty$. Each function is understood to admit an
extension to a neighborhood of $S_W$.
\end{assumption}

\begin{assumption}[Kernel and boundary correction]\label{A3}
Let $\mathcal K_s$ be the class of bounded, symmetric, compactly
supported univariate functions $k$ of bounded variation satisfying
\[
 \int k(u)\,du=1,\qquad
 \int u^j k(u)\,du=0\quad(1\le j<s),
 \qquad \int |u|^s|k(u)|\,du<\infty.
\]
In the interior,
\[
 K_h(w,v)=h^{-r}K((w-v)/h),\qquad
 K(u)=\prod_{\ell=1}^r k_\ell(u_\ell),\quad k_\ell\in\mathcal K_s.
\]
At the boundary, use the tensor product of univariate local-polynomial
equivalent kernels of degree $s-1$, formed from a positive,
compactly supported base kernel on each truncated coordinate interval.
The resulting deterministic equivalent kernels satisfy
\[
 \int_{S_W}K_h(w,v)\,dv=1,\qquad
 \int_{S_W}(v-w)^\alpha K_h(w,v)\,dv=0
 \quad(1\le|\alpha|<s).
\]
Uniformly in $w$, $|K_h(w,v)|\le C h^{-r}
\ind(\norm{w-v}_\infty\le Ch)$ and
$\int|K_h(w,v)|\,dv\le C$. Boundary modification occurs only within
distance $Ch$ of the boundary of $S_W$. The same corrected
kernel is used in \eqref{NW_G} and \eqref{KDE_W}.
\end{assumption}

\begin{assumption}[Bandwidth rates]\label{A4}
Write $a_n=\{\ln n/(nh^r)\}^{1/2}+h^s$. As $n\to\infty$, require
\[
 h\to0,\qquad \frac{nh^r}{\ln n}\to\infty,\qquad
 \sqrt n\,h^s\to0,\qquad
 \frac{n^2h^{3r}}{(\ln n)^3}\to\infty,\qquad
 \frac{h^{2-r}}{\sqrt n}\to0.
\]
\end{assumption}

\begin{remark}\label{rem:bandwidth}
The first two restrictions in Assumption~\ref{A4} ensure shrinking
kernel neighborhoods with sufficiently many local observations. The
condition $\sqrt n h^s\to0$ removes the first-order smoothing bias.
The fourth restriction controls the third-order remainder from the
reciprocal-density expansion; together with the bias restriction it
implies $\sqrt n a_n^3\to0$. The final restriction controls the
uncancelled bias of repeated-index terms under the general null and is
automatic when $r\le2$. For $h=n^{-\gamma}$, a sufficient open range is
\[
 \frac1{2s}<\gamma<\frac2{3r},\qquad
 \gamma<\frac1{2(r-2)}\ \text{if }r>2.
\]
Thus a second-order kernel permits $1/4<\gamma<2/3$ for $r=1$
and $1/4<\gamma<1/3$ for $r=2$. For $r\ge3$, higher-order
smoothing is needed; in general, an even $s$ larger than
$\max\{3r/4,r-2\}$ gives a nonempty range.
\end{remark}

\begin{assumption}[Multipliers]\label{A5}
The multipliers\footnote{Standard Gaussian, Rademacher, and Mammen's two-point multipliers satisfy this condition; see \citet{Mammen1993}.} $V_i$ are i.i.d., independent of the data, with $\E V_i=0$ and $\E V_i^2=1$. For some $\delta>0$,
$\E|V_i|^{2+\delta}<\infty$.
\end{assumption}

\subsection{Two uniform representations}\label{sec:expansions}

For $z=(z_1',z_2')'$, write
\[
 m(z)=\E[\varepsilon\ind(Z\le z)].
\]
Since $W=(X',Z_1')'$, we have, almost surely,
\[
 \E[\ind(Z\le z)\mid W]
 =\ind(Z_1\le z_1)F_{Z_2\mid W}(z_2\mid W).
\]
In particular, at $W_i=(X_i',Z_{1i}')'$, the conditional-indicator
factor is $\ind(Z_{1i}\le z_1)F_{Z_2\mid W}(z_2\mid X_i,Z_{1i})$.
For $w=(x',w_1')'$, its value at $W=w$ is
$\ind(w_1\le z_1)F_{Z_2\mid W}(z_2\mid w)$.
Empty $Z_1$ indicator factors are one. Sample-indexed conditional
moments are written using the same index for the variables and the
conditioning vector, as in
 $\E[\varepsilon_i\ind(Z_i\le z)\mid W_i]$.
Define the auxiliary population-denominator process explicitly by
\[
 \hat T_n^f(z):=
 \frac1{n(n-1)}\sum_{i=1}^n\sum_{j\ne i}
 \frac{K_h(W_i,W_j)(Y_i-Y_j)\ind(Z_i\le z)}{f_W(W_i)}.
\]
The superscript $f$ indicates that this process uses the true density
$f_W(W_i)$ in its denominator. In contrast, the observed process
$\hat T_n$ in \eqref{eq:ts.exog} uses the leave-one-out estimate
$\hat f_i$. Thus $\hat T_n^f$ is an infeasible auxiliary process,
used in Propositions~\ref{prop:density} and \ref{prop:linear};
the actual test uses $\hat T_n$.

\begin{proposition}[Density replacement and its first-order correction]
\label{prop:density}
Under Assumptions~\ref{A1}--\ref{A4}, uniformly in $z\in\Pi$,
\begin{equation}\label{aux1}
 \begin{aligned}
 \hat T_n(z)
 ={}&\hat T_n^f(z)\\
 &-\frac1n\sum_{i=1}^n\Big\{
       \E[\varepsilon_i\ind(Z_i\le z)\mid W_i]-m(z)\Big\}
       +o_p(n^{-1/2}).
 \end{aligned}
\end{equation}
\end{proposition}

\begin{proposition}[Uniform representation of the population-denominator process]
\label{prop:linear}
Under Assumptions~\ref{A1}--\ref{A4}, uniformly in $z\in\Pi$,
\begin{equation}\label{eq:uniform_final}
 \begin{aligned}
 \hat T_n^f(z)
 ={}&\frac1n\sum_{i=1}^n\Big[
      \varepsilon_i\{\ind(Z_i\le z)-\ind(Z_{1i}\le z_1)F_{Z_2\mid W}(z_2\mid W_i)\}
      \\
 &\qquad\qquad\qquad
      +\E[\varepsilon_i\ind(Z_i\le z)\mid W_i]\Big]
      -m(z)+o_p(n^{-1/2}).
 \end{aligned}
\end{equation}
\end{proposition}

The expansions in Propositions~\ref{prop:density} and
\ref{prop:linear} will be useful for constructing the bootstrap
approximation and are analyzed in detail in Section~\ref{sec:altboot}.
Furthermore, combining Propositions~\ref{prop:density} and \ref{prop:linear}
cancels both the conditional-moment average and the deterministic
$m(z)$ terms. Hence the observed residual process satisfies,
uniformly in $z\in\Pi$,
\begin{equation}\label{eq:observed.expansion}
 \begin{aligned}
 \hat T_n(z)
 &=\frac1n\sum_{i=1}^n\varepsilon_i
      \{\ind(Z_i\le z)-\ind(Z_{1i}\le z_1)F_{Z_2\mid W}(z_2\mid W_i)\}
      +o_p(n^{-1/2}).
 \end{aligned}
\end{equation}
Since $\E[\varepsilon\mid W]=0$, the summand in
\eqref{eq:observed.expansion} has mean $m(z)$, not zero in general.
The centering of the indicator therefore does not remove the
population signal under alternatives. Unlike the residual-process
expansion in \citet{Delgado&Manteiga2001}, estimation of $f_W$
produces an additional first-order component. In the Hoeffding
decomposition of $\hat T_n^f$, the projection onto the second
observation contains a conditionally noncentered component; the
density-replacement term in Proposition~\ref{prop:density} cancels
this component, yielding \eqref{eq:observed.expansion}.

\begin{remark}[The density correction is generally nonzero]
\label{rmk:counterexample}
Under $H_0$, $m(z)=0$, but
$q_z(W):=\E[U\ind(Z\le z)\mid W]$ need not vanish. If
$\Var\{q_z(W)\}>0$ for some $z$, the density-replacement term in
Proposition~\ref{prop:density} is of order $n^{-1/2}$, so both
components in the cancellation above are essential. This can occur
even under completeness. For example, let $q_1=0$, let $W$ be uniform
on $[-\pi,\pi)$, and set $Z_2=(W+V)\bmod 2\pi$, where $V$ is
independent wrapped normal noise with Fourier coefficients
$\lambda_k=\exp(-\sigma^2k^2/2)$. With
$a=\lambda_1/(1+\lambda_2)$ and
\[
 g=G=0,\qquad
 U=Y=\cos(W+Z_2)-a\{\cos(2W)+\cos(2Z_2)\},
\]
$\E[U\mid W]=\E[U\mid Z_2]=0$, and completeness follows because
$\lambda_k\ne0$ for every $k$. Nevertheless, $U$ is nonzero, so the
determining property of conditional lower-orthant moments implies
$\Var\{q_z(W)\}>0$ for some $z$.
\end{remark}

\begin{theorem}[Observed process under the null]\label{thm:process.null}
Under $H_0$ and Assumptions~\ref{A1}--\ref{A4},
\[
 \sqrt n\hat T_n\Rightarrow\mathbb G
 \qquad\text{in }\ell^\infty(\Pi),
\]
where $\mathbb G$ is a centered Gaussian process with covariance
\begin{equation}\label{eq:covariance}
 \begin{aligned}
 \Sigma(z,z')=\E\Big[&
 U^2\{\ind(Z\le z)-\ind(Z_1\le z_1)F_{Z_2\mid W}(z_2\mid W)\}\\
 &\times\{\ind(Z\le z')-\ind(Z_1\le z_1')F_{Z_2\mid W}(z_2'\mid W)\}\Big].
 \end{aligned}
\end{equation}
\end{theorem}

Following \citet{Delgado&Manteiga2001}, write the Pitman alternative
as an $n^{-1/2}$ conditional-mean departure. The observations are
i.i.d. within each sample size $n$. Keep the distribution of
$(W,Z)$ and the structural function $g$ fixed. Under the law for the
$n$th sample, the local alternative is
\begin{equation}\label{eq:process.local.alternative}
 H_{1n}:\quad G_n(W):=\E[Y\mid W]
       =g(W)+\frac{\delta(W)}{\sqrt n}\quad\text{a.s.},
\end{equation}
where $\delta$ is a fixed bounded direction with
$P\{\delta(W)\ne0\}>0$. At every $n$, retain
$Y=g(W)+U$ and $\E[U\mid Z]=0$; only directions compatible with
this maintained instrument-validity restriction are considered.
Thus \eqref{eq:process.local.alternative} is equivalently
$\E[U\mid W]=n^{-1/2}\delta(W)$, not a nonzero conditional
mean for $\varepsilon$. Indeed, within the $n$th sample,
\[
 \varepsilon=Y-G_n(W)=U-\frac{\delta(W)}{\sqrt n},
 \qquad \E[\varepsilon\mid W]=0.
\]
Instrument validity then gives the explicit local mean
\[
 m_n(z):=\E[\varepsilon\ind(Z\le z)]
       =-\frac1{\sqrt n}\E[\delta(W)\ind(Z\le z)].
\]

\begin{theorem}[Observed process under Pitman local alternatives]
\label{thm:process.local}
Under $H_{1n}$ in \eqref{eq:process.local.alternative}, suppose
Assumptions~\ref{A1}--\ref{A4} hold uniformly along the sequence
and at the limiting null law. Assume also the residual
second-moment stability condition
\[
 \E\left|\E[\varepsilon^2\mid W,Z_2]
              -\E[U^2\mid W,Z_2]\right|\to0,
\]
where the first conditional moment is evaluated under the law for the
$n$th sample and the second under the limiting null law.
Then, under $H_{1n}$,
\[
 \sqrt n\hat T_n(z)\Rightarrow
       \mathbb G(z)-\E[\delta(W)\ind(Z\le z)]
 \qquad\text{in }\ell^\infty(\Pi),
\]
where $\mathbb G$ is the null Gaussian process in
Theorem~\ref{thm:process.null}, evaluated under the limiting null law.
If Assumption~\ref{ass:complete} also holds, the deterministic
shift is not identically zero.
\end{theorem}

\begin{theorem}[Observed process under fixed alternatives]
\label{thm:process.fixed}
Under a fixed law $P$ satisfying Assumptions~\ref{ass:complete} and
\ref{A1}--\ref{A4}, with $P\{g(W)\ne G(W)\}>0$,
\[
 \sqrt n\{\hat T_n-m\}\Rightarrow\mathbb G_P
 \qquad\text{in }\ell^\infty(\Pi),
\]
where $\mathbb G_P$ is a centered Gaussian process with covariance
\[
 \begin{aligned}
 \operatorname{Cov}\{\mathbb G_P(z),\mathbb G_P(z')\}
 ={}&\E\Big[\varepsilon^2
       \{\ind(Z\le z)-\ind(Z_1\le z_1)F_{Z_2\mid W}(z_2\mid W)\}\\
 &\qquad\quad\times
       \{\ind(Z\le z')-\ind(Z_1\le z_1')F_{Z_2\mid W}(z_2'\mid W)\}\Big]
       -m(z)m(z').
 \end{aligned}
\]
Moreover, $m$ is not identically zero,
\[
 \sup_{z\in\Pi}|\hat T_n(z)-m(z)|\to_p0,
 \qquad
 \sqrt n\sup_{z\in\Pi}|\hat T_n(z)|\to_p\infty.
\]
In particular, the uncentered process $\sqrt n\hat T_n$ does not
have the centered null limit under a fixed alternative.
\end{theorem}

\begin{remark}[KS and CvM statistics under the null and alternatives]
\label{rmk:statistics}
The process limits above give the asymptotic distributions of the
statistics defined in \eqref{eq:ts.s}. Under the conditions of
Theorem~\ref{thm:process.null},
\[
 \begin{aligned}
 T_n^{\rm KS}&\Rightarrow\sup_{z\in\Pi}|\mathbb G(z)|,\quad
 T_n^{\rm CvM}&\Rightarrow\int_\Pi\mathbb G(z)^2\,dF_Z(z).
 \end{aligned}
\]
Under the local alternatives of
Theorem~\ref{thm:process.local},
\[
 \begin{aligned}
 T_n^{\rm KS}&\Rightarrow
       \sup_{z\in\Pi}\left|\mathbb G(z)
                    -\E[\delta(W)\ind(Z\le z)]\right|,\\
 T_n^{\rm CvM}&\Rightarrow
       \int_\Pi\left\{\mathbb G(z)
                    -\E[\delta(W)\ind(Z\le z)]\right\}^2\,dF_Z(z).
 \end{aligned}
\]
Under the fixed alternatives of Theorem~\ref{thm:process.fixed},
\[
 \frac{T_n^{\rm KS}}{\sqrt n}\to_p\sup_{z\in\Pi}|m(z)|>0,
 \qquad
 \frac{T_n^{\rm CvM}}n\to_p\int_\Pi m(z)^2\,dF_Z(z)>0.
\]
Thus both statistics diverge under a fixed alternative. Positivity
of the CvM limit uses the full support of $F_Z$ and the continuity
of $m$ implied by Assumption~\ref{A1}. Because the evaluation points
and the CvM measure are sample dependent, these conclusions do not
follow from the process limits alone. Lemma~\ref{lem:sample.functionals}
shows that the sample points become dense in $\Pi$ and establishes the
required joint convergence with $\hat P_{Z,n}$. The corresponding
bootstrap conclusions are collected in Remark~\ref{rmk:bootstrap.statistics} below.
\end{remark}

\section{Bootstrap and asymptotic properties}\label{sec:altboot}

\subsection{Bootstrap construction and implementation}
\label{sec:bootstrap.construction}\label{sec:implementation}
Because the asymptotic null distribution of our test statistics 
is case-dependent, critical values must be obtained by bootstrap 
simulation. Traditional procedures generate bootstrap samples 
using either residual-based resampling, as in \citet{Koul&Lahiri1994}, 
or wild-bootstrap errors, as in \citet{Mammen1993,Hardle&Mammen1993,Gu_etal2007}; 
these procedures generally require the nonparametric estimates and the 
test statistics to be recomputed for each bootstrap draw. 
An alternative is the multiplier bootstrap based on the empirical-process 
theory of \citet{van1996weak}, which applies independent multipliers directly 
to the estimated first-order representation. Because the nuisance quantities 
and residual marks are computed only once, each replication requires only a 
new set of multipliers and weighted sums, making the procedure simple and 
computationally attractive. In the present setting, however, a direct multiplier
analog would attach $V_i$ to the fitted residual mark and use
$n^{-1}\sum_iV_i\hat\varepsilon_i\ind(Z_i\le z)$. This
construction is generally invalid. Propositions~\ref{prop:density}
and \ref{prop:linear} show that estimating the nonparametric regression
numerator and the regressor density has a first-order effect: after the
two contributions cancel, the influence function of the observed
process contains $\varepsilon\{\ind(Z\le z)-\ind(Z_1\le z_1)F_{Z_2\mid W}(z_2\mid W)\}$, rather than $\varepsilon\ind(Z\le z)$. Direct perturbation reproduces the latter mark and hence the covariance $\E[U^2\ind(Z\le z)\ind(Z\le z')]$, whereas the observed process has the conditionally centered covariance in \eqref{eq:covariance}. The restrictions $\E[U\mid W]=\E[U\mid Z]=0$ do not make these covariance kernels equal, and there is no general ordering between them.

The ``nonparametric estimation effect'' described above prevents the direct
use of the computationally attractive and easy-to-implement multiplier bootstrap. 
A key computational contribution of this paper is to restore its validity 
through the following tailored correction. To reproduce \eqref{eq:observed.expansion}, 
we correct the indicator by its conditional expectation given $W$. 
At the population level,
\[
 \ind(Z\le z)-\ind(Z_1\le z_1)F_{Z_2\mid W}(z_2\mid W)
 =\ind(Z_1\le z_1)
       \{\ind(Z_2\le z_2)-F(z_2\mid W)\}.
\]
The corrected indicator has conditional mean zero given $W$.
Since $\E[\varepsilon\mid W]=0$,
\[
 \E\!\left[\varepsilon
       \{\ind(Z\le z)-\ind(Z_1\le z_1)F_{Z_2\mid W}(z_2\mid W)\}\right]
 =\E[\varepsilon\ind(Z\le z)]=m(z).
\]
We estimate the conditional distribution using the same leave-one-out
kernel and denominator convention as in \eqref{NW_G}:
\begin{equation}\label{eq:conditional.estimates}
 \hat F_i(z_2)
 =\frac{\sum_{j\ne i}K_h(W_i,W_j)\ind(Z_{2j}\le z_2)}
          {\sum_{j\ne i}K_h(W_i,W_j)}.
\end{equation}
The indicator-corrected multiplier process is
\begin{equation}\label{eq:mb}
 \hat T_n^*(z)=\frac1n\sum_{i=1}^n
 V_i\hat\varepsilon_i
       \{\ind(Z_i\le z)-\ind(Z_{1i}\le z_1)\hat F_i(z_2)\}.
\end{equation}
Only $r$-dimensional smoothing is required.

\begin{remark}[Equivalent decomposition and estimation effects]
\label{rmk:bootstrap.decomposition}
Equation~\eqref{eq:mb} admits the algebraically equivalent decomposition
in \eqref{eq:mb.mimic} below. To write it, define the auxiliary marked
conditional mean
\[
 \hat\psi_i(z_2)
 =\frac{\sum_{j\ne i}K_h(W_i,W_j)Y_j\ind(Z_{2j}\le z_2)}
          {\sum_{j\ne i}K_h(W_i,W_j)}.
\]
Then
\begin{equation}\label{eq:mb.mimic}
 \begin{aligned}
 \hat T_n^*(z)=\frac1n\sum_iV_i\Big[
 \hat\varepsilon_i\ind(Z_i\le z)
 &-\ind(Z_{1i}\le z_1)\{Y_i\hat F_i(z_2)-\hat\psi_i(z_2)\}\\
 &-\ind(Z_{1i}\le z_1)
       \{\hat\psi_i(z_2)-\hat G_i\hat F_i(z_2)\}\Big].
 \end{aligned}
\end{equation}
The first subtraction reproduces the regression-numerator estimation
effect appearing in Proposition~\ref{prop:linear}, whereas the second
reproduces the density-replacement correction isolated in
Proposition~\ref{prop:density}. Their sum is exactly
$\hat\varepsilon_i\ind(Z_{1i}\le z_1)\hat F_i(z_2)$.
Keeping only the first subtraction is generally insufficient,
including under $H_0$, because the second has a population counterpart
$\E[\varepsilon\ind(Z\le z)\mid W]$. It need not vanish under
mean exogeneity, as Remark~\ref{rmk:counterexample} demonstrates.
The auxiliary quantity $\hat\psi_i$ is useful only for this decomposition
and is not needed to implement \eqref{eq:mb}.
\end{remark}

Apply the same KS and CvM functionals as in \eqref{eq:ts.s}:
\begin{equation}\label{eq:ts.bootstrap}
 T_n^{*,\rm KS}=\sqrt n\max_{1\le j\le n}|\hat T_n^*(Z_j)|,
 \qquad
 T_n^{*,\rm CvM}=\sum_{j=1}^n\hat T_n^*(Z_j)^2
 =n\int_\Pi\hat T_n^*(z)^2\,d\hat P_{Z,n}(z).
\end{equation}
The observed and bootstrap statistics use the same sample markers and
the same empirical measure $\hat P_{Z,n}$. No subtraction of
$\hat T_n$ is made from
$\hat T_n^*$: its conditional mean is already zero. Reject when
the observed statistic exceeds the conditional $(1-\alpha)$
quantile of the corresponding statistic in
\eqref{eq:ts.bootstrap}.

For implementation, compute the leave-one-out residuals in \eqref{NW_G}, the observed process in \eqref{eq:ts.exog}, and $\hat F_i(z_2)$ once. Hold these fitted quantities fixed throughout the bootstrap. For each draw of the multipliers, form \eqref{eq:mb} and apply the same KS or CvM functional and integration rule. Use a kernel order and bandwidth satisfying Assumption~\ref{A4}, with the boundary correction in Assumption~\ref{A3}. Evaluate both processes at
the observed markers $Z_1,\ldots,Z_n$. With $B$ multiplier draws, report the Monte Carlo $p$-value
\[
 \frac{1+\sum_{b=1}^B\ind(T_n^{*(b)}\ge T_n)}{B+1}.
\]
Its minimum is $1/(B+1)$; this adjustment records simulation
resolution, not finite-sample exactness.

\subsection{Asymptotic validity}\label{sec:bootstrap.validity}

Write $P^*$ and $\E^*$ for probability and expectation conditional
on the entire sample. Conditional weak convergence and conditional
remainder statements are in probability over the data. We first give
the uniform representation of the feasible bootstrap process.

\begin{proposition}[Uniform bootstrap representation]
\label{prop:bootstrap.expansion}
Under Assumptions~\ref{A1}--\ref{A5}, let $\hat T_n^*$ be the
bootstrap process in \eqref{eq:mb}, using the leave-one-out ratio
$\hat F_i$ in \eqref{eq:conditional.estimates}. Then, uniformly
in $z\in\Pi$,
\begin{equation}\label{eq:bootstrap.expansion}
 \hat T_n^*(z)=\frac1n\sum_{i=1}^n V_i\varepsilon_i
       \{\ind(Z_i\le z)-\ind(Z_{1i}\le z_1)F_{Z_2\mid W}(z_2\mid W_i)\}
       +o_{P^*}(n^{-1/2}).
\end{equation}
Here $\varepsilon_i=Y_i-G(W_i)$. The representation does not
require $H_0$; under $H_0$ one may replace $\varepsilon_i$ by
$U_i$. It also holds along \eqref{eq:process.local.alternative}
when Assumptions~\ref{A1}--\ref{A5} hold uniformly, with
$G$ replaced by $G_n$. Conditional second-moment stability is
needed only for the local Gaussian limit.
\end{proposition}

Proposition~\ref{prop:bootstrap.expansion} reduces the feasible
bootstrap to a multiplier average of population marks. Together with
Lemma~\ref{lem:residual.process}, it yields validity under the null and
the corresponding limits under local and fixed alternatives.

\begin{theorem}[Bootstrap under the null]\label{thm:null}
Under $H_0$ and Assumptions~\ref{A1}--\ref{A5},
\begin{equation}\label{eq:mb_11}
 \sqrt n\hat T_n^*\Rightarrow_{P^*}\mathbb G
 \qquad\text{in }\ell^\infty(\Pi),
\end{equation}
where $\mathbb G$ is the null process in
Theorem~\ref{thm:process.null}, with covariance
\eqref{eq:covariance}.
\end{theorem}

\begin{theorem}[Bootstrap under Pitman alternatives]\label{thm:local}
Under the conditions of Theorem~\ref{thm:process.local} and
Assumption~\ref{A5},
\[
 \begin{aligned}
 \sqrt n\hat T_n^*&\Rightarrow_{P^*}\mathbb G,\quad
 \sqrt n\hat T_n(z)&\Rightarrow
            \mathbb G(z)-\E[\delta(W)\ind(Z\le z)].
 \end{aligned}
\]
Here $\mathbb G$ is evaluated under the limiting null law. If
Assumption~\ref{ass:complete} holds, the shift is nonzero for every
nonzero admissible direction $\delta$.
\end{theorem}

\begin{theorem}[Bootstrap under fixed alternatives]\label{thm:consistency}
Under a fixed law $P$ satisfying Assumptions~\ref{A1}--\ref{A5},
\[
 \sqrt n\hat T_n^*\Rightarrow_{P^*}\mathbb G_P^*
 \qquad\text{in }\ell^\infty(\Pi),
\]
where $\mathbb G_P^*$ is centered Gaussian with covariance
\begin{equation}\label{eq:bootstrap.fixed.covariance}
 \begin{aligned}
 \Cov\{\mathbb G_P^*(z),\mathbb G_P^*(z')\}
 =\E\Big[&\varepsilon^2
       \{\ind(Z\le z)-\ind(Z_1\le z_1)F_{Z_2\mid W}(z_2\mid W)\}\\
       &\times\{\ind(Z\le z')-\ind(Z_1\le z_1')F_{Z_2\mid W}(z_2'\mid W)\}\Big].
 \end{aligned}
\end{equation}
In particular,
\[
 \Cov\{\mathbb G_P^*(z),\mathbb G_P^*(z')\}
 =\Cov\{\mathbb G_P(z),\mathbb G_P(z')\}+m(z)m(z'),
\]
so the bootstrap does not estimate the centered
fixed-alternative sampling law.
\end{theorem}

\begin{remark}[Bootstrap KS and CvM tests]\label{rmk:bootstrap.statistics}
The KS and CvM conclusions follow from the preceding process limits
and the sample-functional approximation in
Lemma~\ref{lem:sample.functionals}. Under $H_0$,
\[
 T_n^{*,\rm KS}\Rightarrow_{P^*}\sup_{z\in\Pi}|\mathbb G(z)|,\quad
 T_n^{*,\rm CvM}\Rightarrow_{P^*}\int_\Pi\mathbb G(z)^2\,dF_Z(z).
\]
Let $c_{1-\alpha,\rm KS}$ and $c_{1-\alpha,\rm CvM}$ denote the
unique $(1-\alpha)$ quantiles of the two null functionals, and suppose
their distributions are continuous there. The corresponding bootstrap
tests then have limiting rejection probability $\alpha$. Under the Pitman alternatives
of Theorem~\ref{thm:local}, the bootstrap limit remains unshifted,
whereas the observed process has the shift
$-\E[\delta(W)\ind(Z\le z)]$. Hence, if the shifted functionals have
no atoms at the null critical values, the limiting rejection
probabilities are
\[
 \begin{aligned}
 &P\left\{\sup_{z\in\Pi}\left|\mathbb G(z)
       -\E[\delta(W)\ind(Z\le z)]\right|
                 >c_{1-\alpha,\rm KS}\right\},\\
 &P\left\{\int_\Pi\left\{\mathbb G(z)
       -\E[\delta(W)\ind(Z\le z)]\right\}^2\,dF_Z(z)
                 >c_{1-\alpha,\rm CvM}\right\}.
 \end{aligned}
\]
The symmetric convex acceptance regions imply limiting power at least
$\alpha$. Under a fixed alternative satisfying completeness and
$P\{g(W)\ne G(W)\}>0$, Theorem~\ref{thm:consistency} implies that the
bootstrap critical values are $O_p(1)$, while the observed KS and CvM
statistics diverge at the rates given in
Remark~\ref{rmk:statistics}; both tests are therefore
consistent. Simulated critical values additionally require $B\to\infty$.
\end{remark}

\section{Simulation}\label{sec:simulations}

\subsection{Experimental setup}\label{sec:sim-setup}

We consider three designs with no additional controls, so $W=X$.
In each design, $X$ and $Z$ have uniform marginal distributions on
$[-1,1]$, and $\eta\sim U[-1,1]$ is independent of $(X,Z)$.
The outcome is generated as $Y=g(X)+\rho R+e_0$, where $g$ and
the baseline error $e_0$ vary across designs. We construct $R$ by
subtracting from a function of $X$ its conditional mean given $Z$,
so that
\begin{equation}\label{eq:sim.disturbance}
 R=r(X)-\E[r(X)\mid Z],\qquad
 \E[R\mid Z]=0,\qquad \E[R\mid X]=\delta(\pi X).
\end{equation}
The baseline errors satisfy $\E[e_0\mid X]=\E[e_0\mid Z]=0$.
Consequently, instruments remain valid for every $\rho$, while
$G(X)-g(X)=\rho\delta(\pi X)$ controls departure from mean
exogeneity.

The dependence between $X$ and $Z$ combines independent uniform
variation, a wrapped-normal component, and a smooth sine interaction.
Write $s(t)=(\sin t,\ldots,\sin5t)'$ and $\delta(t)=b's(t)$, where
$b$ is a fixed vector of positive weights.\footnote{Specifically,
$b=(1-e^{-1})(42,12288,177147,524288,390625)'/
\sqrt{458997853886}$. This normalization gives
$\E[\delta(\pi X)^2]=(1-e^{-1})^2/2$.}
Their joint density on $[-1,1]^2$ is
\begin{equation}\label{eq:sim.joint}
 f_{X,Z}(x,z)=\frac14\left\{
 0.8+0.4\pi w_{0.35}\bigl(\pi(z-x)\bigr)
 +0.6\frac{\delta(\pi x)}{\sum_{j=1}^5b_j}\sin(\pi z)\right\},
\end{equation}
where $w_{0.35}$ is the wrapped $N(0,0.35^2)$ density on
$[-\pi,\pi)$. Equivalently, the mixture draws an independent
uniform pair with probability $0.2$, a uniform angle and its
normal perturbation modulo $2\pi$ with probability $0.2$ (dividing
both angles by $\pi$), and
a pair with the sine-modulated conditional density with probability
$0.6$. Both margins remain uniform. The independent component
ensures $f_{X,Z}\ge0.05$, and the wrapped-normal component retains
nonzero Fourier coefficients. The sine interaction changes only
the first sine output coefficient and preserves completeness.

For an explicit construction of $r$, let
\[
 \begin{aligned}
 M_{jk}&=0.2e^{-0.35^2j^2/2}\ind(j=k)
       +\frac{0.3b_k}{\sum_{\ell=1}^5b_\ell}\ind(j=1),
       \quad j,k=1,\ldots,5,\\
 r(x)&=b'(I-M'M)^{-1}s(\pi x).
 \end{aligned}
\]
Here $\E[s(\pi X)\mid Z]=M's(\pi Z)$. The independent mixture
component makes the conditional-expectation operator a strict
contraction on zero-mean functions, so $I-M'M$ is positive
definite. These identities give \eqref{eq:sim.disturbance}.
The smooth positive density and bounded outcomes in all three
designs satisfy the support, moment, and smoothness conditions in
Assumptions~\ref{A1}--\ref{A2}.

\phantomsection\label{sec:sim-bandwidth}
We use the Epanechnikov kernel
$k(u)=\tfrac34(1-u^2)\ind(|u|\le1)$, with the deterministic
first-order boundary correction in Assumption~\ref{A3}, for the
regression, density, and conditional-indicator estimators. The
bandwidth coefficient $c$ is selected by leave-one-out least-squares CV over
$\{0.5,0.8,1.0,1.2,1.5\}$:
\begin{equation}\label{eq:sim.cv}
 \hat c_{\rm CV}=\operatorname*{arg\,min}_{c\in\mathcal C}
 \frac1n\sum_{i=1}^n
 \{Y_i-\hat G_{-i,c n^{-0.30}}(X_i)\}^2.
\end{equation}
To examine bandwidth sensitivity, let $k$ denote a multiplier
relative to the CV-selected bandwidth and use
\begin{equation}\label{eq:sim.cv.undersmooth}
 \hat h_{\rm CV}=\hat c_{\rm CV}n^{-0.30},\qquad
 h_k=k\hat h_{\rm CV}=k\hat c_{\rm CV}n^{-0.30},\quad
 k\in\{0.5,0.8,1.0,1.2,1.5\}.
\end{equation}
With a one-dimensional regressor and a second-order kernel,
bandwidths of the form $h=cn^{-\gamma}$ satisfy the rate conditions
in Assumption~\ref{A4} for any fixed $c>0$ and
$1/4<\gamma<2/3$. The exponent $\gamma=0.30$ used above is
within this range. Selecting the CV coefficient and varying the
bandwidth multiplier over the fixed finite sets above preserves
these rates.\footnote{Candidates with a zero or nonfinite leave-one-out denominator,
nonfinite residuals at any of the five scales, or nonfinite CV
loss are excluded; ties select the smaller coefficient. If none
is admissible, computation stops. Denominators are not clipped.}
CV is repeated for each sample, and fitted quantities remain fixed
within the multiplier draws.

Each configuration uses $1{,}000$ independent samples,
$n\in\{200,400,800\}$, $\rho\in\{0,0.5,1,2\}$, and nominal
levels $\alpha\in\{0.01,0.05,0.10\}$. The case $\rho=0$ measures
size; positive $\rho$ values measure unadjusted power. KS and CvM
are evaluated at the observed instrument values, using the corrected
bootstrap in \eqref{eq:mb} with $B=199$ independent standard-normal
multipliers shared across the two statistics and five bandwidths.
Rejection occurs when the plus-one Monte Carlo $p$-value is at
most $\alpha$.

We compare our tests with \citet{Blundell&Horowitz2007} (BH) and
\citet{Breunig2015} (BS). Each method independently draws samples
from the same DGP. BH transforms both coordinates to $[0,1]$,
selects $\hat c_{\rm CV}$ on its own sample, and uses joint-density
bandwidth $h_{1,H}=\hat c_{\rm CV}n^{-1/6}/2$ and regression
bandwidth $h_{2,H}=\hat c_{\rm CV}n^{-0.30}/2$. It uses the same
boundary correction, 50 normalized shifted-Legendre terms including
the constant, 25 spectrum terms, and sample-node quadrature with
Lebesgue cell weights. These smoothing choices adapt BH to the
present comparison. BS also transforms both coordinates to $[0,1]$,
fits an eight-dimensional Lagrange polynomial regression on
Chebyshev--Lobatto nodes, and uses normalized shifted-Legendre
instrument functions of degrees $1,\ldots,50$ with weights
$\tau_j=j^{-1}$. Both methods use $1{,}000$ weighted-chi-squared
calibration draws, with critical value of rank
$\lceil1{,}000(1-\alpha)\rceil$ and rejection by strict exceedance.
All kernel-based comparisons use $k=1$.

\subsection{Designs and results}\label{sec:sim-designs}

\phantomsection\label{sec:sim-mc}
The first design has a linear structural regression:
\begin{equation}\label{eq:sim.baseline}
 Y=X+\rho R+\eta,\qquad g(X)=X.
\end{equation}
At $\rho=0$, both mean exogeneity and strong conditional exogeneity
hold. Positive $\rho$ introduces endogeneity through $R$, with
$G(X)=X+\rho\delta(\pi X)$. Tables~\ref{tab:linear-robust-5}
and \ref{tab:mc} report bandwidth sensitivity at the $5\%$ level
and the method comparison, respectively. The corresponding
$1\%$ and $10\%$ sensitivity results are in
Appendix~\ref{app:simulation.supplement}.

\phantomsection\label{sec:sim-nonlinear}
The second design replaces the linear structural function by
\begin{equation}\label{eq:sim.nonlinear}
 Y=\sin X+\rho R+\eta,\qquad g(X)=\sin X.
\end{equation}
All other ingredients are unchanged, so this design isolates the
effect of structural nonlinearity. Both exogeneity restrictions
hold at $\rho=0$, while $G(X)=\sin X+\rho\delta(\pi X)$ under
the alternatives. Tables~\ref{tab:nonlinear-robust-5} and
\ref{tab:nonlinear-comparison} report the results.

\phantomsection\label{sec:sim-a8}
The third design retains $g(X)=\sin X$ but allows a component of
the null error to depend jointly on $X$ and $Z$. Specifically, set
\begin{equation}\label{eq:DGP-N}
 \begin{aligned}
 Q={}&\cos(\pi X)\cos(\pi Z)
 -\frac{0.1e^{-0.35^2/2}}{1+0.2e^{-2(0.35)^2}}
       \{\cos(2\pi X)+\cos(2\pi Z)\}
 -0.1e^{-0.35^2/2},\\
 Y={}&\sin X+\rho R+
       \frac{\eta+Q/\sqrt{3\Var(Q)}}{\sqrt2}.
 \end{aligned}
\end{equation}
The cosine terms remove both conditional means of the interaction,
giving $\E[Q\mid X]=\E[Q\mid Z]=0$; $\Var(Q)>0$ denotes its
population variance. The baseline error therefore has variance
$1/3$, as in the first two designs. At $\rho=0$, mean exogeneity
holds, but $\E[U\mid X,Z]=Q/\sqrt{6\Var(Q)}$ is not identically
zero. Thus Tables~\ref{tab:weak-robust-5} and \ref{tab:mcA8}
examine the test without imposing strong conditional exogeneity.

\begin{table}[!htbp]
\centering
\small
\renewcommand{\arraystretch}{0.85}
\caption{Bandwidth sensitivity in Design 1 at the nominal $5\%$ level.}
\label{tab:linear-robust-5}
\setlength{\tabcolsep}{4pt}
\begin{tabular}{cc*{8}{c}}
\toprule
$n$ & $k$ & \multicolumn{2}{c}{$\rho=0$} & \multicolumn{2}{c}{$\rho=0.5$} & \multicolumn{2}{c}{$\rho=1$} & \multicolumn{2}{c}{$\rho=2$}\\
\cmidrule(lr){3-4}\cmidrule(lr){5-6}\cmidrule(lr){7-8}\cmidrule(lr){9-10}
& & KS & CvM & KS & CvM & KS & CvM & KS & CvM\\
\midrule
200 & 0.5 & 0.035 & 0.040 & 0.077 & 0.077 & 0.239 & 0.248 & 0.769 & 0.752\\
 & 0.8 & 0.049 & 0.055 & 0.074 & 0.071 & 0.222 & 0.227 & 0.710 & 0.685\\
 & 1.0 & 0.047 & 0.053 & 0.067 & 0.070 & 0.207 & 0.216 & 0.595 & 0.566\\
 & 1.2 & 0.052 & 0.048 & 0.061 & 0.067 & 0.168 & 0.183 & 0.449 & 0.434\\
 & 1.5 & 0.058 & 0.058 & 0.055 & 0.056 & 0.121 & 0.122 & 0.241 & 0.257\\
\midrule
400 & 0.5 & 0.045 & 0.051 & 0.184 & 0.176 & 0.575 & 0.549 & 0.955 & 0.955\\
 & 0.8 & 0.050 & 0.060 & 0.170 & 0.167 & 0.551 & 0.524 & 0.976 & 0.968\\
 & 1.0 & 0.048 & 0.055 & 0.148 & 0.151 & 0.515 & 0.480 & 0.972 & 0.956\\
 & 1.2 & 0.047 & 0.055 & 0.133 & 0.124 & 0.442 & 0.417 & 0.928 & 0.907\\
 & 1.5 & 0.050 & 0.061 & 0.104 & 0.103 & 0.321 & 0.317 & 0.792 & 0.762\\
\midrule
800 & 0.5 & 0.039 & 0.051 & 0.355 & 0.339 & 0.918 & 0.897 & 0.992 & 0.991\\
 & 0.8 & 0.044 & 0.054 & 0.324 & 0.304 & 0.901 & 0.875 & 0.998 & 0.997\\
 & 1.0 & 0.045 & 0.055 & 0.291 & 0.265 & 0.889 & 0.857 & 1.000 & 1.000\\
 & 1.2 & 0.046 & 0.056 & 0.249 & 0.238 & 0.845 & 0.823 & 1.000 & 1.000\\
 & 1.5 & 0.046 & 0.056 & 0.187 & 0.185 & 0.755 & 0.730 & 1.000 & 1.000\\
\bottomrule
\end{tabular}
\end{table}

\begin{table}[!htbp]
\centering
\small
\renewcommand{\arraystretch}{0.85}
\caption{Bandwidth sensitivity in Design 2 at the nominal $5\%$ level.}
\label{tab:nonlinear-robust-5}
\setlength{\tabcolsep}{4pt}
\begin{tabular}{cc*{8}{c}}
\toprule
$n$ & $k$ & \multicolumn{2}{c}{$\rho=0$} & \multicolumn{2}{c}{$\rho=0.5$} & \multicolumn{2}{c}{$\rho=1$} & \multicolumn{2}{c}{$\rho=2$}\\
\cmidrule(lr){3-4}\cmidrule(lr){5-6}\cmidrule(lr){7-8}\cmidrule(lr){9-10}
& & KS & CvM & KS & CvM & KS & CvM & KS & CvM\\
\midrule
200 & 0.5 & 0.044 & 0.041 & 0.085 & 0.089 & 0.237 & 0.239 & 0.753 & 0.730\\
 & 0.8 & 0.051 & 0.046 & 0.087 & 0.095 & 0.213 & 0.218 & 0.684 & 0.646\\
 & 1.0 & 0.049 & 0.049 & 0.076 & 0.084 & 0.187 & 0.191 & 0.589 & 0.542\\
 & 1.2 & 0.048 & 0.052 & 0.068 & 0.073 & 0.159 & 0.150 & 0.437 & 0.415\\
 & 1.5 & 0.059 & 0.051 & 0.052 & 0.058 & 0.113 & 0.100 & 0.270 & 0.258\\
\midrule
400 & 0.5 & 0.047 & 0.045 & 0.181 & 0.193 & 0.551 & 0.535 & 0.966 & 0.967\\
 & 0.8 & 0.046 & 0.050 & 0.183 & 0.177 & 0.539 & 0.513 & 0.969 & 0.966\\
 & 1.0 & 0.048 & 0.049 & 0.159 & 0.159 & 0.499 & 0.481 & 0.969 & 0.956\\
 & 1.2 & 0.049 & 0.048 & 0.141 & 0.141 & 0.434 & 0.409 & 0.935 & 0.912\\
 & 1.5 & 0.046 & 0.053 & 0.111 & 0.126 & 0.316 & 0.300 & 0.800 & 0.767\\
\midrule
800 & 0.5 & 0.054 & 0.056 & 0.377 & 0.360 & 0.922 & 0.899 & 0.989 & 0.989\\
 & 0.8 & 0.055 & 0.053 & 0.340 & 0.331 & 0.915 & 0.890 & 0.996 & 0.997\\
 & 1.0 & 0.054 & 0.054 & 0.307 & 0.295 & 0.898 & 0.866 & 1.000 & 1.000\\
 & 1.2 & 0.056 & 0.058 & 0.273 & 0.254 & 0.858 & 0.814 & 1.000 & 1.000\\
 & 1.5 & 0.057 & 0.059 & 0.214 & 0.203 & 0.775 & 0.729 & 1.000 & 1.000\\
\bottomrule
\end{tabular}
\end{table}

\begin{table}[!htbp]
\centering
\small
\renewcommand{\arraystretch}{0.85}
\caption{Bandwidth sensitivity in Design 3 at the nominal $5\%$ level.}
\label{tab:weak-robust-5}
\setlength{\tabcolsep}{4pt}
\begin{tabular}{cc*{8}{c}}
\toprule
$n$ & $k$ & \multicolumn{2}{c}{$\rho=0$} & \multicolumn{2}{c}{$\rho=0.5$} & \multicolumn{2}{c}{$\rho=1$} & \multicolumn{2}{c}{$\rho=2$}\\
\cmidrule(lr){3-4}\cmidrule(lr){5-6}\cmidrule(lr){7-8}\cmidrule(lr){9-10}
& & KS & CvM & KS & CvM & KS & CvM & KS & CvM\\
\midrule
200 & 0.5 & 0.041 & 0.042 & 0.082 & 0.074 & 0.286 & 0.262 & 0.767 & 0.728\\
 & 0.8 & 0.042 & 0.045 & 0.089 & 0.078 & 0.283 & 0.249 & 0.695 & 0.649\\
 & 1.0 & 0.047 & 0.048 & 0.084 & 0.070 & 0.255 & 0.225 & 0.618 & 0.566\\
 & 1.2 & 0.050 & 0.053 & 0.075 & 0.063 & 0.196 & 0.185 & 0.475 & 0.456\\
 & 1.5 & 0.055 & 0.058 & 0.055 & 0.059 & 0.150 & 0.141 & 0.283 & 0.271\\
\midrule
400 & 0.5 & 0.047 & 0.051 & 0.177 & 0.157 & 0.591 & 0.537 & 0.957 & 0.956\\
 & 0.8 & 0.055 & 0.056 & 0.169 & 0.155 & 0.573 & 0.504 & 0.966 & 0.965\\
 & 1.0 & 0.057 & 0.058 & 0.151 & 0.137 & 0.524 & 0.466 & 0.978 & 0.968\\
 & 1.2 & 0.061 & 0.059 & 0.124 & 0.116 & 0.459 & 0.404 & 0.934 & 0.906\\
 & 1.5 & 0.061 & 0.061 & 0.099 & 0.098 & 0.359 & 0.311 & 0.828 & 0.770\\
\midrule
800 & 0.5 & 0.062 & 0.058 & 0.385 & 0.337 & 0.922 & 0.888 & 0.991 & 0.992\\
 & 0.8 & 0.060 & 0.058 & 0.352 & 0.300 & 0.915 & 0.881 & 0.999 & 0.999\\
 & 1.0 & 0.056 & 0.056 & 0.315 & 0.273 & 0.886 & 0.850 & 1.000 & 1.000\\
 & 1.2 & 0.059 & 0.059 & 0.277 & 0.241 & 0.845 & 0.807 & 1.000 & 1.000\\
 & 1.5 & 0.062 & 0.061 & 0.213 & 0.183 & 0.759 & 0.686 & 0.998 & 0.997\\
\bottomrule
\end{tabular}
\end{table}

\begin{table}[!htbp]
\centering
\footnotesize
\renewcommand{\arraystretch}{0.80}
\caption{Method comparison in Design 1.}
\label{tab:mc}
\setlength{\tabcolsep}{4pt}
\begin{tabular}{cc*{12}{c}}
\toprule
$\alpha$ & $\rho$ & \multicolumn{4}{c}{$n=200$} & \multicolumn{4}{c}{$n=400$} & \multicolumn{4}{c}{$n=800$}\\
\cmidrule(lr){3-6}\cmidrule(lr){7-10}\cmidrule(lr){11-14}
& & KS & CvM & BH & BS & KS & CvM & BH & BS & KS & CvM & BH & BS\\
\midrule
$1\%$ & $0$ & 0.011 & 0.013 & 0.002 & 0.010 & 0.011 & 0.008 & 0.003 & 0.009 & 0.009 & 0.009 & 0.012 & 0.012\\
 & $0.5$ & 0.014 & 0.015 & 0.005 & 0.008 & 0.050 & 0.043 & 0.011 & 0.014 & 0.124 & 0.114 & 0.090 & 0.010\\
 & $1$ & 0.075 & 0.070 & 0.008 & 0.013 & 0.262 & 0.254 & 0.111 & 0.108 & 0.703 & 0.658 & 0.587 & 0.111\\
 & $2$ & 0.321 & 0.284 & 0.057 & 0.109 & 0.893 & 0.839 & 0.588 & 0.113 & 1.000 & 1.000 & 0.993 & 0.205\\
\midrule
$5\%$ & $0$ & 0.047 & 0.053 & 0.002 & 0.061 & 0.048 & 0.055 & 0.019 & 0.059 & 0.045 & 0.055 & 0.057 & 0.057\\
 & $0.5$ & 0.067 & 0.070 & 0.003 & 0.055 & 0.148 & 0.151 & 0.065 & 0.043 & 0.291 & 0.265 & 0.222 & 0.049\\
 & $1$ & 0.207 & 0.216 & 0.016 & 0.148 & 0.515 & 0.480 & 0.256 & 0.140 & 0.889 & 0.857 & 0.815 & 0.139\\
 & $2$ & 0.595 & 0.566 & 0.141 & 0.248 & 0.972 & 0.956 & 0.798 & 0.252 & 1.000 & 1.000 & 0.999 & 0.238\\
\midrule
$10\%$ & $0$ & 0.100 & 0.100 & 0.007 & 0.110 & 0.110 & 0.099 & 0.053 & 0.112 & 0.111 & 0.106 & 0.111 & 0.116\\
 & $0.5$ & 0.134 & 0.129 & 0.015 & 0.112 & 0.231 & 0.246 & 0.129 & 0.097 & 0.409 & 0.393 & 0.320 & 0.098\\
 & $1$ & 0.301 & 0.290 & 0.043 & 0.199 & 0.618 & 0.603 & 0.364 & 0.185 & 0.942 & 0.923 & 0.881 & 0.200\\
 & $2$ & 0.713 & 0.703 & 0.240 & 0.290 & 0.982 & 0.977 & 0.869 & 0.290 & 1.000 & 1.000 & 0.999 & 0.284\\
\bottomrule
\end{tabular}
\end{table}

\begin{table}[!htbp]
\centering
\footnotesize
\renewcommand{\arraystretch}{0.80}
\caption{Method comparison in Design 2.}
\label{tab:nonlinear-comparison}
\setlength{\tabcolsep}{4pt}
\begin{tabular}{cc*{12}{c}}
\toprule
$\alpha$ & $\rho$ & \multicolumn{4}{c}{$n=200$} & \multicolumn{4}{c}{$n=400$} & \multicolumn{4}{c}{$n=800$}\\
\cmidrule(lr){3-6}\cmidrule(lr){7-10}\cmidrule(lr){11-14}
& & KS & CvM & BH & BS & KS & CvM & BH & BS & KS & CvM & BH & BS\\
\midrule
$1\%$ & $0$ & 0.011 & 0.011 & 0.001 & 0.008 & 0.014 & 0.006 & 0.004 & 0.008 & 0.010 & 0.015 & 0.008 & 0.005\\
 & $0.5$ & 0.018 & 0.014 & 0.003 & 0.008 & 0.044 & 0.055 & 0.008 & 0.015 & 0.126 & 0.121 & 0.075 & 0.008\\
 & $1$ & 0.068 & 0.061 & 0.003 & 0.004 & 0.270 & 0.230 & 0.104 & 0.108 & 0.713 & 0.654 & 0.596 & 0.107\\
 & $2$ & 0.317 & 0.296 & 0.041 & 0.006 & 0.916 & 0.869 & 0.579 & 0.206 & 1.000 & 1.000 & 0.993 & 0.212\\
\midrule
$5\%$ & $0$ & 0.049 & 0.049 & 0.006 & 0.046 & 0.048 & 0.049 & 0.017 & 0.046 & 0.054 & 0.054 & 0.055 & 0.051\\
 & $0.5$ & 0.076 & 0.084 & 0.009 & 0.057 & 0.159 & 0.159 & 0.055 & 0.058 & 0.307 & 0.295 & 0.207 & 0.046\\
 & $1$ & 0.187 & 0.191 & 0.019 & 0.109 & 0.499 & 0.481 & 0.240 & 0.146 & 0.898 & 0.866 & 0.795 & 0.149\\
 & $2$ & 0.589 & 0.542 & 0.128 & 0.140 & 0.969 & 0.956 & 0.772 & 0.241 & 1.000 & 1.000 & 0.997 & 0.244\\
\midrule
$10\%$ & $0$ & 0.096 & 0.100 & 0.010 & 0.093 & 0.096 & 0.102 & 0.047 & 0.095 & 0.101 & 0.104 & 0.093 & 0.108\\
 & $0.5$ & 0.143 & 0.155 & 0.021 & 0.107 & 0.248 & 0.251 & 0.104 & 0.104 & 0.428 & 0.429 & 0.318 & 0.099\\
 & $1$ & 0.305 & 0.304 & 0.047 & 0.126 & 0.621 & 0.618 & 0.367 & 0.190 & 0.942 & 0.927 & 0.880 & 0.204\\
 & $2$ & 0.704 & 0.684 & 0.215 & 0.179 & 0.986 & 0.984 & 0.848 & 0.293 & 1.000 & 1.000 & 0.997 & 0.297\\
\bottomrule
\end{tabular}
\end{table}

\FloatBarrier

\begin{table}[!htbp]
\centering
\footnotesize
\renewcommand{\arraystretch}{0.80}
\caption{Method comparison in Design 3.}
\label{tab:mcA8}
\setlength{\tabcolsep}{4pt}
\begin{tabular}{cc*{12}{c}}
\toprule
$\alpha$ & $\rho$ & \multicolumn{4}{c}{$n=200$} & \multicolumn{4}{c}{$n=400$} & \multicolumn{4}{c}{$n=800$}\\
\cmidrule(lr){3-6}\cmidrule(lr){7-10}\cmidrule(lr){11-14}
& & KS & CvM & BH & BS & KS & CvM & BH & BS & KS & CvM & BH & BS\\
\midrule
$1\%$ & $0$ & 0.007 & 0.008 & 0.001 & 0.011 & 0.012 & 0.018 & 0.005 & 0.007 & 0.022 & 0.014 & 0.008 & 0.012\\
 & $0.5$ & 0.022 & 0.018 & 0.003 & 0.010 & 0.049 & 0.039 & 0.006 & 0.008 & 0.127 & 0.099 & 0.067 & 0.028\\
 & $1$ & 0.098 & 0.086 & 0.005 & 0.016 & 0.291 & 0.219 & 0.066 & 0.018 & 0.714 & 0.596 & 0.525 & 0.112\\
 & $2$ & 0.356 & 0.319 & 0.027 & 0.019 & 0.910 & 0.834 & 0.492 & 0.106 & 1.000 & 1.000 & 0.989 & 0.212\\
\midrule
$5\%$ & $0$ & 0.047 & 0.048 & 0.003 & 0.043 & 0.057 & 0.058 & 0.024 & 0.045 & 0.056 & 0.056 & 0.057 & 0.057\\
 & $0.5$ & 0.084 & 0.070 & 0.005 & 0.053 & 0.151 & 0.137 & 0.036 & 0.050 & 0.315 & 0.273 & 0.196 & 0.042\\
 & $1$ & 0.255 & 0.225 & 0.021 & 0.143 & 0.524 & 0.466 & 0.179 & 0.144 & 0.886 & 0.850 & 0.771 & 0.144\\
 & $2$ & 0.618 & 0.566 & 0.095 & 0.250 & 0.978 & 0.968 & 0.732 & 0.237 & 1.000 & 1.000 & 0.998 & 0.246\\
\midrule
$10\%$ & $0$ & 0.114 & 0.108 & 0.010 & 0.103 & 0.102 & 0.095 & 0.053 & 0.108 & 0.123 & 0.113 & 0.098 & 0.099\\
 & $0.5$ & 0.155 & 0.153 & 0.012 & 0.109 & 0.233 & 0.223 & 0.088 & 0.112 & 0.430 & 0.384 & 0.302 & 0.081\\
 & $1$ & 0.365 & 0.341 & 0.043 & 0.193 & 0.645 & 0.580 & 0.309 & 0.203 & 0.935 & 0.912 & 0.876 & 0.179\\
 & $2$ & 0.735 & 0.684 & 0.178 & 0.297 & 0.991 & 0.987 & 0.844 & 0.290 & 1.000 & 1.000 & 0.998 & 0.295\\
\bottomrule
\end{tabular}
\end{table}

\FloatBarrier

Across the three designs, the baseline KS and CvM rejection rates
at the $5\%$ level range from $0.045$ to $0.058$ under the null.
Power increases substantially with sample size and departure
strength. For example, at $\rho=1$, baseline KS power ranges
from $0.187$--$0.255$ at $n=200$ to $0.886$--$0.898$ at
$n=800$; the corresponding CvM ranges are $0.191$--$0.225$
and $0.850$--$0.866$. The third design exhibits the same broad
pattern, supporting the use of the test under mean exogeneity
without strong conditional exogeneity.

The comparison tables show higher power for KS and CvM than
for BH and BS at most reported alternatives, in both the linear
and nonlinear structural models. These rankings concern the
present DGPs and implementations. BH and BS may have advantages
under designs better suited to their respective constructions;
the results do not establish a uniform power ranking.

The sensitivity tables show relatively stable size across the
five bandwidth scales and increasing power with $n$ throughout
the range. Power is more sensitive to bandwidth, particularly in
smaller samples, where larger bandwidths can reduce rejection
rates appreciably. The results therefore indicate robustness
of null calibration over the bandwidth range considered, while
retaining the usual finite-sample trade-off in power.

\section{Empirical application}\label{sec:application}

We examine the exogeneity of schooling using the NLS young-men
data analyzed by \citet{Card1995}. The outcome is log wage,
$X$ is years of education, and $Z_1$ is experience, so
$W=(\mathrm{educ},\mathrm{exper})$. Two excluded instruments
are considered: the binary proximity indicator \texttt{nearc4},
and mother's years of education, \texttt{motheduc}.
The proximity specification has $n=3{,}010$ observations.
Deleting observations with missing maternal education leaves
$n=2{,}657$ for the second specification. OLS with experience
and its square gives a schooling coefficient
of $0.093$ in both samples. The corresponding 2SLS coefficients
are $0.259$ with \texttt{nearc4} and $0.157$ with
\texttt{motheduc}. These differences motivate a formal test of
the mean exogeneity of schooling.

We apply the KS and CvM statistics in \eqref{eq:ts.s}, evaluated
at all observed instrument vectors, and use $B=1{,}000$ standard-normal
multiplier draws in the indicator-corrected bootstrap
\eqref{eq:mb}. For each instrument specification, education and
experience are separately rescaled to $[0,1]$ using their sample
ranges. We use the product Epanechnikov kernel with the continuous-moment
boundary correction described in Assumption~\ref{A3}.
For the two-dimensional conditioning vector and a second-order kernel,
Assumption~\ref{A4} allows bandwidth sequences $h=c n^{-\gamma}$
with $1/4<\gamma<1/3$. We set $\gamma=0.30$ and select the
coefficient $\hat c_{\rm CV}$ by the leave-one-out CV criterion
in \eqref{eq:sim.cv}, with
$c\in\mathcal C=\{0.6,0.8,1.0,1.2,1.4\}$.\footnote{The CV
search is restricted to candidates for which the estimates required
in the bandwidth sensitivity exercise are computable.}
The baseline bandwidth is $\hat h_{\rm CV}=\hat c_{\rm CV}n^{-0.30}$,
giving $0.12664$
for \texttt{nearc4} and $0.13147$ for \texttt{motheduc}.

We compute the series test of \citet{Breunig2015}
and the test of \citet{Blundell&Horowitz2007}, denoted BS and BH.
The BS implementation estimates $E[Y\mid W]$ with a
polynomial basis and forms instrument moments with weights
$\tau_j=j^{-1}$. The BH implementation includes experience
as a conditioning variable and uses a polynomial
approximation to its covariance operator. Both comparison tests use
$1{,}000$ draws from their fitted weighted chi-squared reference
distributions. For BH, the joint-density bandwidth is
$\hat c_{\rm CV}n^{-1/7}$, the regression bandwidth is
$\hat h_{\rm CV}$, and the control-variable smoothing kernel is
$\ell(v,t)=\exp(vt)$. Its spectral approximation retains all ten
total-degree-three basis terms, with twelve-point Gauss--Legendre
quadrature in each integration direction. These finite-dimensional
choices specify the empirical implementation of the comparison tests.
Table~\ref{tab:application} reports the baseline results, with
KS, CvM, and BH evaluated at the CV baseline ($k=1$).

\begin{table}[tbp]
\centering
\caption{Exogeneity of schooling: baseline comparison}
\label{tab:application}
\small
\setlength{\tabcolsep}{4pt}
\begin{tabular}{lrrrrrrr}
\toprule
$Z_2$ & $n$ & OLS & 2SLS & BS & BH & KS & CvM\\
\midrule
\texttt{nearc4}   & $3010$ & $0.093$ & $0.259$ & $0.001$ & $0.214$ & $0.001$ & $0.001$\\
\texttt{motheduc} & $2657$ & $0.093$ & $0.157$ & $0.001$ & $0.308$ & $0.001$ & $0.003$\\
\bottomrule
\end{tabular}
\end{table}

Table~\ref{tab:application} shows that both KS and CvM reject mean
exogeneity at the $1\%$ level in each instrument specification.
The KS $p$-value is $0.001$ in both samples, while the CvM
$p$-values are $0.001$ for \texttt{nearc4} and $0.003$ for
\texttt{motheduc}. BS also rejects in both samples. In contrast,
BH yields $p$-values of $0.214$ and $0.308$ and does not reject
at conventional significance levels. The KS and CvM results
therefore agree with the series test in providing evidence against
mean exogeneity within the maintained specifications.

To examine the robustness of these findings to bandwidth selection,
we also report empirical results at
$h_k=k\hat h_{\rm CV}=k\hat c_{\rm CV}n^{-0.30}$,
where $k\in\{0.5,0.8,1.0,1.2,1.5\}$.
Table~\ref{tab:application.robustness} presents the resulting KS
and CvM $p$-values. The case $k=1$ corresponds to the recommended
CV bandwidth in Table~\ref{tab:application}; the other multipliers
evaluate smaller and larger bandwidths around that choice.

\begin{table}[tbp]
\centering
\caption{Bandwidth sensitivity of the schooling exogeneity tests}
\label{tab:application.robustness}
\small
\setlength{\tabcolsep}{7pt}
\begin{tabular}{rrrrrrr}
\toprule
 & \multicolumn{3}{c}{\texttt{nearc4}, $n=3010$}
 & \multicolumn{3}{c}{\texttt{motheduc}, $n=2657$}\\
\cmidrule(lr){2-4}\cmidrule(lr){5-7}
$k$ & $h_k$ & KS & CvM & $h_k$ & KS & CvM\\
\midrule
$0.5$ & $0.06332$ & $0.001$ & $0.001$ & $0.06573$ & $0.001$ & $0.001$\\
$0.8$ & $0.10131$ & $0.001$ & $0.001$ & $0.10517$ & $0.001$ & $0.001$\\
$1.0$ & $0.12664$ & $0.001$ & $0.001$ & $0.13147$ & $0.001$ & $0.003$\\
$1.2$ & $0.15196$ & $0.001$ & $0.001$ & $0.15776$ & $0.001$ & $0.001$\\
$1.5$ & $0.18995$ & $0.001$ & $0.001$ & $0.19720$ & $0.001$ & $0.001$\\
\bottomrule
\end{tabular}
\end{table}

Both KS and CvM reject mean exogeneity at the $1\%$ level across
all five bandwidths for both instruments. Every reported KS
$p$-value is $0.001$. The CvM $p$-values are also $0.001$,
except for $0.003$ at the recommended bandwidth in the
\texttt{motheduc} specification. Hence the rejection decisions
of both tests remain unchanged when the bandwidth is decreased
or increased relative to the CV recommendation. These results
support the robustness of the empirical conclusion over the
bandwidth range considered.

These findings are subject to support and identification limitations.
The variables are discrete, whereas the sufficient theory assumes
continuous supports; a formal justification therefore requires a
mixed-support extension or a suitably restricted model. Instrument
validity and completeness also require separate justification,
particularly for the binary instrument. The nonparametric
specification controls only for experience, so richer conditioning
sets may change the conclusion and increase the smoothing dimension.
The two instrument specifications use overlapping samples and
should be interpreted as related empirical checks.

\section{Discussion}\label{sec:discussion}

\subsection{Exogeneity and significance of excluded instruments}

Testing mean exogeneity differs from testing the significance of
the excluded instruments $Z_2$ for the conditional mean of $Y$
given the regressors $W=(X',Z_1')'$. Recall that
$Z=(Z_1',Z_2')'$, $G(W)=\E[Y\mid W]$, and
$\varepsilon=Y-G(W)$. In the nonparametric significance-testing
problem considered by \citet{Delgado&Manteiga2001}, the null is
\begin{equation}\label{eq:cmr.sig}
 \widetilde H_0:\quad
 \E[Y\mid X,Z]=\E[Y\mid X,Z_1]=G(W)
 \quad\text{a.s.}
\end{equation}
Thus $Z_2$ provides no additional information about the conditional
mean of $Y$ once $X$ and $Z_1$ are given. Equivalently,
$\widetilde H_0$ requires $\E[\varepsilon\mid X,Z]=0$,
whereas the exogeneity null in \eqref{eq:cmr} requires
$\E[\varepsilon\mid Z]=0$. Here significance concerns the
conditional mean of $Y$, rather than the first-stage relevance
of $Z_2$ for $X$.

The difference also appears in the unconditional moment
restrictions used to construct the tests. For the significance
null, the equivalent restrictions are
\[
 \E[\varepsilon\ind(X\le x)\ind(Z\le z)]=0
 \quad\text{for all }(x,z)\in\R^p\times\R^q,
\]
whereas the exogeneity test uses
$\E[\varepsilon\ind(Z\le z)]=0$ for all $z$. With the same
leave-one-out residuals $\hat\varepsilon_i=Y_i-\hat G_i$,
an analogous empirical process for significance testing is
\begin{equation}\label{eq:ts.sig}
 \widetilde T_n(x,z)=\frac1n\sum_{i=1}^n
 \hat\varepsilon_i\ind(X_i\le x)\ind(Z_i\le z).
\end{equation}
Compared with $\hat T_n(z)$ in \eqref{eq:ts.exog}, this process
includes the additional indicator $\ind(X_i\le x)$ because the
significance restriction conditions on both $X$ and $Z$.

By iterated expectations, $\widetilde H_0$ implies
$\E[\varepsilon\mid Z]=0$ and hence, under instrument validity
and completeness, the mean-exogeneity null $g=G$. The converse
does not hold: mean exogeneity and instrument validity imply
$\E[U\mid W]=\E[U\mid Z]=0$, but do not require
$\E[U\mid W,Z_2]=0$. The latter is equivalent to
$\widetilde H_0$ under mean exogeneity. Remark~\ref{rmk:counterexample}
and the null design in Section~\ref{sec:sim-a8} illustrate this
distinction. The nonsignificance null is therefore more restrictive
than the mean-exogeneity null, and the two tests address different
questions.

\subsection{Testing strong conditional exogeneity}
\label{sec:projection}\label{sec:strong.exogeneity}

The mean-exogeneity procedure developed above suggests
a corresponding construction for testing strong conditional
exogeneity,
\begin{equation}\label{eq:A8}
 H_0^{\mathrm s}:\quad \E[U\mid W,Z_2]=0\quad\text{a.s.}
\end{equation}
Under the maintained IV restriction and completeness, this is
equivalent to the nonsignificance null in \eqref{eq:cmr.sig}.
Here nonsignificance means that $Z_2$ does not change the conditional
mean of $Y$ once $W=(X',Z_1')'$ is given; it does not refer to the
first-stage relevance of the instruments for $X$.
As in Section~\ref{sec:altboot}, we retain the original indicator in the
observed process and center it in the multiplier bootstrap:
\begin{align}
 \widetilde T_n(x,z)
 &=\frac1n\sum_{i=1}^n\hat\varepsilon_i
       \ind(X_i\le x)\ind(Z_i\le z),
       \label{eq:strong.process}\\
 \widetilde T_n^*(x,z)
 &=\frac1n\sum_{i=1}^n V_i\hat\varepsilon_i
       \ind(X_i\le x)\ind(Z_{1i}\le z_1)
       \{\ind(Z_{2i}\le z_2)-\hat F_i(z_2)\}.
       \label{eq:strong.bootstrap}
\end{align}
Here $\hat\varepsilon_i$ and $\hat F_i$ are the same leave-one-out
estimates used in Section~\ref{sec:altboot}. The additional
indicator $\ind(X_i\le x)$ makes the moments characterize a
restriction conditional on $(X,Z)$, as explained in the preceding
subsection. KS or CvM functionals of these processes would provide
corresponding test statistics and bootstrap critical values.

Under $H_0^{\mathrm s}$, $\varepsilon=U$ and
\[
 \E[\varepsilon\ind(Z_2\le z_2)\mid W]=0.
\]
Thus the conditional-residual-moment component of the
\emph{nonparametric estimation effect} identified in
Proposition~\ref{prop:density} and the last subtraction in
\eqref{eq:mb.mimic} has a zero population counterpart. The effect
of estimating $G$ is still represented by the indicator centering
in \eqref{eq:strong.bootstrap}. 

Strong conditional exogeneity implies mean exogeneity by iterated
expectations, but the converse fails, as the null design in
Section~\ref{sec:sim-a8} illustrates. The two hypotheses therefore
require different determining moment families: moments indexed
only by $Z$ characterize mean exogeneity, whereas the joint
index $(X,Z)$ is needed here to characterize the stronger null.
Theoretical justification and applications of the proposed
strong-exogeneity constructions are beyond the scope of this paper.

Finally, under $H_0^{\mathrm s}$, $\phi(W,Z_2):=\E[Y\mid W,Z_2]=G(W)$. Conditional on $W$, the functions $1$ and $\phi$ are therefore collinear, so their conditional Gram matrix is singular, and its ordinary inverse is undefined. Projection on their span nevertheless reduces to projection on constants, whose indicator residual is
\begin{equation}\label{eq:P0}
 \ind(Z_2\le z_2)-F(z_2\mid W).
\end{equation}
Our centering correction thus admits a projection interpretation
under the stronger null. Related specification tests use orthogonal
projections \citep{SantAnna&Song2019,Yang_etal2024,Song&Yang2025}.
This connection suggests another way to construct a test for exogeneity.

\clearpage
\Appendix
\numberwithin{equation}{section}
\numberwithin{lemma}{section}
\numberwithin{table}{section}
\section{Supplementary simulation results}\label{app:simulation.supplement}

Tables~\ref{tab:linear-robust-1}--\ref{tab:weak-robust-10}
report bandwidth sensitivity at the $1\%$ and $10\%$ nominal
levels for the three designs in Section~\ref{sec:sim-designs}.
All settings are as described in Section~\ref{sec:sim-setup}.

\begin{table}[!htbp]
\centering
\small
\renewcommand{\arraystretch}{0.85}
\caption{Bandwidth sensitivity in Design 1 at the nominal $1\%$ level.}
\label{tab:linear-robust-1}
\setlength{\tabcolsep}{4pt}
\begin{tabular}{cc*{8}{c}}
\toprule
$n$ & $k$ & \multicolumn{2}{c}{$\rho=0$} & \multicolumn{2}{c}{$\rho=0.5$} & \multicolumn{2}{c}{$\rho=1$} & \multicolumn{2}{c}{$\rho=2$}\\
\cmidrule(lr){3-4}\cmidrule(lr){5-6}\cmidrule(lr){7-8}\cmidrule(lr){9-10}
& & KS & CvM & KS & CvM & KS & CvM & KS & CvM\\
\midrule
200 & 0.5 & 0.008 & 0.006 & 0.017 & 0.017 & 0.090 & 0.084 & 0.530 & 0.460\\
 & 0.8 & 0.010 & 0.011 & 0.013 & 0.015 & 0.087 & 0.073 & 0.427 & 0.388\\
 & 1.0 & 0.011 & 0.013 & 0.014 & 0.015 & 0.075 & 0.070 & 0.321 & 0.284\\
 & 1.2 & 0.011 & 0.012 & 0.016 & 0.016 & 0.056 & 0.058 & 0.205 & 0.194\\
 & 1.5 & 0.014 & 0.011 & 0.012 & 0.012 & 0.038 & 0.038 & 0.091 & 0.096\\
\midrule
400 & 0.5 & 0.005 & 0.008 & 0.062 & 0.059 & 0.335 & 0.315 & 0.928 & 0.910\\
 & 0.8 & 0.011 & 0.007 & 0.057 & 0.048 & 0.310 & 0.283 & 0.924 & 0.893\\
 & 1.0 & 0.011 & 0.008 & 0.050 & 0.043 & 0.262 & 0.254 & 0.893 & 0.839\\
 & 1.2 & 0.011 & 0.012 & 0.039 & 0.037 & 0.209 & 0.208 & 0.812 & 0.751\\
 & 1.5 & 0.014 & 0.016 & 0.026 & 0.029 & 0.153 & 0.153 & 0.590 & 0.539\\
\midrule
800 & 0.5 & 0.008 & 0.008 & 0.154 & 0.148 & 0.773 & 0.715 & 0.990 & 0.990\\
 & 0.8 & 0.007 & 0.007 & 0.139 & 0.127 & 0.736 & 0.691 & 0.997 & 0.997\\
 & 1.0 & 0.009 & 0.009 & 0.124 & 0.114 & 0.703 & 0.658 & 1.000 & 1.000\\
 & 1.2 & 0.010 & 0.009 & 0.100 & 0.094 & 0.646 & 0.589 & 1.000 & 1.000\\
 & 1.5 & 0.011 & 0.012 & 0.077 & 0.069 & 0.528 & 0.477 & 0.986 & 0.976\\
\bottomrule
\end{tabular}
\end{table}

\begin{table}[!htbp]
\centering
\small
\renewcommand{\arraystretch}{0.85}
\caption{Bandwidth sensitivity in Design 1 at the nominal $10\%$ level.}
\label{tab:linear-robust-10}
\setlength{\tabcolsep}{4pt}
\begin{tabular}{cc*{8}{c}}
\toprule
$n$ & $k$ & \multicolumn{2}{c}{$\rho=0$} & \multicolumn{2}{c}{$\rho=0.5$} & \multicolumn{2}{c}{$\rho=1$} & \multicolumn{2}{c}{$\rho=2$}\\
\cmidrule(lr){3-4}\cmidrule(lr){5-6}\cmidrule(lr){7-8}\cmidrule(lr){9-10}
& & KS & CvM & KS & CvM & KS & CvM & KS & CvM\\
\midrule
200 & 0.5 & 0.086 & 0.089 & 0.141 & 0.146 & 0.347 & 0.336 & 0.861 & 0.845\\
 & 0.8 & 0.099 & 0.100 & 0.143 & 0.137 & 0.319 & 0.328 & 0.795 & 0.782\\
 & 1.0 & 0.100 & 0.100 & 0.134 & 0.129 & 0.301 & 0.290 & 0.713 & 0.703\\
 & 1.2 & 0.107 & 0.101 & 0.117 & 0.113 & 0.250 & 0.256 & 0.576 & 0.569\\
 & 1.5 & 0.110 & 0.102 & 0.108 & 0.107 & 0.190 & 0.198 & 0.372 & 0.370\\
\midrule
400 & 0.5 & 0.092 & 0.094 & 0.282 & 0.297 & 0.685 & 0.669 & 0.964 & 0.963\\
 & 0.8 & 0.104 & 0.105 & 0.262 & 0.266 & 0.659 & 0.638 & 0.982 & 0.981\\
 & 1.0 & 0.110 & 0.099 & 0.231 & 0.246 & 0.618 & 0.603 & 0.982 & 0.977\\
 & 1.2 & 0.111 & 0.105 & 0.208 & 0.220 & 0.555 & 0.538 & 0.965 & 0.954\\
 & 1.5 & 0.105 & 0.109 & 0.187 & 0.191 & 0.441 & 0.438 & 0.860 & 0.845\\
\midrule
800 & 0.5 & 0.104 & 0.098 & 0.476 & 0.457 & 0.951 & 0.938 & 0.992 & 0.993\\
 & 0.8 & 0.108 & 0.100 & 0.442 & 0.420 & 0.947 & 0.929 & 0.998 & 0.998\\
 & 1.0 & 0.111 & 0.106 & 0.409 & 0.393 & 0.942 & 0.923 & 1.000 & 1.000\\
 & 1.2 & 0.110 & 0.110 & 0.363 & 0.344 & 0.915 & 0.891 & 1.000 & 1.000\\
 & 1.5 & 0.111 & 0.107 & 0.286 & 0.283 & 0.839 & 0.822 & 1.000 & 1.000\\
\bottomrule
\end{tabular}
\end{table}

\begin{table}[!htbp]
\centering
\small
\renewcommand{\arraystretch}{0.85}
\caption{Bandwidth sensitivity in Design 2 at the nominal $1\%$ level.}
\label{tab:nonlinear-robust-1}
\setlength{\tabcolsep}{4pt}
\begin{tabular}{cc*{8}{c}}
\toprule
$n$ & $k$ & \multicolumn{2}{c}{$\rho=0$} & \multicolumn{2}{c}{$\rho=0.5$} & \multicolumn{2}{c}{$\rho=1$} & \multicolumn{2}{c}{$\rho=2$}\\
\cmidrule(lr){3-4}\cmidrule(lr){5-6}\cmidrule(lr){7-8}\cmidrule(lr){9-10}
& & KS & CvM & KS & CvM & KS & CvM & KS & CvM\\
\midrule
200 & 0.5 & 0.009 & 0.008 & 0.016 & 0.019 & 0.088 & 0.075 & 0.528 & 0.464\\
 & 0.8 & 0.007 & 0.008 & 0.019 & 0.017 & 0.079 & 0.076 & 0.414 & 0.365\\
 & 1.0 & 0.011 & 0.011 & 0.018 & 0.014 & 0.068 & 0.061 & 0.317 & 0.296\\
 & 1.2 & 0.010 & 0.009 & 0.011 & 0.015 & 0.054 & 0.042 & 0.224 & 0.204\\
 & 1.5 & 0.011 & 0.008 & 0.006 & 0.009 & 0.032 & 0.029 & 0.106 & 0.115\\
\midrule
400 & 0.5 & 0.011 & 0.007 & 0.057 & 0.060 & 0.319 & 0.276 & 0.936 & 0.917\\
 & 0.8 & 0.013 & 0.009 & 0.046 & 0.064 & 0.293 & 0.263 & 0.928 & 0.902\\
 & 1.0 & 0.014 & 0.006 & 0.044 & 0.055 & 0.270 & 0.230 & 0.916 & 0.869\\
 & 1.2 & 0.013 & 0.007 & 0.035 & 0.037 & 0.219 & 0.189 & 0.805 & 0.750\\
 & 1.5 & 0.014 & 0.010 & 0.019 & 0.030 & 0.143 & 0.127 & 0.559 & 0.506\\
\midrule
800 & 0.5 & 0.010 & 0.012 & 0.165 & 0.145 & 0.785 & 0.705 & 0.987 & 0.987\\
 & 0.8 & 0.008 & 0.015 & 0.145 & 0.134 & 0.762 & 0.695 & 0.993 & 0.993\\
 & 1.0 & 0.010 & 0.015 & 0.126 & 0.121 & 0.713 & 0.654 & 1.000 & 1.000\\
 & 1.2 & 0.010 & 0.013 & 0.103 & 0.100 & 0.653 & 0.585 & 1.000 & 0.999\\
 & 1.5 & 0.010 & 0.013 & 0.073 & 0.065 & 0.544 & 0.485 & 0.992 & 0.977\\
\bottomrule
\end{tabular}
\end{table}

\begin{table}[!htbp]
\centering
\small
\renewcommand{\arraystretch}{0.85}
\caption{Bandwidth sensitivity in Design 2 at the nominal $10\%$ level.}
\label{tab:nonlinear-robust-10}
\setlength{\tabcolsep}{4pt}
\begin{tabular}{cc*{8}{c}}
\toprule
$n$ & $k$ & \multicolumn{2}{c}{$\rho=0$} & \multicolumn{2}{c}{$\rho=0.5$} & \multicolumn{2}{c}{$\rho=1$} & \multicolumn{2}{c}{$\rho=2$}\\
\cmidrule(lr){3-4}\cmidrule(lr){5-6}\cmidrule(lr){7-8}\cmidrule(lr){9-10}
& & KS & CvM & KS & CvM & KS & CvM & KS & CvM\\
\midrule
200 & 0.5 & 0.076 & 0.083 & 0.156 & 0.168 & 0.351 & 0.349 & 0.836 & 0.830\\
 & 0.8 & 0.092 & 0.100 & 0.141 & 0.155 & 0.329 & 0.326 & 0.770 & 0.764\\
 & 1.0 & 0.096 & 0.100 & 0.143 & 0.155 & 0.305 & 0.304 & 0.704 & 0.684\\
 & 1.2 & 0.091 & 0.104 & 0.127 & 0.131 & 0.250 & 0.257 & 0.573 & 0.554\\
 & 1.5 & 0.095 & 0.112 & 0.113 & 0.112 & 0.173 & 0.193 & 0.368 & 0.356\\
\midrule
400 & 0.5 & 0.088 & 0.087 & 0.279 & 0.279 & 0.673 & 0.652 & 0.971 & 0.971\\
 & 0.8 & 0.093 & 0.097 & 0.272 & 0.274 & 0.659 & 0.656 & 0.976 & 0.976\\
 & 1.0 & 0.096 & 0.102 & 0.248 & 0.251 & 0.621 & 0.618 & 0.986 & 0.984\\
 & 1.2 & 0.095 & 0.101 & 0.224 & 0.222 & 0.559 & 0.551 & 0.956 & 0.951\\
 & 1.5 & 0.097 & 0.099 & 0.191 & 0.188 & 0.434 & 0.444 & 0.871 & 0.858\\
\midrule
800 & 0.5 & 0.101 & 0.104 & 0.486 & 0.494 & 0.953 & 0.946 & 0.989 & 0.990\\
 & 0.8 & 0.100 & 0.106 & 0.447 & 0.465 & 0.954 & 0.943 & 0.997 & 0.998\\
 & 1.0 & 0.101 & 0.104 & 0.428 & 0.429 & 0.942 & 0.927 & 1.000 & 1.000\\
 & 1.2 & 0.106 & 0.108 & 0.386 & 0.382 & 0.914 & 0.895 & 1.000 & 1.000\\
 & 1.5 & 0.108 & 0.101 & 0.322 & 0.316 & 0.854 & 0.829 & 1.000 & 1.000\\
\bottomrule
\end{tabular}
\end{table}

\begin{table}[!htbp]
\centering
\small
\renewcommand{\arraystretch}{0.85}
\caption{Bandwidth sensitivity in Design 3 at the nominal $1\%$ level.}
\label{tab:weak-robust-1}
\setlength{\tabcolsep}{4pt}
\begin{tabular}{cc*{8}{c}}
\toprule
$n$ & $k$ & \multicolumn{2}{c}{$\rho=0$} & \multicolumn{2}{c}{$\rho=0.5$} & \multicolumn{2}{c}{$\rho=1$} & \multicolumn{2}{c}{$\rho=2$}\\
\cmidrule(lr){3-4}\cmidrule(lr){5-6}\cmidrule(lr){7-8}\cmidrule(lr){9-10}
& & KS & CvM & KS & CvM & KS & CvM & KS & CvM\\
\midrule
200 & 0.5 & 0.003 & 0.005 & 0.020 & 0.020 & 0.117 & 0.103 & 0.552 & 0.489\\
 & 0.8 & 0.004 & 0.007 & 0.024 & 0.021 & 0.113 & 0.094 & 0.466 & 0.396\\
 & 1.0 & 0.007 & 0.008 & 0.022 & 0.018 & 0.098 & 0.086 & 0.356 & 0.319\\
 & 1.2 & 0.006 & 0.009 & 0.017 & 0.012 & 0.081 & 0.063 & 0.238 & 0.216\\
 & 1.5 & 0.008 & 0.011 & 0.013 & 0.010 & 0.052 & 0.050 & 0.103 & 0.093\\
\midrule
400 & 0.5 & 0.010 & 0.017 & 0.065 & 0.044 & 0.356 & 0.302 & 0.929 & 0.895\\
 & 0.8 & 0.013 & 0.019 & 0.058 & 0.044 & 0.333 & 0.269 & 0.925 & 0.881\\
 & 1.0 & 0.012 & 0.018 & 0.049 & 0.039 & 0.291 & 0.219 & 0.910 & 0.834\\
 & 1.2 & 0.012 & 0.017 & 0.038 & 0.036 & 0.236 & 0.188 & 0.822 & 0.740\\
 & 1.5 & 0.012 & 0.016 & 0.034 & 0.032 & 0.157 & 0.130 & 0.588 & 0.530\\
\midrule
800 & 0.5 & 0.020 & 0.017 & 0.161 & 0.128 & 0.778 & 0.682 & 0.987 & 0.987\\
 & 0.8 & 0.020 & 0.014 & 0.143 & 0.113 & 0.744 & 0.653 & 0.999 & 0.999\\
 & 1.0 & 0.022 & 0.014 & 0.127 & 0.099 & 0.714 & 0.596 & 1.000 & 1.000\\
 & 1.2 & 0.020 & 0.012 & 0.104 & 0.083 & 0.650 & 0.534 & 0.999 & 0.997\\
 & 1.5 & 0.020 & 0.014 & 0.081 & 0.056 & 0.522 & 0.431 & 0.992 & 0.975\\
\bottomrule
\end{tabular}
\end{table}

\begin{table}[!htbp]
\centering
\small
\renewcommand{\arraystretch}{0.85}
\caption{Bandwidth sensitivity in Design 3 at the nominal $10\%$ level.}
\label{tab:weak-robust-10}
\setlength{\tabcolsep}{4pt}
\begin{tabular}{cc*{8}{c}}
\toprule
$n$ & $k$ & \multicolumn{2}{c}{$\rho=0$} & \multicolumn{2}{c}{$\rho=0.5$} & \multicolumn{2}{c}{$\rho=1$} & \multicolumn{2}{c}{$\rho=2$}\\
\cmidrule(lr){3-4}\cmidrule(lr){5-6}\cmidrule(lr){7-8}\cmidrule(lr){9-10}
& & KS & CvM & KS & CvM & KS & CvM & KS & CvM\\
\midrule
200 & 0.5 & 0.087 & 0.093 & 0.148 & 0.146 & 0.404 & 0.395 & 0.836 & 0.813\\
 & 0.8 & 0.107 & 0.106 & 0.167 & 0.149 & 0.389 & 0.376 & 0.780 & 0.750\\
 & 1.0 & 0.114 & 0.108 & 0.155 & 0.153 & 0.365 & 0.341 & 0.735 & 0.684\\
 & 1.2 & 0.121 & 0.110 & 0.139 & 0.142 & 0.306 & 0.285 & 0.607 & 0.574\\
 & 1.5 & 0.120 & 0.113 & 0.135 & 0.118 & 0.225 & 0.219 & 0.412 & 0.410\\
\midrule
400 & 0.5 & 0.091 & 0.098 & 0.273 & 0.259 & 0.706 & 0.643 & 0.961 & 0.963\\
 & 0.8 & 0.095 & 0.101 & 0.256 & 0.244 & 0.681 & 0.614 & 0.971 & 0.971\\
 & 1.0 & 0.102 & 0.095 & 0.233 & 0.223 & 0.645 & 0.580 & 0.991 & 0.987\\
 & 1.2 & 0.105 & 0.103 & 0.201 & 0.199 & 0.563 & 0.522 & 0.968 & 0.956\\
 & 1.5 & 0.108 & 0.109 & 0.178 & 0.162 & 0.466 & 0.418 & 0.887 & 0.843\\
\midrule
800 & 0.5 & 0.119 & 0.109 & 0.503 & 0.440 & 0.954 & 0.938 & 0.992 & 0.992\\
 & 0.8 & 0.124 & 0.115 & 0.477 & 0.413 & 0.950 & 0.931 & 0.999 & 0.999\\
 & 1.0 & 0.123 & 0.113 & 0.430 & 0.384 & 0.935 & 0.912 & 1.000 & 1.000\\
 & 1.2 & 0.126 & 0.122 & 0.380 & 0.348 & 0.905 & 0.882 & 1.000 & 1.000\\
 & 1.5 & 0.126 & 0.123 & 0.322 & 0.291 & 0.836 & 0.803 & 0.999 & 0.998\\
\bottomrule
\end{tabular}
\end{table}

\FloatBarrier

\section{Auxiliary lemmas and proofs}\label{sec:proofs}\label{app:lemmas}
\allowdisplaybreaks[1]

\begin{proof}[Proof of Lemma~\ref{lem:equiv}]
Since $G(W)=g(W)+\E[U\mid W]$, \eqref{eq:null1} and
\eqref{eq:null2} are equivalent. Either implies \eqref{eq:cmr}
by instrument validity. Conversely, if \eqref{eq:cmr} holds,
$\E[(G-g)(W)\mid Z]=0$. Completeness gives $G=g$ a.s., and
then $\E[U\mid W]=0$. Lower-orthant indicators determine finite
signed measures on the instrument support, which proves the
equivalence with \eqref{eq:ucmr}.
\end{proof}

Use the notation $f_h(w)=\E[K_h(w,W)]$ and
$b_h(w)=f_h(w)-f(w)$.
For observations, let $f_i=f(W_i)$ and
\[
 \begin{aligned}
 N_i&=\frac1{n-1}\sum_{j\ne i}K_h(W_i,W_j)(Y_i-Y_j),
 &D_i&=\frac1{n-1}\sum_{k\ne i}d_{ik},\\[3pt]
 d_{ik}&=K_h(W_i,W_k)-f_h(W_i),
 &\Delta_i&=\hat f_i-f_i=D_i+b_h(W_i).
 \end{aligned}
\]

\begin{lemma}[Kernel and absolute-average bounds]\label{lem:kernel}
Under Assumptions~\ref{A1}--\ref{A4}, with
$a_n=\{\log(en)/(nh^r)\}^{1/2}+h^s$ as specified in
Assumption~\ref{A4},
\[
 \begin{aligned}
 \norm{b_h}_\infty
 +\sup_{w\in S_W}|\E[K_h(w,W)\{G(w)-Y\}]|&=O(h^s),\\
 \max_i|\Delta_i|+\max_i|\hat G_i-G(W_i)|&=O_p(a_n),
 \end{aligned}
\]
and
\[
 \sup_{i,z_2}|\hat F_i(z_2)-F(z_2\mid W_i)|
 +\sup_{i,z_2}|\hat\psi_i(z_2)-\psi(z_2,W_i)|=O_p(a_n).
\]
Moreover,
\begin{equation}\label{eq:absolute.N}
 \frac1n\sum_i|N_i|=O_p(1),
\end{equation}
and, with probability tending to one,
$\min_i\hat f_i\ge\underline f/2>\tau_n$.
\end{lemma}

\begin{proof}
The moment conditions of the corrected kernel and Taylor expansion
give $f_h=f+O(h^s)$ and
$\E[K_h(w,W)Y]=f(w)G(w)+O(h^s)$ uniformly in $w$, including
the support boundary. Therefore
$\E[K_h(w,W)\{G(w)-Y\}]=O(h^s)$ uniformly.
The same argument applies to the smooth
functions $fF$ and $f\psi$. Each centered kernel or marked-kernel
average has variance bounded by $C/(nh^r)$. Bernstein bounds for
the conditionally sub-Gaussian marks, followed by the polynomial
covering bounds for bounded-variation kernel translates and lower
orthants (see the VC closure results in \citet{van1996weak}), give the uniform stochastic
rate $O_p(\sqrt{\ln n/(nh^r)}+\ln n/(nh^r))$.
One can first truncate $Y$ at $C\sqrt{\ln n}$; its discarded
probability is smaller than any prescribed power of $n$ after
increasing $C$. The resulting additional bounded-envelope term is
$O_p\{(\ln n)^{3/2}/(nh^r)\}$, also smaller than
$\sqrt{\ln n/(nh^r)}$ under Assumption~\ref{A4}.
Deleting one observation changes the relevant full-sample averages
by at most $O_p\{\sqrt{\ln n}/(nh^r)\}$. Ratio expansion on
$\inf f\ge\underline f$ gives the asserted regression rates.

For the absolute average, the kernel support and conditional first
moment bound imply
\[
 \E\left[\frac1{n(n-1)}\sum_{i\ne j}
 |K_h(W_i,W_j)|(|Y_i|+|Y_j|)\right]\le C.
\]
This expression dominates $n^{-1}\sum_i|N_i|$. Markov's
inequality proves \eqref{eq:absolute.N}. Finally $a_n\to0$
gives the common positive-denominator event.
\end{proof}

\begin{lemma}[Kernel smoothing across indicator and support boundaries]
\label{lem:indicator.boundary}
Under Assumptions~\ref{A1}--\ref{A3}, let $a(w,z_2)$ be
uniformly bounded, with uniformly bounded $w$-derivatives
through order $s$ and the Taylor remainders in Assumption~\ref{A2}.
For an integration point, write $w=(x',v_1')'$. Then both the kernel
operator and its adjoint satisfy
\[
 \begin{aligned}
 &\sup_z\left\|
 \int_{S_W}K_h(W,w)a(w,z_2)\ind(v_1\le z_1)\,dw
       -a(W,z_2)\ind(Z_1\le z_1)\right\|_{P,2}\\
 &\quad+
 \sup_z\left\|
 \int_{S_W}K_h(w,W)a(w,z_2)\ind(v_1\le z_1)\,dw
       -a(W,z_2)\ind(Z_1\le z_1)\right\|_{P,2}\\
 &\qquad\le C(h^{1/2}+h^s).
 \end{aligned}
\]
The same orders hold for finite sums with factors
$a_0(w,z_2)+Y a_1(w,z_2)$, each $a_j$ satisfying these bounds,
where $Y$ is held fixed inside each kernel integral and
$\E[Y^2\mid W]$ is uniformly bounded.
\end{lemma}

\begin{proof}
The exceptional region consists of a strip of width $Ch$
along the support boundary and strips of width $Ch$
along the marker hyperplanes $Z_{1j}=z_{1j}$.
The constant is enlarged to contain all kernel neighborhoods
intersecting these boundaries. Bounded coordinate densities
and the bounded density on the rectangular support give
\[
 P\{\operatorname{dist}(W,\partial S_W)\le Ch\}
 +\sup_{z_1}\sum_{j=1}^{q_1}
       P(|Z_{1j}-z_{1j}|\le Ch)\le C'h.
\]
Outside this region the indicator is constant throughout the
kernel neighborhood. Both operators are then ordinary
interior convolutions, and the order-$s$ moment conditions
give error $O(h^s)$.

Inside the exceptional region each error is bounded by a
constant. For the forward operator this follows from its
bounded absolute row integral. For the adjoint it follows
from the kernel height $Ch^{-r}$ and a column support
volume $O(h^r)$; its column integral need not equal one.
Thus the squared $L^2(P)$ error of each operator is at most
$C(h+h^{2s})$, proving the stated bound.
For a factor involving $Y$, the exceptional-region envelope
is $C(1+|Y|)$ and conditioning on $W$ gives the same squared
bound. The smooth-region error is handled in the same way.

The $h^{1/2}$ contribution is the square root of the
$O(h)$ strip probability, not a uniform pointwise bias rate.
If $q_1=0$, the marker strips are absent; support-boundary
effects, in particular for the adjoint, can still contribute
this sufficient $L^2$ bound. No claim of sharpness in every
special case is made.
\end{proof}

For a possibly nonsymmetric kernel $H$ of order $d$, write
\[
 U_{n,d}(H)=\frac1{(n)_d}
       \sum_{\substack{i_1,\ldots,i_d\\\text{all distinct}}}
       H(\chi_{i_1},\ldots,\chi_{i_d}),\qquad
 (n)_d=n(n-1)\cdots(n-d+1).
\]
Symmetrization leaves this ordered-index average unchanged. Write
$\bar H$ for the symmetrization of an order-$m$ kernel $H$, and set
\[
 \theta_H=P^m\bar H,\qquad
 H^{[d]}=\pi_d\bar H
 = (\delta_{x_1}-P)\cdots(\delta_{x_d}-P)P^{m-d}\bar H,
 \quad 1\le d\le m.
\]
Each $H^{[d]}$ is canonical: integrating any one of its arguments
gives zero. With the ordered-index normalization above, the exact
Hoeffding decomposition is
\[
 U_{n,m}(H)=\theta_H+
       \sum_{d=1}^m\binom md U_{n,d}(H^{[d]}).
\]
In particular, this algebraic identity needs no bandwidth condition.
Bandwidth conditions are needed to discard its higher-order terms.
The next lemma specializes the auxiliary maximal-inequality lemma
to VC-type classes, in the form of Proposition~4 of
\citet{Delgado&Manteiga2001}; see also \citet{Sherman1994}.

\begin{lemma}[VC-type canonical U-process bound]
\label{lem:maximal}
Let $1\le d\le4$. Suppose $\mathcal H_h$ is a measurable class of
canonical order-$d$ kernels with a square-integrable envelope $F_h$.
For each fixed $h$, suppose that, for every product probability
measure $Q$,
\[
 N(u\norm{F_h}_{Q,2},\mathcal H_h,L^2(Q))\le(A/u)^v,
 \qquad 0<u\le1,
\]
where $A,v$ do not depend on $h,Q$. Then
\[
 \left\{\E\sup_{H\in\mathcal H_h}|U_{n,d}(H)|^2\right\}^{1/2}
 \le C_d n^{-d/2}\norm{F_h}_{P^d,2}.
\]
Consequently, if a canonical projection $H^{[d]}$ has an envelope
of $L^2(P^d)$ order $b_{d,h}$ with this entropy property, its
contribution with deterministic coefficient $c_{n,h}$ is
$O_p(|c_{n,h}|n^{-d/2}b_{d,h})$, uniformly in the index.
A sufficient condition for this contribution to be
$o_p(n^{-1/2})$ is
\[
 |c_{n,h}|n^{(1-d)/2}b_{d,h}\longrightarrow0.
\]
If only the weaker entropy bound $(Ah^{-v}/u)^v$ is available,
the displayed maximal bound instead has the factor
$\{1+\log(1/h)\}^{d/2}$. The log-free version therefore requires
the stated uniform VC characteristics, not just a bound on the
largest individual variance.
\end{lemma}

\begin{proof}
Randomly partition the observations into $d$ disjoint groups of sizes $m_j\asymp n/d$. Conditional on the observations, the average over partitions of the normalized complete cross-group sum is $U_{n,d}(H)$: each ordered distinct tuple has the same inclusion probability and the coefficients sum to one. Jensen's inequality therefore reduces the second moment of the supremum to a cross-group sum. Unconditionally the groups are independent samples.

Canonicality permits successive symmetrization in each group,
at cost at most $2^d$. Conditional on all the observations, the
result is a Rademacher chaos of order $d$. If $R(H)$ denotes its
normalized value, repeated Khintchine inequalities give, for $q\ge2$,
\[
 \norm{R(H)-R(H')}_{q,\epsilon}
 \le C_d q^{d/2}(m_1\cdots m_d)^{-1/2}
       \norm{H-H'}_{Q_m,2},
\]
where $Q_m$ is the product of the group empirical measures.
This uses the conditional empirical norm. Chaining finite covers
in that norm, and taking the conditional $L^2$ norm of the supremum,
gives the upper bound
\[
 C_d(m_1\cdots m_d)^{-1/2}\norm{F_h}_{Q_m,2}
 \int_0^1\{1+\log(A/u)\}^{d/2}\,du.
\]
The integral is finite independently of $h$. Independence of the
groups gives $\E\norm{F_h}_{Q_m,2}^2=\norm{F_h}_{P^d,2}^2$.
This proves the maximal bound; Markov's inequality proves the
consequences. Replacing $A$ by $Ah^{-v}$ gives the weaker version.

Here is why the log-free version applies to the projection classes used below. Under Assumption~\ref{A3}, the bounded-variation kernel translates, and the finite-dimensional local-polynomial boundary corrections are VC-type. Lower orthants are a VC class, and the usual closure properties preserve polynomial covering numbers under finite products and sums; see \citet{van1996weak}. At each fixed $h$, a raw summand is a fixed weight $B_h$ times one lower-orthant indicator. Multiplication by $B_h$ preserves the VC covering bound relative to $|B_h|$, with characteristics
independent of that weight. Finite sums and permutations preserve this property. Conditional integration also preserves it, relative to the integrated absolute envelope. Indeed, for a lower-orthant class $\{A_z:z\in\Pi\}$, if $g_z(x)=\int B_h(x,y)\ind\{(x,y)\in A_z\}\,dP(y)$ and $F(x)=\int|B_h(x,y)|\,dP(y)$, then
\[
 |g_z(x)-g_{z'}(x)|^2
  \le F(x)\int|B_h(x,y)|
  \left|\ind\{(x,y)\in A_z\}-\ind\{(x,y)\in A_{z'}\}\right|^2\,dP(y).
\]
Integrating against $Q(dx)$ reduces covering to the weighted
probability measure proportional to
$F(x)|B_h(x,y)|Q(dx)P(dy)$, whose total mass is $QF^2$.
The VC bound for orthants is uniform over that measure.
After any exact zero integrations have been performed, this
argument applies to each remaining marginal and hence to the
finite sums defining its canonical projections. We verify their
envelope norms separately below. There is no supremum over $h$
in this application, so no growing bandwidth-entropy factor is
needed. At no point is a generic high-moment inequality inferred
from the variance of a single kernel.
\end{proof}

Lemma~\ref{lem:indicator.boundary} supplies the variance radius
for the smoothed projections, while the weighted integration
argument in Lemma~\ref{lem:maximal} supplies the entropy bound.
The following lemma combines these ingredients to control
the first-order discrepancies.

\begin{lemma}[Localized first-order discrepancies]\label{lem:localization}
Let $\mathcal D_h$ be any of the centered first-order discrepancy
classes obtained by replacing a kernel-smoothed projection by its
limit. Suppose that, for every discrete probability measure $Q$, it
satisfies
\[
 N(u\norm{D_h}_{Q,2},\mathcal D_h,L^2(Q))\le(A/u)^v,
 \qquad 0<u\le1,
\]
relative to an envelope $D_h$, as established by the weighted
integration argument in the proof of Lemma~\ref{lem:maximal}, and that
its envelope is bounded by $C(1+|Y|)$, and
$\sup_{d\in\mathcal D_h}P d^2\le\delta_h^2$, where
$\delta_h=O(h^{1/2}+h^s)$. Then
\[
 \sup_{d\in\mathcal D_h}|\Gn d|
 =O_p\{\delta_h\sqrt{\ln n}+(\ln n)^{3/2}/\sqrt n\}=o_p(1).
\]
\end{lemma}

\begin{proof}
Truncate the common outcome envelope at $B_n=C\sqrt{\ln n}$;
the discarded probability and expectation are negligible by
Assumption~\ref{A1}. For the truncated VC-type class of variance
radius $\delta_h$, symmetrization and the Bernstein maximal
inequality on dyadic covers give
$C\delta_h\sqrt{\log(Ah^{-v}B_n/\delta_h)}
+C B_n\log(Ah^{-v}B_n/\delta_h)/\sqrt n$.
One obtains the first term by integrating the square root of
the covering logarithm up to the radius $\delta_h$, and the
second by the bounded-envelope part of Bernstein's inequality. Under the bandwidth conditions, these logarithms are $O(\ln n)$. Taking $\delta_h$ to be a constant multiple of $h^{1/2}+h^s$ gives the displayed bound. The undersmoothing restriction
implies $h^{1/2}\sqrt{\ln n}\to0$. The radius is a uniform bound on individual $L^2$ norms. The envelope of all moving boundary strips need not have a vanishing
$L^2$ norm, and is not treated as if it did.
\end{proof}

\begin{lemma}[Residual-mark empirical and multiplier processes]
\label{lem:residual.process}
Under Assumptions~\ref{A1} and \ref{A3}, consider
\[
 \left\{\varepsilon
 [\ind(Z\le z)-\ind(Z_1\le z_1)F_{Z_2\mid W}(z_2\mid W)]:z\in\Pi\right\}.
\]
The following assertions hold.

\textup{(i)} For constants $C,v$ independent of $Q$,
\[
 \begin{aligned}
 &N\left(u\|\varepsilon\|_{Q,2},
 \left\{\varepsilon
 [\ind(Z\le z)-\ind(Z_1\le z_1)F_{Z_2\mid W}(z_2\mid W)]:z\in\Pi\right\},
 L^2(Q)\right)\\
 &\hspace{35mm}\le(C/u)^v,\qquad 0<u\le1.
 \end{aligned}
\]
The class is $P$-Donsker; its product and squared-increment
classes are $P$-Glivenko--Cantelli. Its mean at $z$ is $m(z)$,
so its centered empirical process has a tight Gaussian limit
with covariance
\[
 \begin{aligned}
 &\E\Big[\varepsilon^2
 [\ind(Z\le z)-\ind(Z_1\le z_1)F_{Z_2\mid W}(z_2\mid W)]\\
 &\qquad\times[\ind(Z\le z')-\ind(Z_1\le z_1')F_{Z_2\mid W}(z_2'\mid W)]\Big]
       -m(z)m(z').
 \end{aligned}
\]

\textup{(ii)} If Assumption~\ref{A5} also holds, then
\[
 \frac1{\sqrt n}\sum_i V_i\varepsilon_i
 [\ind(Z_i\le z)-\ind(Z_{1i}\le z_1)F_{Z_2\mid W}(z_2\mid W_i)]
 \Rightarrow_{P^*}\mathbb G_P^*
 \quad\text{in }\ell^\infty(\Pi),
\]
with the second-moment covariance
\eqref{eq:bootstrap.fixed.covariance}, not the centered
covariance in \textup{(i)}.
Assertions \textup{(i)} and \textup{(ii)} also hold for the
uncorrected class $\{\varepsilon\ind(Z\le z):z\in\Pi\}$,
with its corresponding mean and covariance.
Replacing $\varepsilon$ in the multiplier class by a
fixed bounded function of $W$ gives a conditionally tight
process as well.

\textup{(iii)} Consider row-i.i.d. laws with a common
distribution of $(W,Z)$, uniform versions of
Assumptions~\ref{A1} and \ref{A3}, and residuals
$\varepsilon=Y-\E[Y\mid W]$, where $\E$ refers to the current row law.
Suppose $\sup_z|m_n(z)|\to0$ and
\[
 \E\left|\E[\varepsilon^2\mid W,Z_2]
                   -\E[U^2\mid W,Z_2]\right|\to0,
\]
where the first conditional moment is under the current row law and
the second under a limiting null law satisfying the same moment
conditions. Then
\[
 \begin{aligned}
 &\frac1{\sqrt n}\sum_i
 \Big[\varepsilon_i
 [\ind(Z_i\le z)-\ind(Z_{1i}\le z_1)F_{Z_2\mid W}(z_2\mid W_i)]
                         -m_n(z)\Big]\Rightarrow\mathbb G,\\
 &\frac1{\sqrt n}\sum_i V_i\varepsilon_i
 [\ind(Z_i\le z)-\ind(Z_{1i}\le z_1)F_{Z_2\mid W}(z_2\mid W_i)]
                         \Rightarrow_{P^*}\mathbb G,
 \end{aligned}
\]
the second assertion requiring Assumption~\ref{A5}.
Both limits are in $\ell^\infty(\Pi)$, with
\eqref{eq:covariance} evaluated under the limiting null law.
\end{lemma}

\begin{proof}
We give the covering, covariance, and tightness arguments
here once, so the process theorems need only identify the
appropriate mean and covariance.

For any probability measure $Q$ on $W$, conditional
Jensen's inequality gives
\[
 \begin{aligned}
 &\left\|\ind(Z_1\le z_1)F_{Z_2\mid W}(z_2\mid W)-\ind(Z_1\le z_1')F_{Z_2\mid W}(z_2'\mid W)\right\|_{Q,2}^2\\
 &\quad\le\int
 P\{\ind(Z\le z)\ne\ind(Z\le z')\mid W=w\}\,dQ(w).
 \end{aligned}
\]
The right side is the squared lower-orthant indicator
distance under the mixture of the conditional laws of
$Z$ given $W$. Since lower orthants form a VC class, the universal
covering bound in \citet{van1996weak}
therefore also covers their population conditional
expectations. To include the outcome mark, use the
probability measure proportional to $\varepsilon^2\,dQ$
and its $W$ marginal. If $Q\varepsilon^2=0$, the class
is identically zero in $L^2(Q)$.
Combining covers of the two indicator classes proves
the displayed covering bound. This argument concerns
population conditional expectations, not ratios involving
estimated kernel denominators.

For the ordinary empirical-process limit, lower-orthant
brackets and their conditional expectations give polynomial
$L^2(P)$ bracketing numbers. Multiplication by $\varepsilon$
preserves these bounds: the conditional second moments
given $(W,Z_2)$ and given $W$ are uniformly bounded.
Splitting the positive and negative parts of $\varepsilon$
preserves bracket order. The square-root bracketing entropy integral is finite, and the envelope $2|\varepsilon|$ has a finite second moment. The bracketing central limit theorem of \citet{van1996weak} proves the Donsker assertion. Also,
\[
 \begin{aligned}
 &\E\{\varepsilon[\ind(Z\le z)-\ind(Z_1\le z_1)F_{Z_2\mid W}(z_2\mid W)]\}
 =\E[\varepsilon\ind(Z\le z)]\\
 &\qquad-\E\{\E[\varepsilon\mid W]\ind(Z_1\le z_1)F_{Z_2\mid W}(z_2\mid W)\}
 =m(z),
 \end{aligned}
\]
which gives the centered covariance in \textup{(i)}.

For the product and squared-increment classes, first
truncate $|\varepsilon|$ at a fixed constant $M$ and apply
finite covers to the bounded products. The omitted envelope
is bounded by $C\varepsilon^2\ind(|\varepsilon|>M)$.
Its expectation tends to zero, and its empirical average is
controlled by the law of large numbers. Letting $M\to\infty$
gives the claimed uniform laws. In particular,
\[
 \begin{aligned}
 &\sup_{z,z'}\Big|(\Pn-P)\Big[
 \varepsilon^2
 [\ind(Z\le z)-\ind(Z_1\le z_1)F_{Z_2\mid W}(z_2\mid W)]\\
 &\qquad\times[\ind(Z\le z')-\ind(Z_1\le z_1')F_{Z_2\mid W}(z_2'\mid W)]
 \Big]\Big|\to_p0.
 \end{aligned}
\]

For \textup{(ii)}, condition on the whole sample.
The oracle multiplier sums have mean zero and covariance
equal to the empirical second moments in the last display,
without subtracting $m(z)m(z')$. The uniform law just proved
gives covariance convergence.
The class in \textup{(i)}, augmented by the constant function, is
$P$-Donsker with a square-integrable envelope. Moreover,
$\E|V|^{2+\delta}<\infty$ implies the standard multiplier
integrability condition. The conditional multiplier central limit
theorem in \citet{van1996weak}, applied jointly to the centered mark
class and the constant function, therefore gives conditional
finite-dimensional convergence and asymptotic equicontinuity.
Indeed, writing a generic mark as $q_z$,
\[
 \frac1{\sqrt n}\sum_iV_iq_z(\chi_i)
 =\frac1{\sqrt n}\sum_iV_i\{q_z(\chi_i)-\Pn q_z\}
   +(\Pn q_z)\frac1{\sqrt n}\sum_iV_i
\]
shows why the limiting covariance is $P(q_zq_{z'})$, rather than the
centered covariance $P(q_zq_{z'})-(Pq_z)(Pq_{z'})$.
This proves the conditional process limit in \textup{(ii)}.
The uncorrected class is simpler; a bounded function of
$W$ in place of $\varepsilon$ satisfies the same covering
and moment argument, proving the remaining assertions
of \textup{(ii)}.

For \textup{(iii)}, the covers and moment bounds are uniform
in the rows. Truncation followed by finite covers gives the
rowwise product and increment uniform laws, with truncation
tails uniformly controlled by the fourth moments.
The common covariate law fixes the conditional-indicator
function. Since its residual is bounded by one,
the difference of the row second moments from the limiting
null second moments is bounded uniformly in $(z,z')$ by
\[
 \E\left|\E[\varepsilon^2\mid W,Z_2]
                   -\E[U^2\mid W,Z_2]\right|=o(1),
\]
where the two conditional moments are understood under the current
row law and the limiting null law, respectively.
The centered empirical covariance subtracts
$m_n(z)m_n(z')$, which is uniformly negligible.
The ordinary triangular-array Lindeberg condition follows
from the uniform fourth moments. For the multiplier arrays,
truncate the multipliers at a level diverging slower than $\sqrt n$,
apply the bounded-multiplier maximal inequality to the truncated
array, and use $\E|V|^{2+\delta}<\infty$ to make the tail contribution
uniformly negligible.
Uniform bracketing gives tightness of the centered empirical
arrays. For the multiplier arrays, the covariance comparison
also controls squared increments in the fixed limiting-null
semimetric, so the preceding conditional tightness argument
applies unchanged. Both arrays therefore have the asserted
null Gaussian limit. No assumption of
$\E[\varepsilon\mid W,Z_2]=0$ is used.
\end{proof}

\begin{lemma}[Sample-point KS and empirical CvM functionals]
\label{lem:sample.functionals}
Suppose Assumption~\ref{A1} holds, and let
$\mathbb Z_n$ be a process in $\ell^\infty(\Pi)$ such that
$\mathbb Z_n\Rightarrow\mathbb Z$, where $\mathbb Z$ has almost surely
continuous sample paths. Then
\[
 \max_{1\le j\le n}|\mathbb Z_n(Z_j)|
 \Rightarrow \sup_{z\in\Pi}|\mathbb Z(z)|,\quad
 \int_\Pi \mathbb Z_n(z)^2\,d\hat P_{Z,n}(z)
 \Rightarrow \int_\Pi \mathbb Z(z)^2\,dF_Z(z).
\]
If, conditionally on the data,
$\mathbb Z_n^*\Rightarrow_{P^*}\mathbb Z$ in
$\ell^\infty(\Pi)$, then
\[
 \max_{1\le j\le n}|\mathbb Z_n^*(Z_j)|
 \Rightarrow_{P^*} \sup_{z\in\Pi}|\mathbb Z(z)|,\quad
 \int_\Pi \mathbb Z_n^*(z)^2\,d\hat P_{Z,n}(z)
 \Rightarrow_{P^*} \int_\Pi \mathbb Z(z)^2\,dF_Z(z).
\]
Finally, if
$\sup_{z\in\Pi}|\mathbb Z_n(z)-a(z)|\to_p0$ for a continuous function
$a$, then
\[
 \max_{1\le j\le n}|\mathbb Z_n(Z_j)|\to_p
       \sup_{z\in\Pi}|a(z)|,\quad
 \int_\Pi\mathbb Z_n(z)^2\,d\hat P_{Z,n}(z)\to_p
       \int_\Pi a(z)^2\,dF_Z(z).
\]
The Gaussian limits in Theorems~\ref{thm:process.null},
\ref{thm:process.local}, \ref{thm:process.fixed}, and
\ref{thm:null}--\ref{thm:consistency} satisfy the stated path
continuity condition.
\end{lemma}

\begin{proof}
Let $\Delta_n=\sup_{z\in\Pi}\min_{1\le j\le n}\|z-Z_j\|$.
Compactness and full support imply $\Delta_n\to0$ almost surely. To
see this, cover $\Pi$ by finitely many balls of radius $\eta$. Each
ball that intersects $\Pi$ has strictly positive $F_Z$ probability,
so the probability that any one of these balls contains no sample
point is bounded by a finite sum of terms of the form $(1-p)^n$.

Weak convergence to a process with continuous paths implies
asymptotic equicontinuity in the Euclidean metric. Hence
\[
 0\le \sup_{z\in\Pi}|\mathbb Z_n(z)|
       -\max_{1\le j\le n}|\mathbb Z_n(Z_j)|
 \le \sup_{\|z-z'\|\le\Delta_n}
       |\mathbb Z_n(z)-\mathbb Z_n(z')|=o_p(1).
\]
Moreover, $\hat P_{Z,n}\Rightarrow F_Z$ almost surely. Since its limit
is nonrandom, Slutsky's theorem gives the joint convergence
$(\mathbb Z_n,\hat P_{Z,n})\Rightarrow(\mathbb Z,F_Z)$. The map
\[
 (a,Q)\longmapsto\int_\Pi a(z)^2\,dQ(z)
\]
is continuous at every pair for which $a$ is continuous, when the
first coordinate has the uniform topology and the second the weak
topology. This proves the empirical CvM assertion. The conditional
claims follow from the same mesh bound, conditional Slutsky theorem,
and the conditional continuous mapping theorem. Replacing weak
convergence by uniform convergence in probability proves the final
two deterministic-limit assertions.

For the processes used here, the relevant covariance semimetrics are
bounded by $C\|z-z'\|^{1/2}$. Indeed, the difference between two
lower-orthant indicators is nonzero only when at least one coordinate
of $Z$ lies between the corresponding coordinates of $z$ and $z'$;
Assumption~\ref{A1} bounds this probability by
$C\|z-z'\|_1$, and the conditional moment bounds control the residual
factor. The tight Gaussian limits therefore admit versions with
uniformly continuous, and hence Euclidean-continuous, sample paths.
\end{proof}

\section{Proofs of propositions and theorems}\label{app:theory.proofs}

\subsection{Proof of Proposition~\ref{prop:density}}\label{app:proof.density}

\subsubsection{Exact reciprocal expansion and notation}

Recall explicitly that $\Delta_i=\hat f_i-f_i$, so that
$\hat f_i=f_i+\Delta_i$. On the positive-denominator event of
Lemma~\ref{lem:kernel}, the exact identity is
\begin{equation}\label{eq:reciprocal.cubic}
 \frac1{\hat f_i}=\frac1{f_i}
 -\frac{\Delta_i}{f_i^2}
 +\frac{\Delta_i^2}{f_i^3}
 -\frac{\Delta_i^3}{f_i^3\hat f_i}.
\end{equation}
The signs are $-,+,-$ for this definition of $\Delta_i$.
Multiplication of the right-hand side by $f_i+\Delta_i$ gives
$1$ exactly. If instead the difference were defined as
$f_i-\hat f_i$, all three displayed correction signs would be
positive; one must not change the signs while keeping
$\Delta_i=\hat f_i-f_i$.

Since $\hat T_n(z)=n^{-1}\sum_iN_i\ind(Z_i\le z)/\hat f_i$, it follows that
\begin{equation}\label{eq:LQR}
 \hat T_n(z)=\hat T_n^f(z)-R_{n,1}(z)+R_{n,2}(z)+R_{n,3}(z),
\end{equation}
where, throughout this proof,
\[
 \begin{aligned}
 R_{n,1}(z)&=\frac1n\sum_i\frac{N_i\ind(Z_i\le z)\Delta_i}{f_i^2},\quad
 R_{n,2}(z)&=\frac1n\sum_i\frac{N_i\ind(Z_i\le z)\Delta_i^2}{f_i^3},\\
 R_{n,3}(z)&=-\frac1n\sum_i
       \frac{N_i\ind(Z_i\le z)\Delta_i^3}{f_i^3\hat f_i}.
 \end{aligned}
\]
Thus $R_{n,1}$ and $R_{n,2}$ are defined without their outer signs
in \eqref{eq:LQR}; the minus sign of the exact cubic remainder is
included in $R_{n,3}$. We analyze these three terms separately.
All stochastic orders below are uniform in $z$. For integration
points, write $w=(x',v_1')'$. Also write
$K_{ij}=K_h(W_i,W_j)$,
$A_{ij,z}=K_{ij}(Y_i-Y_j)\ind(Z_i\le z)$, and $H_n=nh^r$.
Use $H^{[d]}=\pi_d\bar H$ as defined before
Lemma~\ref{lem:maximal}.

\subsubsection{Analysis of \texorpdfstring{$R_{n,1}$}{Rn1}}

\emph{Summation expansion and exact Hoeffding decomposition.}
First replace $\Delta_i=D_i+b_h(W_i)$ by $D_i$ and denote the
result by $R_{n,1}^0$. Lemma~\ref{lem:kernel} gives
\[
 \sup_z|R_{n,1}(z)-R_{n,1}^0(z)|
 \le C\|b_h\|_\infty\frac1n\sum_i|N_i|
 =O_p(h^s)=o_p(n^{-1/2})
\]
under $\sqrt n h^s\to0$. The complete sum is
\[
 R_{n,1}^0(z)=\frac1{n(n-1)^2}
       \sum_{i=1}^n\sum_{j\ne i}\sum_{k\ne i}
       \frac{A_{ij,z}d_{ik}}{f_i^2}.
\]
Separating $j\ne k$ from $j=k$ yields the exact partition
\begin{equation}\label{eq:L.partition}
 R_{n,1}^0
 =\frac{n-2}{n-1}U_{n,3}(H_L)
       +\frac1{n-1}U_{n,2}(H_D),
\end{equation}
with distinct argument labels in
\[
 H_L(i,j,k)=\frac{A_{ij,z}d_{ik}}{f_i^2},
 \qquad
 H_D(i,j)=\frac{A_{ij,z}d_{ij}}{f_i^2}.
\]
A distinct $d_{ik}$ integrates to zero, so $\theta_L=0$.
Writing $\theta_D(z)=P^2H_D$, the two Hoeffding identities are
\[
 \begin{aligned}
 U_{n,3}(H_L)
 &=3\Pn H_L^{[1]}+3U_{n,2}(H_L^{[2]})+U_{n,3}(H_L^{[3]}),\\[3pt]
 U_{n,2}(H_D)
 &=\theta_D+2\Pn H_D^{[1]}+U_{n,2}(H_D^{[2]}).
 \end{aligned}
\]

\emph{Higher-order terms and the conditions for a first-order reduction.}
The canonical classes $H_L^{[2]},H_L^{[3]},H_D^{[2]}$ have
VC-type envelopes with respective $L^2$ orders
$h^{-r/2}$, $h^{-r}$, and $h^{-3r/2}$.
Here and below the envelope, rather than just a single kernel
variance, is used. To verify these orders, local support and
bounded kernel height imply, uniformly in $w$,
\[
 \int|K_h(w,v)|^a\,dv\le C_a h^{-(a-1)r},\qquad a\ge1.
\]
For $H_L$, integrating out $k$ gives zero. Integrating out $j$
leaves a bounded marked multiple of $d_{ik}$, with squared
envelope integral $O(h^{-r})$. Integrating out $i$ gives a
two-kernel convolution, bounded by
\[
 C(1+|Y_j|+|Y_k|)
 \{h^{-r}\ind(\|W_j-W_k\|_\infty\le Ch)+1\},
\]
also with squared integral $O(h^{-r})$.
The full squared envelope for $H_L$ integrates to $O(h^{-2r})$.
For $H_D$, its leading squared factor is $K_{ij}^4$, whose
integral is $O(h^{-3r})$. Subtracting finitely many lower
projections does not increase these orders. The envelope
classes have bandwidth-independent VC characteristics by the
fixed-weight and integration argument in Lemma~\ref{lem:maximal}.

Consequently, the total contribution of orders at least two in
\eqref{eq:L.partition} is
\[
 O_p\left\{\frac1{nh^{r/2}}+
             \frac1{n^{3/2}h^r}+
             \frac1{n^2h^{3r/2}}\right\}.
\]
After multiplication by $\sqrt n$, the three rates become
$H_n^{-1/2},H_n^{-1},H_n^{-3/2}$.
Thus $nh^r\to\infty$ suffices to discard these higher-order
terms. This is a condition for their negligibility, not for
the existence of the exact Hoeffding decomposition. It says
nothing yet about either first-order projections or
$\theta_D/(n-1)$.

\emph{First-order projection of the distinct-index term.}
The center-$i$ and neighbor-$j$ first projections of $H_L$
are zero because the independent $k$ factor integrates to
zero. Its $k$ first projection, and hence the sum of the
three first projections, is exactly
\begin{equation}\label{eq:alpha.h}
 \begin{aligned}
 3H_L^{[1]}(\chi_k)
 =&\E[H_L(i,j,k)\mid\chi_k]\\
 =&\int_{S_W}\frac1{f(w)}\Big[
       f_h(w)\E[Y\ind(Z\le z)\mid W=w]\\
 &-\E[K_h(w,W)Y]\ind(w_1\le z_1)F_{Z_2\mid W}(z_2\mid w)\Big]
       \{K_h(w,W_k)-f_h(w)\}\,dw .
 \end{aligned}
\end{equation}
This function has expectation zero. In \eqref{eq:alpha.h}, use
\[
 \frac{f_h(w)}{f(w)}=1+O(h^s),\qquad
 \frac{\E[K_h(w,W)Y]}{f(w)}=G(w)+O(h^s).
\]
Lemma~\ref{lem:indicator.boundary}, applied to the adjoint
kernel and the smooth factor $\psi(z_2,w)-G(w)F(z_2\mid w)$,
controls the indicator and support-boundary strips.
The centering integral differs from $m(z)$ by $O(h^s)$.
Consequently, uniformly in $z$,
\[
 \left\|
 \begin{aligned}
 3H_L^{[1]}(\chi_k)
       &-\E[(Y_k-G(W_k))\ind(Z_k\le z)\mid W_k]\\
       &+\E[(Y_k-G(W_k))\ind(Z_k\le z)]
 \end{aligned}
 \right\|_{P,2}=O(h^{1/2}+h^s).
\]
The norm is taken over the distribution of $\chi_k$.
Both the first-order projection and its conditional-moment limit
are evaluated at $W_k$. The last term is an unconditional
expectation, constant in $\chi_k$ and identical for every $k$.
It centers the limiting projection, since
\[
 \E\!\left[\E[(Y_k-G(W_k))\ind(Z_k\le z)\mid W_k]\right]
       =\E[(Y_k-G(W_k))\ind(Z_k\le z)].
\]
Both functions in the discrepancy have mean zero exactly.
Lemma~\ref{lem:localization}, with
$(h^{1/2}+h^s)\sqrt{\ln n}\to0$ and
$(\ln n)^{3/2}/\sqrt n\to0$, gives
\begin{equation}\label{eq:alpha.empirical}
 \begin{aligned}
 &\sup_z\Bigg|\frac1n\sum_{k=1}^n\Big[
       3H_L^{[1]}(\chi_k)
       -\E[(Y_k-G(W_k))\ind(Z_k\le z)\mid W_k]\\
 &\hspace{35mm}+\E[(Y_k-G(W_k))\ind(Z_k\le z)]\Big]\Bigg|
       =o_p(n^{-1/2}).
 \end{aligned}
\end{equation}
These conditions follow from the maintained bandwidth
assumptions. The first-order term therefore cannot in general
be absorbed into $o_p(n^{-1/2})$: for any fixed $z$ with
$\Var\{\E[(Y-G(W))\ind(Z\le z)\mid W]\}>0$, its
$\sqrt n$-scaled version has a
nondegenerate normal limit. Even under $H_0$,
$\E[(Y-G(W))\ind(Z\le z)]=0$ does
not imply $\E[(Y-G(W))\ind(Z\le z)\mid W]=0$.

\begin{samepage}
\emph{First-order projection and mean of the repeated-index term.}
Integrating out the neighbor gives
\[
 \begin{aligned}
 \E[H_D(i,j)\mid\chi_i]
 =\frac{\ind(Z_i\le z)}{f_i^2}\int_{S_W}
 &K_h(W_i,w)\{K_h(W_i,w)-f_h(W_i)\}\\
 &{}\times\{Y_i-G(w)\}f(w)\,dw .
 \end{aligned}
\]
\end{samepage}
Integrating out the center instead gives
\[
 \begin{aligned}
 \E[H_D(i,j)\mid\chi_j]
 =\int_{S_W}
 &\frac{K_h(w,W_j)\{K_h(w,W_j)-f_h(w)\}}{f(w)}\\
 &{}\times\ind(v_1\le z_1)
       \{\psi(z_2,w)-Y_jF(z_2\mid w)\}\,dw .
 \end{aligned}
\]
Each conditional expectation has mean $\theta_D$.
Their centered sum, evaluated at the same observation, is
\[
 2H_D^{[1]}(\chi_i)
 =\E[H_D(i,j)\mid\chi_i]
       +\E[H_D(j,i)\mid\chi_i]-2\theta_D .
\]
The centered class has an envelope of $L^2(P)$ order
$h^{-r}$, since integrating a squared kernel costs $O(h^{-r})$.
Lemma~\ref{lem:maximal} with $d=1$ therefore gives
\[
 \sup_z\left|\frac{2}{n-1}\Pn H_D^{[1]}\right|
       =O_p(n^{-3/2}h^{-r})=o_p(n^{-1/2})
       \quad\text{if }nh^r\to\infty .
\]
This centered bound does not remove the mean. Direct conditioning
gives its full expression:
\[
 \begin{aligned}
 \theta_D(z)
 =&\int_{S_W}\frac1{f(w)}\Big[
       \{\E[K_h(w,W)^2]-f_h(w)^2\}
       \E[Y\ind(Z\le z)\mid W=w]\\
 &-\{\E[K_h(w,W)^2Y]-f_h(w)\E[K_h(w,W)Y]\}
       \ind(w_1\le z_1)F_{Z_2\mid W}(z_2\mid w)\Big]\,dw\\
 =&O(h^{-r}).
 \end{aligned}
\]
Hence a sufficient condition to discard this mean \emph{on its own}
is $\sqrt n h^r\to\infty$, since
\[
 \sup_z\frac{\sqrt n|\theta_D(z)|}{n-1}
       =O\{(\sqrt n h^r)^{-1}\}.
\]
The maintained assumptions do not in general imply this stronger
condition. Nor does the null make the displayed mean zero.
We therefore retain it.

Combining only the justified reductions within this subsection,
including $(n-2)/(n-1)=1+O(n^{-1})$, gives
\[
 R_{n,1}(z)=(\Pn-P)\E[(Y-G(W))\ind(Z\le z)\mid W]
       +\frac{\theta_D(z)}{n-1}
                 +o_p(n^{-1/2}).
\]
Thus $R_{n,1}$ itself is not generally negligible.
For example, if, for every $z$,
\[
 \E[(Y-G(W))\ind(Z\le z)\mid W]
       =\E[(Y-G(W))\ind(Z\le z)]\quad\text{a.s.},
\]
then this condition, together with
$\sqrt n h^r\to\infty$ and the preceding bias and localization
conditions, is sufficient to make it $o_p(n^{-1/2})$.
Without such restrictions the displayed terms must remain explicit.

\subsubsection{Analysis of \texorpdfstring{$R_{n,2}$}{Rn2}}

\emph{Summation expansion and exact Hoeffding decomposition.}
First replace $\Delta_i^2$ by $D_i^2$ inside $R_{n,2}$.
The absolute-average bound controls the resulting difference:
\[
 \begin{aligned}
 &\sup_z\left|R_{n,2}(z)
       -\frac1n\sum_i\frac{N_i\ind(Z_i\le z)D_i^2}{f_i^3}\right|\\
 &\qquad\le C\|b_h\|_\infty
       (2\max_i|D_i|+\|b_h\|_\infty)\frac1n\sum_i|N_i|
 =O_p\{h^s(a_n+h^s)\}=o_p(n^{-1/2}).
 \end{aligned}
\]
The sufficient condition is
$\sqrt n h^s(a_n+h^s)\to0$, as follows from
Assumption~\ref{A4}. Expanding the square inside the sum gives
\[
 R_{n,2}(z)=\frac1{n(n-1)^3}
       \sum_i\sum_{j\ne i}\sum_{k\ne i}\sum_{l\ne i}
       \frac{A_{ij,z}d_{ik}d_{il}}{f_i^3}
       +o_p(n^{-1/2}).
\]
Partition the quadruple sum by the equalities among $j,k,l$.
Keeping the exact coefficients, this gives
\begin{equation}\label{eq:Q.partition}
 \begin{aligned}
 R_{n,2}(z)
 ={}&\frac{(n-2)(n-3)}{(n-1)^2}
       U_{n,4}\left(\frac{A_{12,z}d_{13}d_{14}}{f_1^3}\right)
       +\frac{n-2}{(n-1)^2}
       U_{n,3}\left(\frac{A_{12,z}d_{13}^2}{f_1^3}\right)\\
 &+\frac{2(n-2)}{(n-1)^2}
       U_{n,3}\left(\frac{A_{12,z}d_{12}d_{13}}{f_1^3}\right)
       +\frac1{(n-1)^2}
       U_{n,2}\left(\frac{A_{12,z}d_{12}^2}{f_1^3}\right)
       +o_p(n^{-1/2}).
 \end{aligned}
\end{equation}
The four terms correspond, respectively, to all $j,k,l$ distinct,
$k=l\ne j$, either $j=k\ne l$ or $j=l\ne k$, and $j=k=l$.
There is no index equal to the center $i$.
The $o_p(n^{-1/2})$ term comes only from replacing $\Delta_i^2$
by $D_i^2$; the partition of the quadruple sum itself is exact.

Using the symmetrization and projection operators already
defined before Lemma~\ref{lem:maximal}, the exact Hoeffding decompositions
of these four averages are
\[
 \begin{aligned}
 &U_{n,4}\left(\frac{A_{12,z}d_{13}d_{14}}{f_1^3}\right)
 =\sum_{d=2}^4\binom4d
       U_{n,d}\left(\pi_d\overline{
                         A_{12,z}d_{13}d_{14}/f_1^3}\right),\\[4pt]
 &U_{n,3}\left(\frac{A_{12,z}d_{13}^2}{f_1^3}\right)
 =P^3\left(\frac{A_{12,z}d_{13}^2}{f_1^3}\right)
       +3\Pn\pi_1\overline{A_{12,z}d_{13}^2/f_1^3}
       +\sum_{d=2}^3\binom3d
       U_{n,d}\left(\pi_d\overline{
                         A_{12,z}d_{13}^2/f_1^3}\right),\\[4pt]
 &U_{n,3}\left(\frac{A_{12,z}d_{12}d_{13}}{f_1^3}\right)
 =3\Pn\pi_1\overline{A_{12,z}d_{12}d_{13}/f_1^3}
 +\sum_{d=2}^3\binom3d
       U_{n,d}\left(\pi_d\overline{
                         A_{12,z}d_{12}d_{13}/f_1^3}\right),\\[4pt]
 &U_{n,2}\left(\frac{A_{12,z}d_{12}^2}{f_1^3}\right)
 =P^2\left(\frac{A_{12,z}d_{12}^2}{f_1^3}\right)
       +2\Pn\pi_1\overline{A_{12,z}d_{12}^2/f_1^3}
       +U_{n,2}\left(\pi_2\overline{
                         A_{12,z}d_{12}^2/f_1^3}\right).
 \end{aligned}
\]
The mean and all first-order projections of the order-four
kernel are zero. The mean of the third kernel is also zero.
In each case, a distinct centered $d$ factor can be integrated
out. The two displayed nonzero means must be retained.

\emph{Higher-order terms and their negligibility conditions.}
For the first kernel in \eqref{eq:Q.partition}, its canonical
projections of orders $2,3,4$ have envelope $L^2$ orders
$h^{-r/2}$, $h^{-r}$, and $h^{-3r/2}$,
respectively. For each order-three kernel, the projections of
orders $2,3$ have envelope orders $h^{-3r/2},h^{-2r}$.
For the last kernel, its order-two canonical projection has
envelope order $h^{-5r/2}$.

Here are the kernel counts establishing these bounds.
For the first kernel, the only nonzero two-coordinate marginal
retains $k,l$ and integrates $i,j$. Its envelope is a two-kernel
convolution of $L^2$ order $h^{-r/2}$. Its three-coordinate
marginals have two localized factors, of $L^2$ order $h^{-r}$.
The full squared envelope integrates to $O(h^{-3r})$.
For the $k=l\ne j$ term, the full leading squared factor is
$K_{ij}^2K_{ik}^4$, with integral $O(h^{-4r})$.
Integrating $j$ leaves a squared edge $ik$; integrating $k$
leaves $O(h^{-r})$ times the edge $ij$; integrating $i$
gives a convolution bounded by $Ch^{-2r}$ on
$\|W_j-W_k\|_\infty\le Ch$, plus lower terms.
All three two-coordinate envelopes have squared integral
$O(h^{-3r})$. The same counts apply to $j=k\ne l$ or
$j=l\ne k$, with the squared edge $ij$ instead of $ik$.
For the $j=k=l$ term the leading squared factor is $K_{ij}^6$,
whose integral is $O(h^{-5r})$.
Finite canonical centering preserves these envelope bounds,
and the conditional outcome second moments are uniformly bounded.

Apply Lemma~\ref{lem:maximal} to the projections of degrees
$2,3,4$ of the first kernel, degrees $2,3$ of the next two
kernels, and degree $2$ of the last kernel. Multiplying by
their respective coefficients in \eqref{eq:Q.partition}, of
orders $1,n^{-1},n^{-1},n^{-2}$, bounds the total higher-order
contribution by
\[
 O_p\left\{
 \frac1{nh^{r/2}}+\frac1{n^{3/2}h^r}
 +\frac1{n^2h^{3r/2}}+\frac1{n^{5/2}h^{2r}}
 +\frac1{n^3h^{5r/2}}\right\}.
\]
Their $\sqrt n$-scaled rates are, respectively,
$H_n^{-1/2},H_n^{-1},H_n^{-3/2},H_n^{-2},H_n^{-5/2}$.
Thus $nh^r\to\infty$ suffices for the higher-order reduction.
This does not yet discard the first-order projections or means.

\emph{Low-order projections, starting with order one.}
For the all-distinct term, every first-order projection is zero.
Its only potentially nonzero two-coordinate marginal retains
$k,l$, and symmetrization gives the following full expression:
\[
 \begin{aligned}
 &6\left(\pi_2\overline{
                 A_{12,z}d_{13}d_{14}/f_1^3}\right)(\chi_k,\chi_l)
 =\int_{S_W}\frac1{f(w)^2}\Big[
       f_h(w)\E[Y\ind(Z\le z)\mid W=w]\\
 &\qquad-\E[K_h(w,W)Y]\ind(w_1\le z_1)F_{Z_2\mid W}(z_2\mid w)\Big]\\
 &\qquad{}\times
       \{K_h(w,W_k)-f_h(w)\}
       \{K_h(w,W_l)-f_h(w)\}\,dw .
 \end{aligned}
\]
The right-hand side is canonical in $W_k,W_l$ and has an
envelope of $L^2$ order $h^{-r/2}$. Its contribution is
$O_p(n^{-1}h^{-r/2})$ by Lemma~\ref{lem:maximal} with $d=2$,
and hence is $o_p(n^{-1/2})$ if $nh^r\to\infty$.

For the $k=l\ne j$ term, the three one-coordinate conditional
expectations are
\begin{align*}
 &\E\left[\frac{A_{ij,z}d_{ik}^2}{f_i^3}\,\middle|\,\chi_i\right]
 =\frac{\ind(Z_i\le z)}{f_i^3}
       \left[\int_{S_W}K_h(W_i,w)\{Y_i-G(w)\}f(w)\,dw\right]\\
 &\qquad{}\times
       \{\E[K_h(W_i,W)^2\mid W_i]-f_h(W_i)^2\},\\[4pt]
 &\E\left[\frac{A_{ij,z}d_{ik}^2}{f_i^3}\,\middle|\,\chi_j\right]
 =\int_{S_W}\frac{K_h(w,W_j)}{f(w)^2}
       \ind(v_1\le z_1)\{\psi(z_2,w)-Y_jF(z_2\mid w)\}\\
 &\qquad{}\times\{\E[K_h(w,W)^2]-f_h(w)^2\}\,dw ,\\[4pt]
 &\E\left[\frac{A_{ij,z}d_{ik}^2}{f_i^3}\,\middle|\,\chi_k\right]
 =\int_{S_W}\frac1{f(w)^2}\Big[
       f_h(w)\E[Y\ind(Z\le z)\mid W=w]\\
 &\qquad-\E[K_h(w,W)Y]\ind(w_1\le z_1)F_{Z_2\mid W}(z_2\mid w)\Big]
       \{K_h(w,W_k)-f_h(w)\}^2\,dw . 
\end{align*}
Their common mean is explicitly
\[
 \begin{aligned}
 &P^3\left(\frac{A_{12,z}d_{13}^2}{f_1^3}\right)
 =\int_{S_W}\frac1{f(w)^2}\Big[
       f_h(w)\E[Y\ind(Z\le z)\mid W=w]\\
 &\qquad-\E[K_h(w,W)Y]\ind(w_1\le z_1)F_{Z_2\mid W}(z_2\mid w)\Big]\\
 &\qquad{}\times\{\E[K_h(w,W)^2]-f_h(w)^2\}\,dw
 =O(h^{-r}).
 \end{aligned}
\]
Subtracting this mean from each conditional expectation and
adding the three centered functions at the same argument gives
$3\pi_1\overline{A_{12,z}d_{13}^2/f_1^3}$.
Their envelope norms are $O(h^{-r})$, because one squared
kernel integrates to $O(h^{-r})$ and each ordinary kernel
integrates at bounded cost. Lemma~\ref{lem:maximal} with $d=1$
therefore gives
\[
 \sup_z\left|\frac{3(n-2)}{(n-1)^2}
       \Pn\pi_1\overline{A_{12,z}d_{13}^2/f_1^3}\right|
 =O_p(n^{-3/2}h^{-r})=o_p(n^{-1/2})
\]
if $nh^r\to\infty$. The mean multiplied by
$(n-2)/(n-1)^2$ has not been centered away.

For the $j=k\ne l$ or $j=l\ne k$ term, the center-$i$
and neighbor-$j$ first projections are zero. The remaining
one-coordinate conditional expectation is
\[
 \begin{aligned}
 &\E\left[\frac{A_{ij,z}d_{ij}d_{ik}}{f_i^3}
                                      \,\middle|\,\chi_k\right]
 =\int_{S_W}\frac1{f(w)^2}\Big[
       \{\E[K_h(w,W)^2]-f_h(w)^2\}
       \E[Y\ind(Z\le z)\mid W=w]\\
 &\qquad-\{\E[K_h(w,W)^2Y]-f_h(w)\E[K_h(w,W)Y]\}
       \ind(w_1\le z_1)F_{Z_2\mid W}(z_2\mid w)\Big]\\
 &\qquad{}\times\{K_h(w,W_k)-f_h(w)\}\,dw .
 \end{aligned}
\]
This function is centered exactly, since averaging over $W_k$
annihilates the final kernel difference.
Its envelope has $L^2$ order $h^{-r}$: the expectations involving
a squared kernel are $O(h^{-r})$, while
$\int_{S_W}|K_h(w,v)-f_h(w)|\,dw\le C$.
The sum of the first three projections equals this single
nonzero projection. Applying Lemma~\ref{lem:maximal} with $d=1$,
including the coefficient of the third term in
\eqref{eq:Q.partition}, gives, under $nh^r\to\infty$,
\[
 \sup_z\left|\frac{6(n-2)}{(n-1)^2}
       \Pn\pi_1\overline{A_{12,z}d_{12}d_{13}/f_1^3}\right|
 =O_p(n^{-3/2}h^{-r})=o_p(n^{-1/2})
\]
For the $j=k=l$ term, its two one-coordinate conditional
expectations are
\[
 \begin{aligned}
 &\E\left[\frac{A_{ij,z}d_{ij}^2}{f_i^3}\,\middle|\,\chi_i\right]
 =\frac{\ind(Z_i\le z)}{f_i^3}\int_{S_W}
       K_h(W_i,w)\{K_h(W_i,w)-f_h(W_i)\}^2
       \{Y_i-G(w)\}f(w)\,dw ,
 \end{aligned}
\]
and
\[
 \begin{aligned}
 &\E\left[\frac{A_{ij,z}d_{ij}^2}{f_i^3}\,\middle|\,\chi_j\right]
 =\int_{S_W}
       \frac{K_h(w,W_j)\{K_h(w,W_j)-f_h(w)\}^2}{f(w)^2}\\
 &\qquad{}\times\ind(v_1\le z_1)
       \{\psi(z_2,w)-Y_jF(z_2\mid w)\}\,dw .
 \end{aligned}
\]
Their common mean is $P^2(A_{12,z}d_{12}^2/f_1^3)$.
Subtracting that mean from each and adding at the same argument
gives $2\pi_1\overline{A_{12,z}d_{12}^2/f_1^3}$.
The envelope norms are $O(h^{-2r})$, because the leading
factor is an integrated cube of the kernel.
Lemma~\ref{lem:maximal} with $d=1$ therefore gives, under $nh^r\to\infty$,
\[
 \sup_z\left|\frac{2}{(n-1)^2}
       \Pn\pi_1\overline{A_{12,z}d_{12}^2/f_1^3}\right|
 =O_p(n^{-5/2}h^{-2r})=o_p(n^{-1/2})
\]
The mean must be bounded separately:
\[
 \sup_z\left|P^2\left(\frac{A_{12,z}d_{12}^2}{f_1^3}\right)\right|
 \le \E\sup_z\left|\frac{A_{12,z}d_{12}^2}{f_1^3}\right|
 \le Ch^{-2r}.
\]
Its coefficient $1/(n-1)^2$ therefore gives order
$O(n^{-2}h^{-2r})$. A sufficient condition for this mean
to be negligible at root-$n$ scale is
$n^{3/2}h^{2r}\to\infty$ or, equivalently,
$H_n/n^{1/4}\to\infty$.
This follows from the stochastic cubic restriction in
Assumption~\ref{A4}, which implies
$H_n/\{n^{1/3}\ln n\}\longrightarrow\infty$.

Before discarding any means, the justified expansion is thus
\[
 \begin{aligned}
 R_{n,2}(z)
 ={}\frac{n-2}{(n-1)^2}
       P^3\left(\frac{A_{12,z}d_{13}^2}{f_1^3}\right)+\frac1{(n-1)^2}
       P^2\left(\frac{A_{12,z}d_{12}^2}{f_1^3}\right)
       +o_p(n^{-1/2}).
 \end{aligned}
\]
Under the maintained cubic condition, this reduces to
\[
 R_{n,2}(z)=\frac{n-2}{(n-1)^2}
       P^3\left(\frac{A_{12,z}d_{13}^2}{f_1^3}\right)
       +o_p(n^{-1/2}).
\]
The remaining mean is $O((nh^r)^{-1})$ and need not be
$o(n^{-1/2})$. As for $R_{n,1}$, a sufficient condition to discard
it separately is $\sqrt n h^r\to\infty$; alternatively one must
establish a sharper bound on the expectation displayed above.
Neither centering of the Hoeffding projections nor the null
hypothesis by itself establishes that stronger conclusion.
Thus we do not assert $R_{n,2}=o_p(n^{-1/2})$ under the
maintained conditions alone.

\subsubsection{Direct bound for \texorpdfstring{$R_{n,3}$}{Rn3}}

No Hoeffding reduction is needed here. On the event with
denominators bounded away from zero,
\begin{equation}\label{eq:cubic.bound}
 \sup_z|R_{n,3}(z)|
 \le C\max_i|\Delta_i|^3\frac1n\sum_i|N_i|
 =O_p(a_n^3)=o_p(n^{-1/2}).
\end{equation}
The last equality requires exactly the maintained sufficient
condition $\sqrt n a_n^3\to0$. In particular, its stochastic part
is $n^2h^{3r}/(\ln n)^3\to\infty$.
The absolute average in \eqref{eq:absolute.N} is $O_p(1)$.
We use no claimed cancellation of
$n^{-1}\sum_iN_i\ind(Z_i\le z)$ when bounding its random multipliers.

\medskip
\noindent\emph{Completion of the proof.}
The separate analyses have retained every term that was not
shown to be $o_p(n^{-1/2})$. Inserting those analyses into
\eqref{eq:LQR} gives
\[
 \begin{aligned}
 \hat T_n-\hat T_n^f
 ={}&-(\Pn-P)\E[(Y-G(W))\ind(Z\le z)\mid W]
       -\frac{\theta_D}{n-1}\\
 &+\frac{n-2}{(n-1)^2}
       P^3\left(\frac{A_{12,z}d_{13}^2}{f_1^3}\right)
       +o_p(n^{-1/2}).
 \end{aligned}
\]
Only now do we compare the two retained means. Their full
expressions above, $b_h=O(h^s)$, and $\sup_w|\E[K_h(w,W)\{G(w)-Y\}]|=O(h^s)$ give
\[
 \begin{aligned}
 &\sup_z\left|\theta_D(z)
       -P^3\left(\frac{A_{12,z}d_{13}^2}{f_1^3}\right)\right|\\
 &\quad\le C\left\{h^{s-r}+\int_{S_W}
       \left|\E\{K_h(w,W)^2(G(w)-Y)\}\right|\,dw\right\}
 =O(h^{2-r}).
 \end{aligned}
\]
Here $\E[K_h(w,W)^2]-f_h(w)^2=O(h^{-r})$.
For the integrated expectation, in the interior $K^2$ is
symmetric, so the linear Taylor term of $G(w)-G(W)$ vanishes
and the quadratic remainder is $O(h^{2-r})$.
On a support-boundary strip the pointwise bound is
$O(h^{1-r})$ and the strip volume is $O(h)$.
This uses two derivatives, not the unsupported assertion that
the squared kernel has order $s$; $s\ge2$ also gives
$h^{s-r}=O(h^{2-r})$.
It follows that the \emph{combined} mean contribution is
\[
 \begin{aligned}
 &-\frac{\theta_D}{n-1}
       +\frac{n-2}{(n-1)^2}
             P^3\left(\frac{A_{12,z}d_{13}^2}{f_1^3}\right)\\
 &\quad=-\frac1{n-1}\left\{\theta_D
             -P^3\left(\frac{A_{12,z}d_{13}^2}{f_1^3}\right)\right\}
       -\frac1{(n-1)^2}
             P^3\left(\frac{A_{12,z}d_{13}^2}{f_1^3}\right)\\
 &\quad=O(n^{-1}h^{2-r}+n^{-2}h^{-r})
       =o(n^{-1/2}).
 \end{aligned}
\]
Here the sufficient conditions are
$h^{2-r}/\sqrt n\to0$ and $n^{3/2}h^r\to\infty$, both implied
by Assumption~\ref{A4}. This joint cancellation does not say
that either retained mean was individually negligible.
Thus \eqref{aux1} follows without imposing the additional
separate-mean condition $\sqrt n h^r\to\infty$.
If $\Var\{\E[(Y-G(W))\ind(Z\le z)\mid W]\}>0$ for some $z$,
its empirical first-order
correction is not $o_p(n^{-1/2})$. No null hypothesis was used.
\hfill$\square$

\subsection{Proof of Proposition~\ref{prop:linear}}\label{app:proof.linear}

Recall the auxiliary process $\hat T_n^f$ explicitly defined in
Section~\ref{sec:expansions}. Using the notation already introduced
in the proof of Proposition~\ref{prop:density}, first average over
the two orders $(i,j)$ and $(j,i)$. The symmetrization and projection
operators defined before Lemma~\ref{lem:maximal} then give the exact Hoeffding decomposition
\begin{equation}\label{eq:hajek.exact}
 \begin{aligned}
 \hat T_n^f(z)
 ={}&\frac{2}{n(n-1)}\sum_{1\le i<j\le n}
       \frac12\left(\frac{A_{ij,z}}{f_i}
                         +\frac{A_{ji,z}}{f_j}\right)\\
 ={}&\E\left[\frac{A_{12,z}}{f_1}\right]
       +2\Pn\pi_1\overline{A_{12,z}/f_1}
       +U_{n,2}\left(\pi_2\overline{A_{12,z}/f_1}\right).
 \end{aligned}
\end{equation}
In particular, the first-order projection in this formula satisfies
\[
 \begin{aligned}
 2\left(\pi_1\overline{A_{12,z}/f_1}\right)(\chi_i)
 ={}&\E\left[\frac{A_{ij,z}}{f_i}
                         +\frac{A_{ji,z}}{f_j}
                    \,\middle|\,\chi_i\right]
 -2\E\left[\frac{A_{12,z}}{f_1}\right].
 \end{aligned}
\]
Here $j\ne i$ is integrated out in the conditional expectation.
The factor $2$ comes from the order-two Hoeffding decomposition;
the factor $1/2$ in the permutation average is already included in
the symmetrization. The second-order projection is
\[
 \begin{aligned}
 \left(\pi_2\overline{A_{12,z}/f_1}\right)(\chi_i,\chi_j)
 =&\frac12\left(\frac{A_{ij,z}}{f_i}
                         +\frac{A_{ji,z}}{f_j}\right)
                         -\frac12\E\left[\frac{A_{ij,z}}{f_i}
                         +\frac{A_{ji,z}}{f_j}
                    \,\middle|\,\chi_i\right]\\
 &-\frac12\E\left[\frac{A_{ij,z}}{f_i}
                         +\frac{A_{ji,z}}{f_j}
                    \,\middle|\,\chi_j\right]
       +\E\left[\frac{A_{12,z}}{f_1}\right].
 \end{aligned}
\]
This kernel has conditional mean zero given either observation.
Its $L^2(P^2)$ envelope is $O(h^{-r/2})$:
the original kernel has this order by Assumptions~\ref{A1} and
\ref{A3}, and conditional expectation does not increase the
$L^2$ envelope norm. Lemma~\ref{lem:maximal}, with $d=2$, therefore
bounds the last term in \eqref{eq:hajek.exact} uniformly in $z$ by
$O_p(n^{-1}h^{-r/2})=o_p(n^{-1/2})$, provided $nh^r\to\infty$.
This condition follows from Assumption~\ref{A4}. The exact
decomposition itself requires no bandwidth condition.

For the first conditional projection, direct integration gives
\begin{align*}
 \E\left[\frac{A_{ij,z}}{f_i}\,\middle|\,\chi_i\right]
 &=\frac{\ind(Z_i\le z)}{f_i}
       \int_{S_W}K_h(W_i,w)\{Y_i-G(w)\}f(w)\,dw\\
 &=\varepsilon_i \ind(Z_i\le z)+O(h^s)(1+|Y_i|),
\end{align*}
uniformly in $z$ and $W_i$. Since
$n^{-1}\sum_i(1+|Y_i|)=O_p(1)$ and $\sqrt n h^s\to0$,
replacing this projection by $\varepsilon_i \ind(Z_i\le z)$
in the first-order term costs $o_p(n^{-1/2})$ uniformly.
Taking expectations also yields
\[
 \sup_z\left|
 \E\left[\frac{A_{12,z}}{f_1}\right]
 -m(z)\right|=O(h^s)=o(n^{-1/2}).
\]

For the second conditional projection,
\begin{align*}
 &\E\left[\frac{A_{ji,z}}{f_j}\,\middle|\,\chi_i\right]
 =\int_{S_W}K_h(w,W_i)\Big[
       \E[Y\ind(Z\le z)\mid W=w]
       -Y_i\ind(w_1\le z_1)F_{Z_2\mid W}(z_2\mid w)\Big]\,dw.
\end{align*}
Its difference from
$\E[Y_i\ind(Z_i\le z)\mid W_i]
-Y_i\ind(Z_{1i}\le z_1)F_{Z_2\mid W}(z_2\mid W_i)$
has $L^2(P)$ norm $O(h^{1/2}+h^s)$ uniformly in $z$ by
Lemma~\ref{lem:indicator.boundary}, using its adjoint and
marked versions with smooth factor $\psi(z_2,w)-Y_iF(z_2\mid w)$.
As in \eqref{eq:alpha.empirical}, apply
Lemma~\ref{lem:localization} to this difference in the
\emph{centered} empirical process, rather than bounding its
uncentered average by the probability of the strips. Thus
\[
 \sup_z\left|\Gn\left\{
 \begin{aligned}
 &\int_{S_W}K_h(w,W)\Big[
       \E[Y\ind(Z\le z)\mid W=w]\\
 &\qquad\qquad-Y\ind(w_1\le z_1)F_{Z_2\mid W}(z_2\mid w)\Big]\,dw\\
 &-\E[Y\ind(Z\le z)\mid W]
       +Y\ind(Z_1\le z_1)F_{Z_2\mid W}(z_2\mid W)
 \end{aligned}
 \right\}\right|=o_p(1).
\]
The second projection cannot be discarded as conditionally
mean zero: by iterated expectations,
\[
 \begin{aligned}
 &\E\Big[
 \E[Y_i\ind(Z_i\le z)\mid W_i]
       -Y_i\ind(Z_{1i}\le z_1)F_{Z_2\mid W}(z_2\mid W_i)
       \,\Big|\,W_i\Big]\\
 &\qquad=\E[(Y_i-G(W_i))\ind(Z_i\le z)\mid W_i],
 \end{aligned}
\]
which need not vanish under $H_0$.

Insert the projection and mean approximations into
\eqref{eq:hajek.exact} and use the identity
\[
 \begin{aligned}
 &\varepsilon \ind(Z\le z)
       +\E[Y\ind(Z\le z)\mid W]
       -Y\ind(Z_1\le z_1)F_{Z_2\mid W}(z_2\mid W)\\
 =&\varepsilon\{\ind(Z\le z)-\ind(Z_1\le z_1)F_{Z_2\mid W}(z_2\mid W)\}
 +\E[(Y-G(W))\ind(Z\le z)\mid W],
 \end{aligned}
\]
to obtain, uniformly in $z$,
\begin{equation}\label{eq:population.denominator.expansion}
 \begin{aligned}
 &\hat T_n^f(z)=m(z)+(\Pn-P)\Big[
 \varepsilon\{\ind(Z\le z)-\ind(Z_1\le z_1)F_{Z_2\mid W}(z_2\mid W)\}\\
 &\qquad\qquad\quad+\E[(Y-G(W))\ind(Z\le z)\mid W]\Big]
       +o_p(n^{-1/2}).
 \end{aligned}
\end{equation}
This is the centered representation in \eqref{eq:uniform_final}.
Since $\E[\varepsilon\mid W]=0$,
$\E[\varepsilon\{\ind(Z\le z)-\ind(Z_1\le z_1)F_{Z_2\mid W}(z_2\mid W)\}]=m(z)$.
Also, by iterated expectations,
$\E\{\E[(Y-G(W))\ind(Z\le z)\mid W]\}=m(z)$,
so expanding $\Pn-P$ gives the sample-average representation
in the proposition. No null hypothesis has been used;
only replacing $\varepsilon$ by $U$ requires $H_0$.
\hfill$\square$

\subsection{Proofs of the observed-process limits}\label{app:proof.process}

\begin{proof}[Proof of Theorem~\ref{thm:process.null}]
Combining \eqref{aux1} and \eqref{eq:uniform_final}
cancels their conditional-moment averages and deterministic
$m(z)$ terms, giving \eqref{eq:observed.expansion}.
Under $H_0$, $\varepsilon=U$ and $m(z)=0$, so
\[
 \sqrt n\hat T_n(z)
 =\frac1{\sqrt n}\sum_iU_i
       \{\ind(Z_i\le z)-\ind(Z_{1i}\le z_1)F_{Z_2\mid W}(z_2\mid W_i)\}
       +o_p(1)
\]
uniformly in $z$. The summand has mean zero:
$\E[U\mid W]=0$ removes the conditional-indicator term,
and $\E[U\mid Z]=0$ gives $\E[U\ind(Z\le z)]=0$.
Its covariance is \eqref{eq:covariance}.
Lemma~\ref{lem:residual.process}\textup{(i)} therefore gives
the Gaussian limit $\mathbb G$ for the displayed sum, and
Slutsky proves the claim.
Neither the multiplier assumption nor the stronger restriction
$\E[U\mid W,Z_2]=0$ is used.
\end{proof}

\begin{proof}[Proof of Theorem~\ref{thm:process.local}]
Under \eqref{eq:process.local.alternative},
$\varepsilon=U-n^{-1/2}\delta(W)$. Instrument validity implies
\[
 m_n(z)=\E[\varepsilon\ind(Z\le z)]
       =-\frac1{\sqrt n}\E[\delta(W)\ind(Z\le z)].
\]
Since $\E[\varepsilon\mid W]=0$, this is also the mean of
$\varepsilon\{\ind(Z\le z)-\ind(Z_1\le z_1)F_{Z_2\mid W}(z_2\mid W)\}$.
The conditional-indicator function is the same at every $n$ because
the covariate distribution is fixed. The uniform assumptions make
the remainders in Propositions~\ref{prop:density} and
\ref{prop:linear} uniform along this sequence.
Centering \eqref{eq:observed.expansion} therefore gives
\[
 \begin{aligned}
 \sqrt n\{\hat T_n(z)-m_n(z)\}
 ={}&\frac1{\sqrt n}\sum_{i=1}^n\Big[
       \varepsilon_i\{\ind(Z_i\le z)\\
 &\qquad-\ind(Z_{1i}\le z_1)F_{Z_2\mid W}(z_2\mid W_i)\}-m_n(z)\Big]+o_p(1),
 \end{aligned}
\]
uniformly in $z$, under $H_{1n}$. The summands in square brackets
are centered under the $n$th alternative.

For any $z,z'$, their covariance is
\[
 \begin{aligned}
 &\E\Big[\varepsilon^2
       \{\ind(Z\le z)-\ind(Z_1\le z_1)F_{Z_2\mid W}(z_2\mid W)\}\\
 &\qquad\quad\times
       \{\ind(Z\le z')-\ind(Z_1\le z_1')F_{Z_2\mid W}(z_2'\mid W)\}\Big]
       -m_n(z)m_n(z').
 \end{aligned}
\]
Each indicator difference is bounded by one. The second-moment
stability condition therefore bounds the difference between the
first term and its null counterpart uniformly in $z,z'$ by
\[
 \E\left|\E[\varepsilon^2\mid W,Z_2]
              -\E[U^2\mid W,Z_2]\right|=o(1),
\]
where the two conditional moments are evaluated under the law for the
$n$th sample and the limiting null law, respectively.
Also $\sup_z|m_n(z)|\le n^{-1/2}\E|\delta(W)|$.
Thus the centered covariances converge uniformly to the null
covariance \eqref{eq:covariance}, evaluated under the limiting null law.
Together with $\sup_z|m_n(z)|\to0$, these facts verify
Lemma~\ref{lem:residual.process}\textup{(iii)}.
The centered sums therefore converge to $\mathbb G$ in
$\ell^\infty(\Pi)$. Finally,
\[
 \sqrt n\hat T_n(z)
 =\sqrt n\{\hat T_n(z)-m_n(z)\}
       -\E[\delta(W)\ind(Z\le z)],
\]
which yields the stated Pitman limit with its explicit deterministic
shift. If that shift vanished for every $z$, the determining
property of lower orthants would give $\E[\delta(W)\mid Z]=0$.
Under completeness this would imply $\delta(W)=0$ a.s., contrary
to the stipulated nonzero direction.
\end{proof}

\begin{proof}[Proof of Theorem~\ref{thm:process.fixed}]
Under the fixed law, Lemma~\ref{lem:residual.process}\textup{(i)}
applies to $\varepsilon=Y-G(W)$; one must not replace it by $U$.
Centering \eqref{eq:observed.expansion} gives, uniformly in $z$,
\[
 \sqrt n\{\hat T_n(z)-m(z)\}
 =\frac1{\sqrt n}\sum_{i=1}^n\Big[
   \varepsilon_i\{\ind(Z_i\le z)
       -\ind(Z_{1i}\le z_1)F_{Z_2\mid W}(z_2\mid W_i)\}-m(z)\Big]+o_p(1).
\]
The lemma gives the centered Gaussian limit $\mathbb G_P$.
The product of the two residual marks has expectation equal to
the first term in the displayed covariance formula; subtracting
their means gives the term $-m(z)m(z')$. In particular, tightness
of this centered process implies
$\sup_z|\hat T_n(z)-m(z)|=O_p(n^{-1/2})$.

Instrument validity implies
$m(z)=\E[(g(W)-G(W))\ind(Z\le z)]$. If $m$ were identically
zero, the determining property of lower orthants for finite
signed measures would give
$\E[g(W)-G(W)\mid Z]=0$. Completeness would then imply
$g(W)=G(W)$ a.s., contradicting the fixed-alternative assumption.
Thus $\sup_z|m(z)|>0$. The uniform convergence just established
gives $\sup_z|\hat T_n(z)|\to_p\sup_z|m(z)|>0$, and multiplying
by $\sqrt n$ proves the claimed divergence.
\end{proof}

\subsection{Proofs of the bootstrap process limits}\label{app:proof.bootstrap}

We first prove Proposition~\ref{prop:bootstrap.expansion} by
expanding the random denominators and the three U-process terms.
Theorem-specific proofs then start from the actual bootstrap
construction, use that proposition, and apply
Lemma~\ref{lem:residual.process} after identifying the appropriate
mean and covariance. The covering and process-tightness arguments
are contained in the lemma and are not repeated here.

For these bootstrap proofs only, abbreviate the population
indicator correction by the scalar weight
\[
 \omega_{i,z}:=\ind(Z_i\le z)
       -\ind(Z_{1i}\le z_1)F_{Z_2\mid W}(z_2\mid W_i).
\]
This is a shorthand for the displayed difference, not a
projection operator; in particular, $|\omega_{i,z}|\le1$.

\subsubsection{Proof of Proposition~\ref{prop:bootstrap.expansion}}
\label{app:proof.bootstrap.expansion}

Start with the exact decomposition
\begin{equation}\label{eq:bootstrap.plugin.split}
 \begin{aligned}
 &\sqrt n\hat T_n^*(z)
 -\frac1{\sqrt n}\sum_i V_i\varepsilon_i
       \{\ind(Z_i\le z)-\ind(Z_{1i}\le z_1)F_{Z_2\mid W}(z_2\mid W_i)\}\\
 ={}&\frac1{\sqrt n}\sum_iV_i
       \{G(W_i)-\hat G_i\}
       \{\ind(Z_i\le z)-\ind(Z_{1i}\le z_1)\hat F_i(z_2)\}\\
 &-\frac1{\sqrt n}\sum_iV_i\varepsilon_i\ind(Z_{1i}\le z_1)
       \{\hat F_i(z_2)-F(z_2\mid W_i)\}.
 \end{aligned}
\end{equation}
Split the first term once more, replacing its fitted indicator
weight by the population weight plus their difference. On the
unscaled $n^{-1}$ level, this gives three terms:
\[
 \begin{aligned}
 \hat T_n^*(z)-\frac1n\sum_iV_i\varepsilon_i
                 \omega_{i,z}
 =&\frac1n\sum_iV_i\{G(W_i)-\hat G_i\}
                  \omega_{i,z}\\
 &+\frac1n\sum_iV_i\{G(W_i)-\hat G_i\}
          \ind(Z_{1i}\le z_1)\{F(z_2\mid W_i)-\hat F_i(z_2)\}\\
 &+\frac1n\sum_iV_i\varepsilon_i
          \ind(Z_{1i}\le z_1)\{F(z_2\mid W_i)-\hat F_i(z_2)\}.
 \end{aligned}
\]
We prove that each of the three unscaled terms is
$o_p(n^{-1/2})$, uniformly in $z$. We work initially under the
joint law of the data and multipliers. Apply the Hoeffding
operators defined before Lemma~\ref{lem:maximal} to the independent augmented
observations $(\chi_i,V_i)$, integrating over both coordinates.
No new normalization of $U_{n,d}$ is used. Within its kernels,
$i,j,k,l$ denote distinct argument slots.

The factor $V_i$ is always attached to the center $i$.
Every kernel below has mean zero. Moreover, a marginal that
integrates out this center is exactly zero, because
$\E[V_i]=0$ and the other factors do not involve any multiplier.
This fact must be distinguished from conditional mean
restrictions on the outcome: we never assume
$\E[\varepsilon\mid W,Z_2]=0$.
Multiplication by a multiplier preserves the VC bounds relative
to the corresponding absolute envelope. Its independent finite
second moment preserves all envelope orders used below.
Thus Lemma~\ref{lem:maximal} applies on the augmented
observation space. Population conditional-indicator factors
also preserve these bounds by the weighted integration
argument in that lemma. There is no fitted ratio left inside
a kernel to which the lemma is applied.

\paragraph{First term: regression estimation with the population weight.}

Use $G(W_i)-\hat G_i=N_i/\hat f_i-\varepsilon_i$ and
apply \eqref{eq:reciprocal.cubic} with
$\Delta_i=D_i+b_h(W_i)$. Uniformly in $z$, the expansion is
\[
 \begin{aligned}
 \frac1n\sum_i\frac{V_iN_i}{\hat f_i}
                       \omega_{i,z}
 =
 &U_{n,2}\left(
 \frac{V_iK_{ij}(Y_i-Y_j)}{f_i}
                   \omega_{i,z}\right)-\frac1{n(n-1)^2}
       \sum_i\sum_{j\ne i}\sum_{k\ne i}
 \frac{V_iK_{ij}(Y_i-Y_j)
            \omega_{i,z}d_{ik}}{f_i^2}\\
 &+\frac1{n(n-1)^3}
       \sum_i\sum_{j\ne i}\sum_{k\ne i}\sum_{l\ne i}
 \frac{V_iK_{ij}(Y_i-Y_j)
            \omega_{i,z}d_{ik}d_{il}}{f_i^3}+O_p(h^s+a_n^3).
 \end{aligned}
\]
The error here is justified by absolute averages, not by
multiplying a supremum error by a signed process order.
In fact, the argument of Lemma~\ref{lem:kernel}, independence
of the multipliers, and the conditional moment bounds give
\[
 \frac1n\sum_i|V_i||N_i|
 +\frac1{n(n-1)}\sum_{i\ne j}|V_i\varepsilon_i||K_{ij}|
 =O_p(1).
\]
Consequently, the cubic remainder is $O_p(a_n^3)$.
Replacing $\Delta_i$ by $D_i$ in the linear correction costs
$O_p(h^s)$; replacing $\Delta_i^2$ by $D_i^2$ in the
quadratic correction costs $O_p(h^sa_n)$, already included
in the displayed error. These bounds use
$\max_i|D_i|+\max_i|\Delta_i|=O_p(a_n)$ and
$\|b_h\|_\infty=O(h^s)$ on the common positive-denominator
event.

The linear correction has the exact index partition
\begin{equation}\label{eq:bootstrap.linear.partition}
 \begin{aligned}
 &\frac1{n(n-1)^2}
       \sum_i\sum_{j\ne i}\sum_{k\ne i}
 \frac{V_iK_{ij}(Y_i-Y_j)
               \omega_{i,z}d_{ik}}{f_i^2}\\
 ={}&\frac{n-2}{n-1}U_{n,3}\left(
 \frac{V_iK_{ij}(Y_i-Y_j)
               \omega_{i,z}d_{ik}}{f_i^2}\right)
               +\frac1{n-1}U_{n,2}\left(
 \frac{V_iK_{ij}(Y_i-Y_j)
               \omega_{i,z}d_{ij}}{f_i^2}\right).
 \end{aligned}
\end{equation}
The distinct-index kernel has zero first-order projection:
if the center is retained, integration of the distinct
$d_{ik}$ gives zero; if the center is integrated out,
integration of $V_i$ gives zero. Its remaining canonical
degrees are two and three. The repeated-index kernel
need not have a zero first-order projection, and is
controlled with its coefficient $1/(n-1)$.

For the quadratic correction, use exactly the four
partitions in \eqref{eq:Q.partition}, with
$A_{ij,z}$ replaced by $V_iK_{ij}(Y_i-Y_j)\omega_{i,z}$.
Thus the exact coefficients of the all-distinct,
$k=l\ne j$, $j=k\ne l$ or $j=l\ne k$, and
$j=k=l$ averages are, respectively,
\[
 \frac{(n-2)(n-3)}{(n-1)^2},\qquad
 \frac{n-2}{(n-1)^2},\qquad
 \frac{2(n-2)}{(n-1)^2},\qquad
 \frac1{(n-1)^2}.
\]
The all-distinct order-four kernel has zero projections
of degrees one and two: retaining the center and only
one other argument still leaves at least one distinct
centered $d$ factor to integrate out.

For reference, the envelope orders needed in the exact
Hoeffding decompositions are as follows. The deterministic
coefficient column gives its order in $n$; constants such
as the coefficient two in the preceding display are retained
in the exact partitions above.
\begin{center}
\small
\begin{tabular}{llll}
\toprule
Correction and pattern & Coefficient & Degrees & $L^2$ envelope orders\\
\midrule
Linear, $j\ne k$ & $1$ & $2,3$
 & $h^{-r/2},\,h^{-r}$\\
Linear, $j=k$ & $n^{-1}$ & $1,2$
 & $h^{-r},\,h^{-3r/2}$\\
Quadratic, all distinct & $1$ & $3,4$
 & $h^{-r},\,h^{-3r/2}$\\
Quadratic, $k=l\ne j$ & $n^{-1}$ & $1,2,3$
 & $h^{-r},\,h^{-3r/2},\,h^{-2r}$\\
Quadratic, $j=k\ne l$ or $j=l\ne k$
 & $n^{-1}$ & $2,3$ & $h^{-3r/2},\,h^{-2r}$\\
Quadratic, $j=k=l$ & $n^{-2}$ & $1,2$
 & $h^{-2r},\,h^{-5r/2}$\\
\bottomrule
\end{tabular}
\end{center}
These are envelope bounds, not just bounds on individual
kernel variances. To check them, each independent local
kernel factor has $L^2$ cost $h^{-r/2}$; squaring a
kernel on a repeated edge has cost $h^{-3r/2}$, and
cubing it has cost $h^{-5r/2}$.
For example, retaining the center and one neighbor in
the all-distinct linear kernel leaves a single local
$d$ factor, while its full kernel contains two local
factors. In the $k=l\ne j$ quadratic kernel, integration
of $d_{ik}^2$ costs $h^{-r}$; retaining both arguments
of that squared factor has $L^2$ cost $h^{-3r/2}$.
In the $j=k=l$ kernel, the full local factor is cubic.
Bounded conditional outcome moments and
$\E V_i^2=1$ control its marks. Conditional integration
and finite canonical subtractions preserve these bounds,
as in the proof of Lemma~\ref{lem:maximal}.

Lemma~\ref{lem:maximal} bounds the linear and quadratic
corrections, uniformly in $z$, by
\[
 \begin{aligned}
 O_p\left\{
 \frac1{nh^{r/2}}+\frac1{n^{3/2}h^r}
                         +\frac1{n^2h^{3r/2}}\right\}
 &=
 O_p\!\left[n^{-1/2}
       \{H_n^{-1/2}+H_n^{-1}+H_n^{-3/2}\}\right],\\
 O_p\left\{
 \frac1{n^{3/2}h^r}+\frac1{n^2h^{3r/2}}
 +\frac1{n^{5/2}h^{2r}}+\frac1{n^3h^{5r/2}}\right\}
 &=
 O_p\!\left[n^{-1/2}
       \{H_n^{-1}+H_n^{-3/2}+H_n^{-2}+H_n^{-5/2}\}\right].
 \end{aligned}
\]
Here $H_n=nh^r$ is the notation already used in the proof
of Proposition~\ref{prop:density}. Hence both corrections
are $o_p(n^{-1/2})$ if $H_n\to\infty$.
In particular, no repeated-index contribution has been
discarded on the assumption that its mean or first
projection vanishes.

It remains to identify the main term of the leading
order-two process. Its only nonzero first-order marginal
retains the center and is
\[
 V_i\omega_{i,z}
 \frac{f_h(W_i)Y_i-\E[K_h(W_i,W)Y]}{f_i}.
\]
The other marginal integrates out $V_i$ and is zero.
The displayed marginal differs from
$V_i\varepsilon_i\omega_{i,z}$ by
$O(h^s)|V_i|(1+|Y_i|)$, uniformly in $z$.
The canonical degree-two remainder has envelope order
$h^{-r/2}$ and is $O_p(n^{-1}h^{-r/2})$.
Thus
\[
 \begin{aligned}
 &U_{n,2}\left(
 \frac{V_iK_{ij}(Y_i-Y_j)}{f_i}
                 \omega_{i,z}\right)
 -\frac1n\sum_iV_i\varepsilon_i\omega_{i,z}=O_p\{h^s+n^{-1}h^{-r/2}\}=o_p(n^{-1/2}).
 \end{aligned}
\]
Combining the denominator expansion and Hoeffding calculations gives
\[
 \frac1n\sum_i\frac{V_iN_i}{\hat f_i}
                    \omega_{i,z}
 -\frac1n\sum_iV_i\varepsilon_i\omega_{i,z}
 =o_p(n^{-1/2}).
\]
The second sum is exactly the explicit term subtracted
when $G(W_i)-\hat G_i=N_i/\hat f_i-\varepsilon_i$ is
inserted into the first contribution. Its two main
terms therefore cancel, proving
\[
 \sup_z\left|\frac1n\sum_iV_i\{G(W_i)-\hat G_i\}
                    \omega_{i,z}\right|
 =o_p(n^{-1/2}).
\]
The conditions used here are
$H_n\to\infty$, $\sqrt n h^s\to0$, and
$\sqrt n a_n^3\to0$, all included in Assumption~\ref{A4}.

\paragraph{Second term: the product of the two estimation errors.}

Write both ratios out before removing the denominator:
\[
 \begin{aligned}
 &\frac1n\sum_iV_i\{G(W_i)-\hat G_i\}\ind(Z_{1i}\le z_1)
                \{F(z_2\mid W_i)-\hat F_i(z_2)\}\\
 ={}&\frac1{n(n-1)^2}
 \sum_i\sum_{j\ne i}\sum_{k\ne i}
 \frac{V_i\ind(Z_{1i}\le z_1)K_{ij}K_{ik}}{\hat f_i^2}
       \{G(W_i)-Y_j\}
       \{F(z_2\mid W_i)-\ind(Z_{2k}\le z_2)\}.
 \end{aligned}
\]
The squared reciprocal has the following exact expansion,
obtained from the same identity as
\eqref{eq:reciprocal.cubic}:
\[
 \frac1{\hat f_i^2}
 =\frac1{f_i^2}-\frac{2\Delta_i}{f_i^3}
      +\frac{3\Delta_i^2}{f_i^4}
      -\frac{\Delta_i^3(4f_i+3\Delta_i)}
                   {f_i^4\hat f_i^2}.
\]
The two numerator averages satisfy
\[
 \begin{aligned}
 \max_i\left|
 \frac1{n-1}\sum_{j\ne i}K_{ij}\{G(W_i)-Y_j\}\right|
 &=O_p(a_n),\\
 \sup_{i,z_2}\left|
 \frac1{n-1}\sum_{k\ne i}K_{ik}
       \{F(z_2\mid W_i)-\ind(Z_{2k}\le z_2)\}\right|
 &=O_p(a_n).
 \end{aligned}
\]
These are exact numerator identities:
\[
 \begin{aligned}
 \frac1{n-1}\sum_{j\ne i}K_{ij}\{G(W_i)-Y_j\}
 &=\hat f_i\{G(W_i)-\hat G_i\},\\[3pt]
 \frac1{n-1}\sum_{k\ne i}K_{ik}
       \{F(z_2\mid W_i)-\ind(Z_{2k}\le z_2)\}
 &=\hat f_i\{F(z_2\mid W_i)-\hat F_i(z_2)\}.
 \end{aligned}
\]
Their rates follow from Lemma~\ref{lem:kernel}
and $\max_i\hat f_i=O_p(1)$.

Consequently, the three denominator corrections, uniformly in $z$,
are explicitly
\[
 \begin{aligned}
 S_{n,1}(z)
 &:=-\frac2n\sum_iV_i\ind(Z_{1i}\le z_1)
       \frac{\Delta_i}{f_i^3}
       \left\{\frac1{n-1}\sum_{j\ne i}K_{ij}
                    \{G(W_i)-Y_j\}\right\}\\
 &\hspace{38mm}\times
       \left\{\frac1{n-1}\sum_{k\ne i}K_{ik}
          \{F(z_2\mid W_i)-\ind(Z_{2k}\le z_2)\}\right\}
   =O_p(a_n^3),\\
 S_{n,2}(z)
 &:=\frac3n\sum_iV_i\ind(Z_{1i}\le z_1)
       \frac{\Delta_i^2}{f_i^4}
       \left\{\frac1{n-1}\sum_{j\ne i}K_{ij}
                    \{G(W_i)-Y_j\}\right\}\\
 &\hspace{38mm}\times
       \left\{\frac1{n-1}\sum_{k\ne i}K_{ik}
          \{F(z_2\mid W_i)-\ind(Z_{2k}\le z_2)\}\right\}
   =O_p(a_n^4),\\
 S_{n,3}(z)
 &:=-\frac1n\sum_iV_i\ind(Z_{1i}\le z_1)
       \frac{\Delta_i^3(4f_i+3\Delta_i)}
             {f_i^4\hat f_i^2}
       \left\{\frac1{n-1}\sum_{j\ne i}K_{ij}
                    \{G(W_i)-Y_j\}\right\}\\
 &\hspace{38mm}\times
       \left\{\frac1{n-1}\sum_{k\ne i}K_{ik}
          \{F(z_2\mid W_i)-\ind(Z_{2k}\le z_2)\}\right\}
   =O_p(a_n^5).
 \end{aligned}
\]
Indeed, these bounds use the two $O_p(a_n)$ numerator factors,
$\max_i|\Delta_i|^d=O_p(a_n^d)$, the common
positive-denominator event, and
$n^{-1}\sum_i|V_i|=O_p(1)$.
No signed-process order is used. All three corrections
are $o_p(n^{-1/2})$ under
$\sqrt n a_n^3\to0$.

The population-denominator sum has the exact partition
\[
 \begin{aligned}
 &\frac1{n(n-1)^2}
 \sum_i\sum_{j\ne i}\sum_{k\ne i}
 \frac{V_i\ind(Z_{1i}\le z_1)K_{ij}K_{ik}}{f_i^2}
       \{G(W_i)-Y_j\}
       \{F(z_2\mid W_i)-\ind(Z_{2k}\le z_2)\}\\
 ={}&\frac{n-2}{n-1}U_{n,3}\left(
 \frac{V_i\ind(Z_{1i}\le z_1)K_{ij}K_{ik}}{f_i^2}
       \{G(W_i)-Y_j\}
       \{F(z_2\mid W_i)-\ind(Z_{2k}\le z_2)\}\right)\\
 &+\frac1{n-1}U_{n,2}\left(
 \frac{V_i\ind(Z_{1i}\le z_1)K_{ij}^2}{f_i^2}
       \{G(W_i)-Y_j\}
       \{F(z_2\mid W_i)-\ind(Z_{2j}\le z_2)\}\right).
 \end{aligned}
\]
Only the first-order marginal retaining the center can be nonzero:
\[
 \begin{aligned}
 \frac{V_i\ind(Z_{1i}\le z_1)}{f_i^2}
 &\E[K_h(W_i,W)\{G(W_i)-Y\}\mid W_i]\\
 {}\times&
 \E[K_h(W_i,W)\{F(z_2\mid W_i)
                       -\ind(Z_2\le z_2)\}\mid W_i].
 \end{aligned}
\]
Both expectations are uniformly $O(h^s)$ by the kernel
moment conditions and Assumption~\ref{A2}.
Thus the sample average of this marginal is
$O_p(h^{2s})$, using the absolute average of $|V_i|$.
The same cancellation can be seen by expanding the product in this
order-three kernel into the four terms
\[
 G(W_i)F(z_2\mid W_i),\quad
 -G(W_i)\ind(Z_{2k}\le z_2),\quad
 -Y_jF(z_2\mid W_i),\quad
 Y_j\ind(Z_{2k}\le z_2).
\]
Their first-order projections each have the same leading
sample average
\[
 \frac1n\sum_i V_iG(W_i)\ind(Z_{1i}\le z_1)
                                      F(z_2\mid W_i),
\]
with signs $+,-,-,+$. Those leading terms cancel.
The product of the two kernel biases displayed above is therefore
the remaining first-order marginal, which also gives its
$O_p(h^{2s})$ bound.

The remaining canonical degrees two and three have envelope
orders at most $h^{-r/2}$ and $h^{-r}$.
Indeed, any nonzero two-coordinate marginal retains the
center; integration of the other neighbor leaves only
one local kernel and a bounded integrated mark.
The full kernel contains two local factors.
Lemma~\ref{lem:maximal} bounds these contributions by
\[
 O_p\{n^{-1}h^{-r/2}+n^{-3/2}h^{-r}\}.
\]
For the repeated-neighbor kernel $j=k$, the first-order
marginal retaining $i$ is
\[
 \begin{aligned}
 \frac{V_i\ind(Z_{1i}\le z_1)}{f_i^2}
 \E\Big[ K_h(W_i,W)^2\{G(W_i)-Y\}
       \{F(z_2\mid W_i)-\ind(Z_2\le z_2)\}
                       \,\Big|\,W_i\Big].
 \end{aligned}
\]
It need not vanish under $H_0$.
Its envelope order is $h^{-r}$, and that of the canonical
degree-two kernel is $h^{-3r/2}$. After multiplication
by $1/(n-1)$ their orders are, respectively,
$n^{-3/2}h^{-r}$ and $n^{-2}h^{-3r/2}$.
This is why the repeated-neighbor contribution is
negligible under $nh^r\to\infty$ without imposing the
stronger condition $\sqrt n h^r\to\infty$.

Together these calculations prove
\[
 \begin{aligned}
 &\sup_z\left|\frac1n\sum_iV_i\{G(W_i)-\hat G_i\}
       \ind(Z_{1i}\le z_1)\{F(z_2\mid W_i)-\hat F_i(z_2)\}\right|\\
 =&
 O_p\!\left[
 h^{2s}+a_n^3+
 n^{-1/2}\{H_n^{-1/2}+H_n^{-1}+H_n^{-3/2}\}\right]
 =o_p(n^{-1/2}).
 \end{aligned}
\]
The conditions are $\sqrt n h^{2s}\to0$,
$\sqrt n a_n^3\to0$, and $H_n\to\infty$;
the first follows from $\sqrt n h^s\to0$ and $h\to0$.

\paragraph{Third term: conditional-distribution estimation.}

Write out its two parts:
\[
 \begin{aligned}
 &\frac1n\sum_iV_i\varepsilon_i\ind(Z_{1i}\le z_1)
                    \{F(z_2\mid W_i)-\hat F_i(z_2)\}\\
 ={}&\frac1n\sum_iV_i\varepsilon_i\ind(Z_{1i}\le z_1)
                                      F(z_2\mid W_i)
                                      -\frac1{n(n-1)}\sum_{i\ne j}
       \frac{V_i\varepsilon_i\ind(Z_{1i}\le z_1)
                     K_{ij}\ind(Z_{2j}\le z_2)}{\hat f_i}.
 \end{aligned}
\]
Apply \eqref{eq:reciprocal.cubic} to the last denominator.
The resulting decomposition is
\[
 \begin{aligned}
 &\frac1{n(n-1)}\sum_{i\ne j}
       \frac{V_i\varepsilon_i\ind(Z_{1i}\le z_1)
                    K_{ij}\ind(Z_{2j}\le z_2)}{\hat f_i}\\
 ={}&U_{n,2}\left(
       \frac{V_i\varepsilon_i\ind(Z_{1i}\le z_1)
                     K_{ij}\ind(Z_{2j}\le z_2)}{f_i}\right)
   +C^F_{n,1}(z)+C^F_{n,2}(z)+C^F_{n,3}(z),
 \end{aligned}
\]
where the linear correction is
\[
 \begin{aligned}
 C^F_{n,1}(z)
 &:=-\frac1{n(n-1)}\sum_{i\ne j}
       \frac{V_i\varepsilon_i\ind(Z_{1i}\le z_1)
                    K_{ij}\ind(Z_{2j}\le z_2)\Delta_i}{f_i^2}\\
 &=-\frac1{n(n-1)^2}
       \sum_i\sum_{j\ne i}\sum_{k\ne i}
       \frac{V_i\varepsilon_i\ind(Z_{1i}\le z_1)
                    K_{ij}\ind(Z_{2j}\le z_2)d_{ik}}{f_i^2}
       +O_p(h^s).
 \end{aligned}
\]
The $O_p(h^s)$ term is the contribution of
$b_h(W_i)$ in $\Delta_i=D_i+b_h(W_i)$: it is bounded by
$\|b_h\|_\infty=O(h^s)$ times the corresponding absolute
kernel average, which is $O_p(1)$.  The $D_i$ part has exactly
the distinct- and repeated-index partition in
\eqref{eq:bootstrap.linear.partition}, after the mark replacement
\[
 (Y_i-Y_j)\omega_{i,z}
 \quad\longmapsto\quad
 \varepsilon_i\ind(Z_{1i}\le z_1)\ind(Z_{2j}\le z_2),
\]
which preserves the zero-projection statements and envelope orders
used in the $U_{n,3}+U_{n,2}$ decomposition following
\eqref{eq:bootstrap.linear.partition}.  Applying that same
decomposition and the same maximal bounds therefore gives
\[
 \sup_z|C^F_{n,1}(z)|
 =O_p\!\left[
 h^s+n^{-1/2}
 \{H_n^{-1/2}+H_n^{-1}+H_n^{-3/2}\}\right].
\]

The quadratic correction is
\[
 \begin{aligned}
 C^F_{n,2}(z)
 &:=\frac1{n(n-1)}\sum_{i\ne j}
       \frac{V_i\varepsilon_i\ind(Z_{1i}\le z_1)
                    K_{ij}\ind(Z_{2j}\le z_2)\Delta_i^2}{f_i^3}\\
 &=\frac1{n(n-1)^3}
       \sum_i\sum_{j\ne i}\sum_{k\ne i}\sum_{l\ne i}
       \frac{V_i\varepsilon_i\ind(Z_{1i}\le z_1)
                    K_{ij}\ind(Z_{2j}\le z_2)d_{ik}d_{il}}{f_i^3}
       +O_p(h^sa_n).
 \end{aligned}
\]
Indeed,
$\Delta_i^2-D_i^2=2D_i b_h(W_i)+b_h(W_i)^2$ is uniformly
$O_p(h^sa_n)$.  Applying the four index partitions in
\eqref{eq:Q.partition}, with
\[
 A_{ij,z}\quad\hbox{replaced by}\quad
 V_i\varepsilon_i\ind(Z_{1i}\le z_1)
 K_{ij}\ind(Z_{2j}\le z_2),
\]
gives exactly the same $U_{n,4}$, $U_{n,3}$, and $U_{n,2}$
decomposition.  The projection and envelope calculation following
\eqref{eq:Q.partition} therefore yields
\[
 \sup_z|C^F_{n,2}(z)|
 =O_p\!\left[
 h^sa_n+n^{-1/2}
 \{H_n^{-1}+H_n^{-3/2}+H_n^{-2}+H_n^{-5/2}\}\right].
\]

Finally, the cubic reciprocal remainder is
\[
 \begin{aligned}
 C^F_{n,3}(z)
 &:=-\frac1{n(n-1)}\sum_{i\ne j}
       \frac{V_i\varepsilon_i\ind(Z_{1i}\le z_1)
             K_{ij}\ind(Z_{2j}\le z_2)\Delta_i^3}
            {f_i^3\hat f_i},\quad \sup_z|C^F_{n,3}(z)|&=O_p(a_n^3).
 \end{aligned}
\]
The last bound follows from $\max_i|\Delta_i|=O_p(a_n)$,
the common positive-denominator event, and the $O_p(1)$
absolute kernel average.  Hence all three corrections are
$o_p(n^{-1/2})$ under $H_n\to\infty$,
$\sqrt n h^s\to0$, and $\sqrt n a_n^3\to0$.

The leading order-two kernel has only the following
nonzero first-order marginal:
\[
 \frac{V_i\varepsilon_i\ind(Z_{1i}\le z_1)}{f_i}
       \E[K_h(W_i,W)\ind(Z_2\le z_2)\mid W_i].
\]
Smoothness gives
\[
 \E[K_h(W_i,W)\ind(Z_2\le z_2)\mid W_i]
       =f_iF(z_2\mid W_i)+O(h^s)
\]
uniformly in $(i,z_2)$. The sample average of the projection
error is therefore $O_p(h^s)$, using
$n^{-1}\sum_i|V_i\varepsilon_i|=O_p(1)$.
The canonical degree-two remainder has envelope order
$h^{-r/2}$ and is $O_p(n^{-1}h^{-r/2})$.
It follows that
\[
 \begin{aligned}
 &U_{n,2}\left(
       \frac{V_i\varepsilon_i\ind(Z_{1i}\le z_1)
                     K_{ij}\ind(Z_{2j}\le z_2)}{f_i}\right)
 -\frac1n\sum_iV_i\varepsilon_i\ind(Z_{1i}\le z_1)
                                      F(z_2\mid W_i)\\
 =&O_p\{h^s+n^{-1}h^{-r/2}\}=o_p(n^{-1/2}).
 \end{aligned}
\]
The main term cancels the explicit first sum in the
third contribution. Including its denominator corrections
therefore yields
\[
 \sup_z\left|\frac1n\sum_iV_i\varepsilon_i\ind(Z_{1i}\le z_1)
                    \{F(z_2\mid W_i)-\hat F_i(z_2)\}\right|
 =o_p(n^{-1/2}).
\]
Again, the sufficient conditions are
$H_n\to\infty$, $\sqrt n h^s\to0$, and
$\sqrt n a_n^3\to0$.

\paragraph{Combining the three conclusions and conditioning.}

The preceding three expansions identify the main terms
before cancellation and bound every remaining projection,
repeated-index contribution, and denominator remainder.
They prove the uniform $o_p(n^{-1/2})$ reduction under the
joint law. Along the Pitman sequence, all calculations use
$G_n$ and uniform Assumptions~\ref{A1}--\ref{A5};
second-moment stability is not needed for this reduction.
No step replaces
$\varepsilon$ by $U$ outside $H_0$.

For each $\eta>0$, joint convergence implies that the
expectation over the data of the conditional probability
of a scaled error exceeding $\eta$ tends to zero.
Markov's inequality for this nonnegative conditional
probability then gives
\begin{equation}\label{eq:bootstrap.plugin}
 \sup_{z\in\Pi}\left|
 \sqrt n\hat T_n^*(z)
 -\frac1{\sqrt n}\sum_iV_i\varepsilon_i
       \{\ind(Z_i\le z)-\ind(Z_{1i}\le z_1)F_{Z_2\mid W}(z_2\mid W_i)\}
 \right|=o_{P^*}(1)
\end{equation}
in probability over the data. This conditioning step
does not assume independence among overlapping
leave-one-out estimators. Dividing by $\sqrt n$ proves
\eqref{eq:bootstrap.expansion}, including its Pitman-sequence version.

\subsubsection{Proof of Theorem~\ref{thm:null}}

\begin{proof}
Under $H_0$, $\varepsilon_i=U_i$. Starting from the construction
\eqref{eq:mb}, Proposition~\ref{prop:bootstrap.expansion} gives
\[
 \begin{aligned}
 \sqrt n\hat T_n^*(z)
 &=\frac1{\sqrt n}\sum_iV_i\hat\varepsilon_i
       \{\ind(Z_i\le z)-\ind(Z_{1i}\le z_1)\hat F_i(z_2)\}\\
 &=\frac1{\sqrt n}\sum_iV_iU_i
       \{\ind(Z_i\le z)-\ind(Z_{1i}\le z_1)F_{Z_2\mid W}(z_2\mid W_i)\}
       +o_{P^*}(1).
 \end{aligned}
\]
For general residuals, the conditional covariance of the
population-mark multiplier averages is
\begin{equation}\label{eq:bootstrap.oracle.covariance}
 \begin{aligned}
 \frac1n\sum_i\varepsilon_i^2
       \{\ind(Z_i\le z)-\ind(Z_{1i}\le z_1)F_{Z_2\mid W}(z_2\mid W_i)\}
       \{\ind(Z_i\le z')-\ind(Z_{1i}\le z_1')F_{Z_2\mid W}(z_2'\mid W_i)\}.
 \end{aligned}
\end{equation}
Under $H_0$, Lemma~\ref{lem:residual.process}\textup{(i)}
makes this converge uniformly to $\Sigma(z,z')$ in
\eqref{eq:covariance}. Part \textup{(ii)} of that lemma
and conditional Slutsky therefore yield
\[
 \frac1{\sqrt n}\sum_iV_iU_i\omega_{i,z}
       \Rightarrow_{P^*}\mathbb G,
 \qquad
 \sqrt n\hat T_n^*\Rightarrow_{P^*}\mathbb G.
\]
Thus the limit is the same as the observed null process,
without imposing $\E[U\mid W,Z_2]=0$.

For the direct residual-mark multiplier construction described in
Section~\ref{sec:bootstrap.construction}, use
the first-term argument of Proposition~\ref{prop:bootstrap.expansion}
with weight $\ind(Z_i\le z)$ instead of the corrected
weight $\omega_{i,z}$. It gives
\[
 \sqrt n\hat T_n^{*,\mathrm{naive}}(z)
 =\frac1{\sqrt n}\sum_iV_iU_i\ind(Z_i\le z)+o_{P^*}(1).
\]
The uncorrected version of Lemma~\ref{lem:residual.process}
then gives covariance $\E[U^2\ind(Z\le z)\ind(Z\le z')]$,
proving the mismatch stated in Section~\ref{sec:altboot}.

Applying Lemma~\ref{lem:sample.functionals} conditionally gives the
bootstrap KS and CvM limits based on the sample markers and empirical
CvM measure. Theorem~\ref{thm:process.null} and the same lemma give the
matching observed limits in Remark~\ref{rmk:statistics}. The stated
uniqueness and continuity conditions
give convergence of conditional critical values; independent
simulation with $B\to\infty$ preserves this convergence.
The rejection probability consequently tends to $\alpha$.
\end{proof}

\subsubsection{Proof of Theorem~\ref{thm:local}}

\begin{proof}
Under \eqref{eq:process.local.alternative}, apply
Proposition~\ref{prop:bootstrap.expansion} with $G_n$.
Using $\varepsilon_i=U_i-n^{-1/2}\delta(W_i)$ gives
\[
 \begin{aligned}
 \sqrt n\hat T_n^*(z)
 &=\frac1{\sqrt n}\sum_iV_i\varepsilon_i
                       \omega_{i,z}+o_{P^*}(1)\\
 &=\frac1{\sqrt n}\sum_iV_iU_i\omega_{i,z}
   -\frac1n\sum_iV_i\delta(W_i)\omega_{i,z}
   +o_{P^*}(1).
 \end{aligned}
\]
The bounded-mark assertion in
Lemma~\ref{lem:residual.process}\textup{(ii)} implies
\[
 \sup_z\left|\frac1n\sum_iV_i\delta(W_i)
                    \omega_{i,z}\right|
       =O_{P^*}(n^{-1/2})=o_{P^*}(1).
\]
In particular, there is no deterministic drift in the
bootstrap: every multiplier average is conditionally centered.

To identify its covariance, write $\Sigma$ for
\eqref{eq:covariance} evaluated under the limiting null law.
The stability assumption gives
\[
 \begin{aligned}
 &\sup_{z,z'}\left|
 \E\!\left[\begin{aligned}
       &\varepsilon^2\{\ind(Z\le z)-\ind(Z_1\le z_1)F_{Z_2\mid W}(z_2\mid W)\}\\
       &\quad\times\{\ind(Z\le z')-\ind(Z_1\le z_1')F_{Z_2\mid W}(z_2'\mid W)\}
       \end{aligned}\right]
                      -\Sigma(z,z')\right|\\
 \le&
 \E\left|\E[\varepsilon^2\mid W,Z_2]
                   -\E[U^2\mid W,Z_2]\right|=o(1),
 \end{aligned}
\]
where the row-law and limiting-null conditional moments are understood
as in Theorem~\ref{thm:process.local}.
Also,
\[
 \sup_z|m_n(z)|
 =\frac1{\sqrt n}\sup_z|\E[\delta(W)\ind(Z\le z)]|
 \le\frac{\E|\delta(W)|}{\sqrt n}\to0.
\]
All conditions of Lemma~\ref{lem:residual.process}\textup{(iii)}
are therefore satisfied. It follows that
\[
 \frac1{\sqrt n}\sum_iV_i\varepsilon_i
                 \omega_{i,z}
      \Rightarrow_{P^*}\mathbb G,
 \qquad
 \sqrt n\hat T_n^*\Rightarrow_{P^*}\mathbb G.
\]
In contrast, Theorem~\ref{thm:process.local} gives
$\sqrt n\hat T_n\Rightarrow
 \mathbb G-\E[\delta(W)\ind(Z\le z)]$.
Lemma~\ref{lem:sample.functionals} gives unshifted bootstrap KS and CvM
limits and shifted observed limits for the sample statistics. Quantile convergence
and the no-atom assumptions then give the rejection
probabilities in Theorem~\ref{thm:local}.
The symmetric convex acceptance regions give the stated
Gaussian shift comparison.
\end{proof}

\subsubsection{Proof of Theorem~\ref{thm:consistency}}

\begin{proof}
Under the fixed law retain $\varepsilon=Y-G(W)$ rather
than replacing it by $U$.
The construction and Proposition~\ref{prop:bootstrap.expansion}
give
\[
 \begin{aligned}
 \sqrt n\hat T_n^*(z)
 &=\frac1{\sqrt n}\sum_iV_i\hat\varepsilon_i
       \{\ind(Z_i\le z)-\ind(Z_{1i}\le z_1)\hat F_i(z_2)\}\\
 &=\frac1{\sqrt n}\sum_iV_i\varepsilon_i
                \omega_{i,z}+o_{P^*}(1)\\
 &\Rightarrow_{P^*}\mathbb G_P^*,
 \end{aligned}
\]
where the last step is
Lemma~\ref{lem:residual.process}\textup{(ii)}.
Part \textup{(i)} identifies the uniform limit of
\eqref{eq:bootstrap.oracle.covariance} as
\eqref{eq:bootstrap.fixed.covariance}.
Since the population mark has mean $m(z)$,
\[
 \Cov\{\mathbb G_P^*(z),\mathbb G_P^*(z')\}
   -\Cov\{\mathbb G_P(z),\mathbb G_P(z')\}
       =m(z)m(z').
\]
This establishes the distinction between the two
fixed-alternative Gaussian laws.

Lemma~\ref{lem:sample.functionals} gives the asserted bootstrap
statistic limits, in particular
\[
 (T_n^{*,\rm KS},T_n^{*,\rm CvM})=O_{P^*}(1)
\]
in probability. Their conditional quantiles are consequently
$O_p(1)$, without requiring continuity of the alternative
limit distributions; simulated quantiles are tight when
$B\to\infty$. Meanwhile, completeness and a fixed alternative
give, by Remark~\ref{rmk:statistics},
\[
 \frac{T_n^{\rm KS}}{\sqrt n}\to_p\sup_{z\in\Pi}|m(z)|>0,
 \qquad
 \frac{T_n^{\rm CvM}}n\to_p\int_\Pi m(z)^2\,dF_Z(z)>0.
\]
Each observed statistic therefore exceeds its tight
bootstrap critical value with probability tending to one.
\end{proof}

\clearpage
\bibliographystyle{ecta}
\bibliography{robust_testing}
\end{document}